\newif\ifFull
\Fulltrue

\ifFull
\documentclass[11pt]{article}
\else
\documentclass[10pt]{IEEEtran}
\usepackage{times}

\fi
\usepackage{paralist,amssymb}
\usepackage[ruled]{algorithm2e}
\usepackage{color}

\ifFull
\usepackage[margin=1in]{geometry}
\fi
\usepackage{multirow}
\usepackage{graphicx}

\newcommand{\ignore}[1]{}
\newtheorem{definition}{Definition}
\newtheorem{theorem}{Theorem}

\ifFull
\else
\fi

\begin{document}

\title{Secure Fingerprint Alignment and Matching
}

\ifFull
\else
\fi

\ifFull
\author{Fattaneh Bayatbabolghani\\Computer Science and Engineering\\
  University of Notre Dame\\ fbayatba@nd.edu
\and Marina Blanton\\ Computer Science and Engineering\\ University at
  Buffalo (SUNY)\\ mblanton@buffalo.edu
\and Mehrdad Aliasgari\\ Computer Engineering and Computer Science\\
  California State University, Long Beach \\ mehrdad.aliasgari@csulb.edu
\and Michael Goodrich\\ Computer Science\\ University of California,
  Irvine\\ goodrich@uci.edu}
\else
\author{\IEEEauthorblockN{Fattaneh Bayatbabolghani\IEEEauthorrefmark{1},
Marina Blanton\IEEEauthorrefmark{2}, Mehrdad Aliasgari\IEEEauthorrefmark{3} and
Michael Goodrich\IEEEauthorrefmark{4}}\\
\IEEEauthorblockA{\IEEEauthorrefmark{1}Computer Science and Engineering, University of Notre Dame, Email: fbayatba@nd.edu\\
\IEEEauthorrefmark{2}Computer Science and Engineering, University at Buffalo (SUNY), Email: mblanton@buffalo.edu\\
\IEEEauthorrefmark{3}Computer Engineering and Computer Science, California State University, Long Beach, Email: mehrdad.aliasgari@csulb.edu\\
\IEEEauthorrefmark{4}Computer Science, University of California, Irvine, Email: goodrich@uci.edu}}
\fi


\maketitle


\begin{abstract}
We present three private fingerprint alignment and matching protocols,
based on precise and 
efficient fingerprint recognition
algorithms 
that use \emph{minutia} points. Our protocols allow two or more semi-honest
parties to compare privately-held fingerprints in a secure way such that
nothing more than an accurate score of how well the fingerprints match is
revealed to output recipients. To the best of our knowledge, this is the
first time fingerprint alignment based on minutiae is considered in a secure
computation framework. We build secure fingerprint alignment and matching
protocols in both the two-party setting using garbled circuits and in the
multi-party setting using secret sharing. In addition to providing precise
and efficient secure fingerprint comparisons, our contributions include the
design of a number of secure sub-protocols for complex operations such as
sine/cosine, arctangent, and selection, which are likely to be
of independent interest.
\end{abstract}

\ifFull
\else

\maketitle
\fi

\section{Introduction} 

Computing securely with biometric data is challenging because
biometric
identification applications often require accurate metrics and recognition
algorithms, but the data involved is so sensitive that if it is
ever revealed or stolen the victim may be vulnerable to impersonation
attacks for the rest of their life.
This risk is particularly true for fingerprints, which have been used
since the 19th century for identification purposes and are being used
for such purposes ubiquitously today, including for cellphones,
laptops, digital storage devices, 
safes, immigration, and physical building/room access, as well as the classic
application of identifying criminals.
Thus, there is a strong motivation for fast and secure fingerprint
recognition protocols that protect fingerprint data but nevertheless
allow for highly accurate scores for fingerprint comparison.



The setting we consider in this paper is one where the parties
involved are honest-but-curious (although, in some cases,
malicious parties can also be tolerated using known hardening techniques).
That is, two or more parties hold private fingerprint data that
they would like to compare in a fast way that provides an accurate score
for the comparison but does not reveal the actual fingerprint data.
\ifFull
For example,
the computation could involve comparing multiple pairs of 
fingerprint representations 
stored in two privacy-sensitive 
databases (e.g., one owned by the FBI and the other owned by a
university),
or it could involve a single comparison of a suspected criminal's
fingerprint to a fingerprint found at a crime scene.
\fi


\ifFull
\begin{figure}[t]
  \begin{center}
\includegraphics[width=3in]{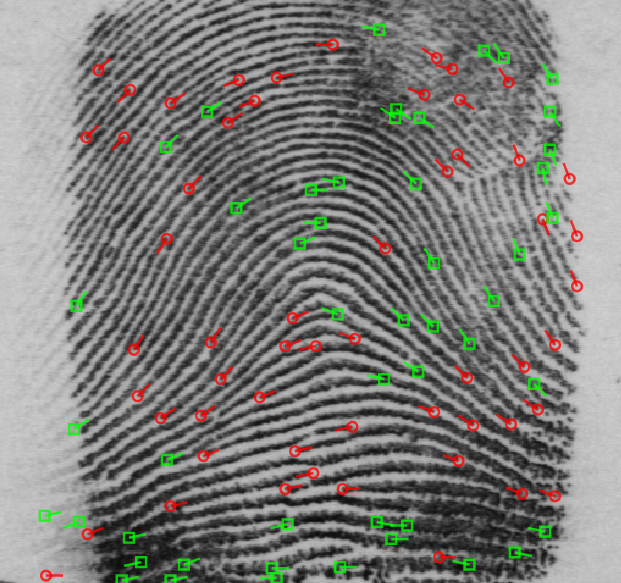}
  \end{center}
\caption{A set of minutiae with orientations. The image is generated by
  NIST's Fingerprint Minutiae Viewer (FpMV) software \cite{soft} using a fingerprint
  from NIST's Special Database 4 \cite{image}.}
\label{fig:nist}
\end{figure}
\fi

According to accepted best practices 
(e.g., see~\cite{fingerbook,gain,haiyan}) 
the most accurate \emph{fingerprint recognition} algorithms are 
based on the use of \emph{minutia} points, which are fingerprint
``landmarks,'' such as where two ridges merge or split.
\ifFull
(See Figure~\ref{fig:nist}.)
\fi
Such \emph{geometric} recognition algorithms generally involve
two main steps: \emph{alignment} and \emph{matching}.  
Alignment involves computing a geometric transformation to 
best ``line up'' two sets
of minutiae and matching involves scoring the similarity of two sets
of (hopefully overlapping) minutia points.
Alignment is necessary for
the meaningful and accurate application of the matching step,
yet, to the best of our knowledge, 
all previous publications on secure fingerprint recognition, except
for a paper by Kerschbaum {\it et al.}~\cite{atallah1}, focus
exclusively on the matching step, possibly because of the 
computational difficulty of finding a good alignment.
Instead, our approach is on the design 
of efficient protocols for
the entire fingerprint recognition process, including both the alignment and 
matching steps, so as to provide secure protocols for 
accurate fingerprint recognition algorithms used in practice.

In this work, we focus on three algorithms that compare fingerprints using
both alignment and matching:

\ifFull
\begin{enumerate}
\else
\begin{compactenum}
\fi
\item
A simple geometric
transformation algorithm that finds the best matching by
tryng each pair of minutiae~\cite{fingerbook}.
\item
An algorithm that
aligns fingerprints based on high curvature points~\cite{gain}.
\item
An algorithm based on spectral minutia representation~\cite{haiyan}.
\ifFull
\end{enumerate}
\else
\end{compactenum}
\fi
These
algorithms were selected due to their precision, speed, and/or popularity,
with the goal of building efficient and practical 
security solutions. 

\ignore{
To be able to create such solution, we first needed to design
new secure sub-protocols for computing sine and cosine for fixed-point
representation (suitable for both two-party and multi-party computation),
and the square root for fixed-point representation (suitable for two-party
computation). Once we obtain the building blocks and compose them to
minimize the overall overhead, we implement the solutions using garbled
circuit evaluation techniques in the two-party setting and linear secret
sharing in the multi-party setting.
}

Our main contributions can be summarized as follows:

\ifFull
\begin{itemize}
\else
\begin{compactitem}
\fi
\item {\color{black}We design secure sub-protocols for trigonometric and
    inverse trigonometric functions for fixed-point values; sine and cosine
    protocols have structure similar to the solution in \cite{kerik16}.} 


\item We design a new secure sub-protocol for selecting the $f$th
  smallest element in a set of comparable elements. It
is based on an efficient square-root sampling algorithm.

\item We build three secure and efficient protocols for fingerprint alignment
  and matching, using of the above new building blocks
  applied to the three well-known fingerprint recognition algorithms 
  mentioned above. 

\item We implement one of the secure fingerprint recognition protocols and
  show its efficiency in practice.
\ifFull
\end{itemize}
\else
\end{compactitem}
\fi
All constructions work in both two-party and multi-party settings and
are presented in the semi-honest model. However, a number
of available techniques can be used to achieve security in the malicious
model (e.g., \cite{gen98,dam10,ash11} and others in the multi-party secret
sharing setting and \cite{kre12} among others in the two-party garbled
circuits setting). 

\ifFull
In what follows, we describe related work in
Section~\ref{sec:relwork}. Section~\ref{sec:fprelim} provides necessary
fingerprint background, including the three fingerprint comparison
algorithms which are the focus of this work. Section~\ref{sec:sm} describes
the problem statement, security model, and frameworks on which we rely to
realize secure fingerprint comparisons. Then Section~\ref{sec:bb} gives an
overview of existing secure building blocks used in this work and describes
new secure sub-protocols not available in the prior literature that we design to
enable secure fingerprint comparisons. Our main protocols for the three
selected fingerprint alignment and matching algorithms are given in
Section~\ref{sec:protocols}. Experimental evaluation of our secure
realization of the third spectral minutia representation algorithm 
in two different secure computation frameworks 
is given in Section~\ref{sec:perf}.
Lastly, Section~\ref{sec:conclusions} concludes this work.
\else
\fi
\section{Related Work}
\label{sec:relwork}

Work on secure two- and multi-party computation is extensive and
such protocols have been shown to be effective for privacy-preserving
evaluation of various functions (e.g., see~\cite{Du:2001,bog08,lin09,dam12}). 
Following the seminal work of Yao~\cite{yao86}, it is generally known 
today that any computable function can be securely evaluated, albeit
with varying degrees of practicality.
\ifFull
For instance, there are a variety
of general approaches that typically represent the function to be evaluated
as a Boolean or arithmetic circuit 
(e.g., see~\cite{bog08,lin09,dam12}). 
\fi
To date, such general approaches have typically been the most
efficient in practice. 
Thus, there are also a wide variety of custom function-specific
protocols that target the design of more efficient secure solutions
than what the generic approaches provide, including protocols for
biometric identification for different modalities, such as
iris, face, etc. (see., e.g., ~\cite{erk09,bla11} among others).
{\color{black}Because each biometric modality has a unique representation and drastically different algorithms for their comparisons, techniques for other biometric modalities are out of scope of this work.}

The first work to treat secure fingerprint comparisons is due to Barni {\it
et al.}~\cite{bar10}. The method uses the FingerCode representation of
fingerprints (which uses texture information from a fingerprint) and 
utilizes a homomorphic encryption scheme. FingerCode-based fingerprint
comparisons can be implemented efficiently using Euclidean distance; hence,
the solution of \cite{bar10} is relatively fast. Unfortunately, 
FingerCode-based comparisons are not as discriminative as other fingerprint
recognition algorithms (in particular, minutia-based algorithms) and is not
considered sufficient for fingerprint-based identification suitable for
criminal trials.

Huang {\it et al.}~\cite{hung11} propose a privacy-preserving protocol for
biometric identification using fingerprint matching which uses homomorphic
encryption and garbled circuit (GC) evaluation. The construction utilizes
FingerCodes and the work provides some optimizations, such as using off-line
execution and smaller circuits, which make the solution more efficient than
prior work. Nevertheless, the solution still suffers from the lack of
accuracy derived from being based on FingerCodes.

Blanton and Gasti~\cite{bla11,bla15chapter} also improve the performance of
secure fingerprint comparisons based on FingerCodes and additionally provide
the first privacy-preserving solution for minutia-based fingerprint
matching. The solutions use a combination of homomorphic encryption and GC
evaluation, and assume that fingerprints are independently pre-aligned. That
is, the work treats the matching step only, not the more difficult alignment
step. 
\ifFull
To compare two 
fingerprints, $T$ and $S$, consisting of pre-aligned sets of 
$m$ and $n$ minutia points,
respectively, their algorithm considers each point $t_i$ of $T$ in turn,
determines the list of points in $S$ within a certain distance and
orientation from $s_i$ that have not yet been paired up with another point
in $T$. If this list is not empty, $t_i$ is paired up with the closest point
on its list. The total number of paired up points is the size of matching,
which can consequently be compared to a threshold to determine whether the
fingerprints are related or not. 
\fi
Although their method is lacking in the alignment step,
we use similar logic for computing
matching between two transformed fingerprints in our protocols.

Shahandashti {\it et al.}~\cite{sha12} also propose a privacy-preserving protocol
for minutia-based fingerprint matching, which uses homomorphic encryption
and is based on evaluation of polynomials in encrypted form. However, the complexity
of their protocol is substantially higher than that of~\cite{bla15chapter}
and their method can also introduce an error
when a minutia point from one fingerprint has more than one minutia
point from the
other fingerprint within a close distance and orientation from it.

More recently, Blanton and Saraph~\cite{bla15esorics} introduce a privacy-preserving
solution for minutia-based fingerprint matching that formulates the problem
as determining the size of the maximum flow in a bipartite graph and
provides a data-oblivious algorithm for it.
\ifFull
The
algorithm is guaranteed to pair the minutiae from $S$ with the minutiae from
$T$ in such a way that the size of the pairing is maximal (which previous
solutions could not achieve). 
\fi
The algorithm can be used in both two-party
and multi-party settings, but only the two-party protocol based on GC
evaluation was implemented.

\ignore{In \cite{bri07}, a biometric authentication protocol is proposed
based on Goldwasser-Micali public-key encryption scheme as a homomorphic
encryption. This protocol can be used for fingerprint images as biometric
data, but it is rather slow because computation is based on homomorphic
encryption that is time consuming.}
\ignore{In addition, \cite{cans} introduces the concept of Extended Private
Information Retrieval (EPIR) by generalizing the concept of PIR, and it can
be applied for biometric computation like fingerprint recognition.} 

Lastly,
Kerschbaum {\it et al.}~\cite{atallah1} propose a private fingerprint
verification protocol between two parties that includes both alignment and
matching steps. Unfortunately, the solution leaks information about fingerprint
images used and uses a simplified
alignment and matching computation that is not as robust to fingerprint
variations as other algorithms. 
\ifFull
\fi

\section{Fingerprint Background}
\label{sec:fprelim}

A fingerprint represents an exterior surface of a finger. All features
extracted from a fingerprint image allow one to uniquely identify the
individual who owns the fingerprint. One of the features most commonly used
for fingerprint recognition is \emph{minutiae},
which are represented as a set of points in
the two-dimensional plane with an angle orientation.
Minutiae points generally
correspond to fingerprint ridge
endings or branches, and are typically represented by the following elements
\cite{fingerbook}: 1) an $x$-coordinate, 2) a $y$-coordinate, 3) 
an orientation, $\theta$,
{\color{black} that corresponds to the angle between the minutiae ridge and the 
horizontal line} measured in degrees, and 4) an optional minutia type. Some
algorithms might store additional information in the fingerprint
representation, e.g., the relationship of each minutia point
to a reference point
such as the core point representing the center of a fingerprint. 

All of the algorithms that we describe are based on minutia points. One of
them, however, uses an alternative representation in the form of a spectrum
of a fixed size. Thus, the minutiae might be not represented as points, but
instead use a spectral representation as detailed later.
Furthermore, one of the fingerprint recognition algorithms upon which we
build relies on an additional type of fingerprint features, which are high
curvature points extracted from a fingerprint image. Then in addition to
storing minutia points, each fingerprint contains a set of high curvature
points, each of which is represented as $(x, y, w)$. Here, first two
elements indicate the location of the point and the last element is its
curvature value in the range 0--2 (see \cite{gain} for the details of its
computation).

In the remainder of this section, we describe the three selected fingerprint
recognition algorithms without security considerations. All of them take two
fingerprints $T$ and $S$ as their input, align them, and output the number
of matching minutiae between the aligned fingerprints. In all algorithms, a
fingerprint contains a set of minutiae, $(t_1, {\ldots}, t_m)$ for $T$ and
$(s_1, {\ldots}, s_n)$ for $S$, or an alternative representation derived
from the minutiae. Additionally, the second algorithm uses a number of high
curvature points stored in a fingerprint, and we denote them by $(\hat{t}_1,
{\ldots}, \hat{t}_{\hat{m}})$ for $T$ and $(\hat{s}_1, {\ldots},
\hat{s}_{\hat{n}})$ for $S$. 

\subsection{Fingerprint Recognition using Brute Force Search}
\label{sec:alg1}

The first algorithm searches for the best alignment between $T$ and $S$ by
considering that minutia $t_i \in T$ corresponds to minutia $s_j \in S$ for
all possible pairs $(t_i, s_j)$ (called a reference pair)~\cite{fingerbook}.
Once a new reference pair is chosen, the algorithm transforms the
minutiae
from $S$ using a geometrical transformation and also rotates them, after
which it counts the number of matched minutiae. After trying all possible
reference pairs, the algorithm chooses an alignment that maximizes the
number of matched minutiae and thus increases the likelihood of the two
fingerprints to be matched. Algorithm~\ref{alg1} lists the details of this
approach. {\color{black}The parameters $\lambda$ and $\lambda_{\theta}$ are
used by the matching step as detailed below.}

\begin{algorithm}[t] \small
\caption{Fingerprint recognition based on geometrical transformation}
\label{alg1}
\textbf{Input:} Two fingerprints $T = \{t_i = (x_i, y_i,
  \theta_i)\}_{i=1}^m$ and $S = \{s_i = (x'_i, y'_i, \theta'_i)\}_{i=1}^n$,
  {\color{black}and thresholds $\lambda$ and $\lambda_{\theta}$}.\\
\textbf{Output:} The largest number of matching minutiae, $C_{max}$, and the
corresponding alignment. 
\begin{compactenum}
  \item Initialize $C_{max} = 0$.
  \item For $i = 1, {\ldots}, m$ and $j = 1, {\ldots}, n$, use $(t_i, s_j)$ as a
    reference pair and perform:
  \begin{compactenum}    
  \item Compute transformation 
    %
    $\Delta x = x'_j - x_i$, $\Delta y = y'_j - y_i$, and $\Delta \theta =
    \theta'_j - \theta_i$. 

  \item Transfer each minutia $s_k \in S$ as 
   $x''_k = \cos(\Delta \theta) \cdot x'_k+ $ $\sin(\Delta \theta)\cdot
   y'_k - \Delta x$, $y''_k = -\sin(\Delta \theta)\cdot x'_k  + \cos(\Delta
   \theta)\cdot$ $ y'_k- \Delta y$, and $\theta''_k = \theta'_k - \Delta
   \theta$ and save the result as $S''= \{s''_i = (x''_i, y''_i, \theta''_i)\}_{i=1}^n$.

  \item Compute the number $C$ of matched minutiae between $T$ and $S''$
     using Algorithm~\ref{alg-matching} {\color{black}and parameters $\lambda$ and
     $\lambda_{\theta}$.}

  \item If $C_{max} < C$, then set $C_{max} = C$ and $(\Delta x_{max},
    \Delta y_{max}, \Delta \theta_{max}) = (\Delta x, \Delta y, \Delta \theta)$. 
  \end{compactenum}

 \item Return $C_{max}$ and its
 corresponding alignment $(\Delta x_{max}, \Delta y_{max}, \Delta \theta_{max})$.
\end{compactenum}
\end{algorithm}

To implement the matching step, we use an algorithm that iterates through
all points in $t_i \in T$ and pairs $t_i$ with a minutia $s_j \in S$ (if
any) within a close spatial and directional distance from $t_i$. When
multiple points satisfy the matching criteria, the closest to $t_i$ is
chosen, as described in \cite{fingerbook,bla15chapter}. This computation of
the matching step is given in Algorithm~\ref{alg-matching}.

\begin{algorithm}[t] \small
\caption{Fingerprint matching}
\label{alg-matching}
\textbf{Input:} Two fingerprints $T = \{t_i = (x_i, y_i,
  \theta_i)\}_{i=1}^m$,  $S = \{s_i = (x'_i, y'_i, \theta'_i)\}_{i=1}^n$ and
thresholds $\lambda$ and $\lambda_{\theta}$ for distance and orientation.\\ 
\textbf{Output:} The number $C$ of matched minutiae between $T$ and $S$.
\begin{compactenum}
  \item Set $C = 0$.
  \item Mark each $s_j \in S$ as available.

  \item For $i = 1, {\ldots}, m$, do:
  \begin{compactenum}
  
  \item Create a list $L_i$ consisting of all available points $s_j \in S$
    that satisfy $\min(|\theta_i-\theta'_j|$, $360 - |\theta_i - \theta'_j|) <
    \lambda_{\theta}$ and $\sqrt{(x_i- x'_j)^2 +(y_i- y'_j)^2} < \lambda$.

    \item If $L_i$ is not empty, select the closest minutia $s_k$ to $t_i$
      from $L_i$, mark $s_k$ as unavailable and set $C = C+1$.
    \end{compactenum}

  \item return $C$.
\end{compactenum}
\end{algorithm}

The time complexity of Algorithm~\ref{alg-matching} is ${O}(nm)$, and the time complexity of
Algorithm~\ref{alg1} is ${O}(n^2m^2)$ when Algorithm~\ref{alg-matching} is used as its subroutine.  

\ifFull
\else
\fi
\subsection{Fingerprint Recognition using High Curvature Points for
Alignment}
\label{sec:alg2}

The second fingerprint recognition algorithm \cite{gain} uses fingerprints'
high curvature information to align them. The alignment is
based on the iterative closest point (ICP) algorithm \cite{icp} that
estimates the rigid transformation $F$ between $T$ and $S$. Once this
transformation is determined, it is applied to the minutia points of $S$,
after which the algorithm computes the number of matched minutiae between
$T$ and transformed $S$. The ICP algorithm assumes that the fingerprints are
roughly pre-aligned, 
which can be done, e.g., by aligning each fingerprint independently using
its core point, and iteratively finds point correspondences and the
transformation between them. To eliminate alignment errors when the overlap
between the two sets is partial, \cite{gain} suggests using the trimmed ICP
(TICP) algorithm \cite{chet}. The TICP algorithm ignores a proportion of the
points in $T$ whose distances to the corresponding points in $S$ are large,
which makes it robust to outliers. 

\begin{algorithm}[t] \small
\textbf{Input:} Two fingerprints consisting of minutiae and high curvature
points $T = (\{t_i = (x_i, y_i, \theta_i)\}_{i=1}^m$, $\{\hat{t}_i =
  (\hat{x}_i, \hat{y}_i, \hat{w}_i)\}_{i=1}^{\hat{m}})$ and $S = (\{s_i =
  (x'_i, y'_i, \theta'_i)\}_{i=1}^n$, $\{\hat{s}_i = (\hat{x}'_i, \hat{y}'_i,
 \hat{w}'_i)\}_{i=1}^{\hat{n}})$; {\color{black}matching thresholds $\lambda$
 and $\lambda_{\theta}$; scaling parameter $\beta$;} minimum number of
 matched high curvature points $f$; algorithm termination parameters
 {\color{black}$\delta$} and $\gamma$.\\
\textbf{Output:} The largest number of matching minutiae, $C$, and the
corresponding alignment.
\begin{compactenum}
 \item Set $S_{LTS} = 0$.
 \item Store a copy of $\hat{s}_i$'s as $\bar{s}_i = (\bar{x}_i, \bar{y}_i,
   \bar{w}_i)$ for $i = 1, \ldots, \hat{n}$.

 \item For $i = 1, {\ldots}, \hat{n}$, find the closest to $\hat{s}_i$
   point $\hat{t}_j$ in $T$ and store their distance $d_i$. Here, the
   distance between any $\hat{s}_i$ and $\hat{t}_j$ is defined as
   $\sqrt{(\hat{x}'_i - \hat{x}_j)^2 + (\hat{y}'_i - \hat{y}_j)^2} +
   \beta|\hat{w}'_i - \hat{w}_j|$.
 \item Find $f$ smallest $d_i$'s and calculate their sum $S'_{LTS}$.
 \item If $S'_{LTS} \leq {\color{black}\delta}$ or $\gamma = 0$, proceed with step 11.
 \item Set $S_{LTS} = S'_{LTS}$.
 \item Compute the optimal motion $(R, v)$ for the selected $f$
\ifFull
   pairs using Algorithm~\ref{alg2-trans}. 
\else
  pairs.
\fi
 \item For $i = 1, \ldots, \hat{n}$, transform point $\hat{s}_i$ according
   to $(R, v)$ as $\hat{s}_i = R \hat{s}_i + v$.
\ifFull
 \item Set $\gamma = \gamma -1$.
 \item Repeat from step 3.
\else
 \item Set $\gamma = \gamma -1$ and repeat from step 3.
\fi
  \item Compute the optimal motion $(R, v)$ for $\hat{n}$ pairs $(\bar{s}_i, 
\ifFull
    \hat{s}_i)$ using Algorithm~\ref{alg2-trans}.
\else
    \hat{s}_i)$.
\fi
   Let $r_{i,j}$ denote $R$'s cell at row $i$ and column $j$ and $v_i$
   denote the $i$th \mbox{element of $v$.}
  
 \item Compute $c_1 = (\bar{y}_2-\bar{y}_1)/(\bar{x}_2-\bar{x}_1)$, $c_2 =
   (\hat{y}'_2-\hat{y}'_1)(\hat{x}'_2-\hat{x}'_1)$, and $\Delta \theta =
   \arctan((c_1 - c_2)/(1 + c_1 c_2))$.
  \item For $i = 1, {\ldots}, n$, apply the transformation to
  minutia $s_i$
    by computing $x'_i = r_{1,1} x'_i+ r_{1,2} y'_i + v_1$, $y'_i = r_{2,1}
    x'_i+ r_{2,2} y'_i + v_2$, and $\theta'_i = \theta'_i - \Delta \theta$. 
  
 \item Compute the number, $C$, of matched minutiae between $t_i$'s and
   transformed $s_i$'s (e.g., using Algorithm~\ref{alg-matching}).
 \item Return $C$ and the transformation $(R, v)$. 
\end{compactenum}
\caption{Fingerprint recognition based on high curvature points for
alignment}
\label{alg2}
\end{algorithm}

\ifFull
\begin{algorithm}[t] \small
\textbf{Input:} $n$ pairs $\{(t_i = (x_i, y_i, z_i), s_i = (x'_i, y'_i,
 z'_i))\}_{i=1}^n$.\\
\textbf{Output:} Optimal motion $(R, v)$.
\begin{compactenum}
 \item For $i = 1, \ldots, n$, compute unit quaternion $q_i = (q_{(i,1)},
   q_{(i,2)}, q_{(i,3)}, q_{(i,4)}) = \left(\sqrt{\frac{1+k_i}{2}}, u_i
   \sqrt{\frac{1-k_i}{2}}\right)$, where $k_i =
   \frac{x_i \cdot x'_i + y_i \cdot y'_i + z_i \cdot
     z'_i}{\sqrt{x^2_i + y^2_i + z^2_i} \cdot \sqrt{x'^2_i +
	y'^2_i + z'^2_i}}$, $u_i = 
     \left(\frac{y_i \cdot z'_i - z_i \cdot y'_i} {|| t_i \times
	    s_i ||}, \frac{z_i x'_i - x_i z'_i}{|| t_i
	    \times s_i ||}, \frac{x_i \cdot y'_i - y_i \cdot x'_i} {|| t_i
	    \times s_i ||}\right)$, and $|| t_i \times s_i || =
	  \sqrt{(y_i \cdot z'_i - z_i \cdot y'_i)^2 + (z_i
	    \cdot x'_i - x_i \cdot z'_i)^2 + (x_i \cdot y'_i - y_i
	    \cdot x'_i)^2}$.
 \item Compute the overall unit quaternions $q = [q_1, q_2, q_3, q_4] =
   q_1q_2 \ldots q_{n-1} q_n$ by executing multiplication from left to
   right, where $q_jq_{j+1} = [q_{(j,1)}q_{(j+1, 1)} - v_j \cdot v_{j+1},
   q_{(j,1)} v_{j+1}+ q_{(j+1, 1)} v_j + v_j \times v_{j+1}]$, $v_j =
   (q_{(j,2)}, q_{(j,3)}, q_{(j,4)})$, and $v_{j+1} = (q_{(j+1,2)},
   q_{(j+1,3)}, q_{(j+1,4)})$ for $j =1,\ldots, n-1$.
 \item Compute rotation matrix $R = \left[{\begin{array}{ccc}
     q_1^2+q_2^2-q_3^2-q_4^2 &  2(q_2 q_3-q_1 q_4)& 2(q_2 q_4+q_1 q_3)\\
     2(q_3 q_2+q_1 q_4) & q_1^2-q_2^2+q_3^2-q_4^2 & 2(q_3 q_4-q_1 q_2)\\
     2(q_4 q_2-q_1 q_3) & 2(q_4 q_3-q_1 q_2) & q_1^2-q_2^2-q_3^2+q_4^2
 \end{array} }\right]$.
 \item Compute transformation vector $v = t-Rs$, where
 $t = (\sum^{n}_{i = 1} t_i)/n$ and $s = (\sum^{n}_{i = 1}
 s_i)/n$.
 \item Return $(R, v)$.
\end{compactenum}
\caption{Union quaternion method for computing optimal motion}
\label{alg2-trans}
\end{algorithm}
\fi

The details of this fingerprint recognition approach are given in
Algorithm~\ref{alg2}. Steps 1--10 correspond to the TICP algorithm that
proceeds with iterations, aligning the fingerprints closer to each other in each
successive iteration. The termination parameters {\color{black}$\delta$} and
$\gamma$ correspond to the threshold for the total distance between the
matched points of $T$ and $S$ and the limit on the number of iterations,
respectively. For robustness, the alignment is performed using only a subset
of the points ($f$ closest pairs computed in step 4). The optimal motion for
transforming one set of the points into the other (treated as two 3-D shapes) is
computed using the union quaternion method due to Horn~\cite{horn87},
\ifFull
and is given in Algorithm~\ref{alg2-trans}.
\else
which we omit due to space considerations; we defer description of that to the full version of the paper \cite{FingerprintFull}. We only mention that it is a solution for the least-squares problem for three or more points and uses
operations such as square root, division, multiplication, addition, and
subtraction on real values.
\fi
The transformation is represented
as a $3 \times 3$ matrix $R$ and a vector $v$ of size 3, which are
consequently used to align the points in $S$ (step 8).

Once the fingerprints are sufficiently aligned, the overall transformation
between the original and aligned $S$ is computed (\ifFull step 11\else step 10\fi). It is then
applied to the minutia points of $S$: The new $x$ and $y$ coordinates of the
minutiae can be computed directly using the transformation $(R, v)$, while
orientation $\theta$ requires special handling (since it is not used in
computing the optimal motion). We compute the difference in the orientation
between the original and aligned $S$ as the angle between the slopes of two
lines drawn using two points from the original and transformed $S$,
respectively (\ifFull step 12\else step 11\fi).

The original ICP algorithm \cite{icp} lists expected and worst case
complexities of computing the closest points (step 3 of
Algorithm~\ref{alg2}) to be $O(\hat{m} \log \hat{n})$ and $O(\hat{m}
\hat{n})$, respectively. The approach, however, becomes prone to errors in
the presence of outliers, and thus in the TICP solution \cite{chet} a bounding
box is additionally used to eliminate errors. In step 4, we can use $f$th
smallest element selection to find the smallest distances, which runs in
linear time (the TICP paper suggests sorting, but sorting results in higher
asymptotic complexities). The complexity of 
\ifFull
Algorithm \ref{alg2-trans} 
\else
optimal motion computation
\fi
is ${O}(k)$ when run on $k$ input pairs. Thus, the overall complexity of
Algorithm~\ref{alg2} 
\ifFull
(including Algorithms \ref{alg2-trans} and~\ref{alg-matching}) 
\else
(including sub-algorithms)
\fi
is ${O}(\gamma \hat{m} \hat{n} + mn)$, where
$\gamma$ is the upper bound on the number of iterations in the algorithm. 
\ifFull
\else 
\fi
\subsection{Fingerprint Recognition based on Spectral Minutiae Representation}
\label{sec:alg3}

The last fingerprint recognition algorithm \cite{haiyan} uses
minutia-based representation of fingerprints, but offers greater efficiency
than other algorithms because of an alternative form of minutia
representation. The spectral minutia representation \cite{xu09} that it
uses is a fixed-length feature vector, in which rotation and scaling can be
easily compensated for, resulting in efficient alignment of two
fingerprints.

In the original spectral minutia representation \cite{xu09}, a set of
minutiae corresponds to a real-valued vector of a fixed length $D$, which is
written in the form of a matrix of dimensions $M \times N$ ($=D$). The vector
is normalized to have zero mean and unit energy. Thus, we now represent $T$
and $S$ as matrices of dimensions $M \times N$ and denote their individual
elements as $t_{i,j}$'s and $s_{i,j}$'s, respectively. \ignore{In \cite{xu09}, there
are two types of minutia spectra (location-based and orientation-based),
each with $M = 128$ and $N = 256$, and a fingerprint represented \mbox{by their
combination consists of 65,536 real numbers.}}

To compare two fingerprints in the spectral representation, two-dimensional
correlation is used as a measure of their similarity. Furthermore, to
compensate for fingerprint image rotations, which in this representation
become circular shifts in the horizontal direction, the algorithm tries a
number of rotations in both directions and computes the highest score among
all shifts. This alignment and matching algorithm is presented as
Algorithm~\ref{alg3}. It is optimized to perform shifts from $-15$ to $15$
units (which correspond to rotations from $-10^{\circ}$ to $+10^{\circ}$) and computes only
9 similarity scores instead of all 31 of them. \ifFull Later in this section, we show
how the algorithm can be generalized to work with any amount of shift
$\lambda$ resulting in only $O(\log \lambda)$ score computations.\fi

\begin{algorithm}[t] \small
\caption{Fingerprint recognition based on spectral minutia representation}
\label{alg3}
\textbf{Input:} Two real-valued matrices $T = \{t_{i,j}\}_{i=1,j=1}^{M,N}$
and $S = \{s_{i,j}\}_{i = 1, j = 1}^{M,N}$ and parameter $\lambda=15$
indicating the maximum amount of rotation.\\ 
\textbf{Output:} The best matching score $C_{max}$ and the corresponding
alignment. 
\begin{compactenum}
 \item Set $C_{max} = 0$.
 \item For $\alpha = -12, -6, 0, 6, 12$, do:
   \begin{compactenum}
     \item Compute the similarity score between $T$ and $S$ horizontally
       shifted by $\alpha$ positions as $C_\alpha = \frac{1}{MN}
       \sum^{M}_{i=1} \sum^{N}_{j=1} t_{i,j} \cdot s_{i,(j+\alpha) \bmod
       N}$.
     \item If $C_\alpha > C_{max}$, set $C_{max} = C$, $k = \alpha$, and
       $\alpha_{max} = \alpha$.
   \end{compactenum}
 \item For $\alpha = k-2, k+2$, do:
   \begin{compactenum}
     \item Compute the similarity score between $T$ and $S$ shifted by $\alpha$
       positions as in step 2(a).
     \item If $C > C_{max}$, set $C_{max} = C$, $k' = \alpha$, and
       $\alpha_{max} = \alpha$.
   \end{compactenum}
 \item For $\alpha = k'-1, k'+1$, do:
   \begin{compactenum}
     \item Compute the similarity score between $T$ and $S$ shifted by $\alpha$
       positions as in step 2(a).
     \item If $C > C_{max}$, set $C_{max} = C$ and $\alpha_{max} = \alpha$.
   \end{compactenum}
 \item Return $C_{max}$ and $\alpha_{max}$.
\end{compactenum}
\end{algorithm}

Xu et al. \cite{haiyan} apply feature reduction to the spectral minutia
representation to reduce the size of the feature vector and consequently
improve the time of fingerprint comparisons without losing precision. That
work describes two types of feature reduction: Column Principle Component
Analysis (CPCA) and Line Discrete Fourier Transform (LDFT). Then one or both
of them can be applied to the feature vector, with the latter option
providing the greatest savings.

The first form of feature reduction, CPCA, reduces the minutia spectrum
feature in the vertical direction. After the transformation, the signal is
concentrated in the top rows and the remaining rows that carry little or no
energy are removed. As a result, a minutia spectrum is represented as a
matrix of dimension $M' \times N$ for some $M' < M$. Because the rotation
operation commutes with this transformation, the score computation in
Algorithm~\ref{alg3} \ifFull \else\mbox{\fi remain unchanged after CPCA feature reduction.\ifFull \else}\fi

The second form of feature reduction, LDFT, reduces the minutia spectrum in
the horizontal direction and returns a feature matrix of dimension $M \times
N'$ (or $M' \times N'$ if applied after the CPCA feature reduction) for some
$N' < N$. The matrix consists of complex numbers as a result of applying
Fourier transform. After the LDFT, the energy is concentrated in the middle
columns and other columns are consequently removed. Shifting matrix $S$ in
the resulting representation by $\alpha$ positions now translates into
setting its cell at location $k,j$ to $e^{-i\frac{2\pi}{N}j\alpha} \cdot
s_{k,j}$, where $i$ indicates the imaginary part of a complex number. Note
that circular shifts are no longer applied to matrix rows (including when
the LDFT is applied after the CPCA feature reduction). 

The resulting matrices of complex numbers $T$ and $S$ are then converted to
real-valued matrices and processed using Algorithm~\ref{alg3}. If we express
a cell of $T$ or $S$ at location $k,j$ as $a_{k,j} + ib_{k,j}$ by separating
its real and imaginary parts, then the $k$th row of $T$ and $S$ is now
expressed as a real-valued vector 
\ifFull
$$
\sqrt{\frac{1}{N}} a_{k,1},
\sqrt{\frac{2}{N}} a_{k,2}, \ldots, \sqrt{\frac{2}{N}} a_{k,N’},\\
\sqrt{\frac{2}{N}} b_{k,2}, \ldots, \sqrt{\frac{2}{N}} b_{k,N'}.
$$
\fi
\ifFull
\else
$
\sqrt{\frac{1}{N}} a_{k,1}, \sqrt{\frac{2}{N}} a_{k,2},$ $ \ldots, \sqrt{\frac{2}{N}} a_{k,N’},
\sqrt{\frac{2}{N}} b_{k,2}, \ldots, \sqrt{\frac{2}{N}} b_{k,N'}.
$
\fi
Computing
the score between $T$ and $S$ then corresponds to computing the dot product
of each $k$th row of $T$ and $S$ and adding the values across all $k$. When
we use Algorithm~\ref{alg3} to compute the best similarity score, we need to
adjust the computation in step 2(a) for the new matrix dimensions and also
implement rotation as a multiplication instead of a shift. If we expand the
quantity $e^{-i\frac{2\pi}{N}j\alpha}$ using the formula $e^{i\varphi} =
\cos(\varphi) + i \sin(\varphi)$, the real part of each score between $T$ and
$S$ rotated by $\alpha$ \mbox{positions now becomes:}
\ifFull
\begin{eqnarray}\label{eq:score}
C_\alpha & = & \frac{1}{MN^2} \sum_{k=1}^{M'} \left(a_{k,1} \left(a'_{k,1}
\cos\left(-\frac{2\pi\alpha}{N}\right) - b'_{k,1}
\sin\left(-\frac{2\pi\alpha}{N}\right)\right) + \right. \\ \nonumber & & +
\left. 2\sum_{j=2}^{N'} \left(\cos\left(-\frac{2\pi j \alpha}{N}\right)
\left(a_{k,j}a_{k,j}' + b_{k,j}b'_{k,j}\right) + \sin\left(-\frac{2\pi j
\alpha}{N}\right) \left(a'_{k,j}b_{k,j} - a_{k,j}b'_{k,j}\right)\right)\right) 
\end{eqnarray}
where $t_{k,j} = a_{k,j} + ib_{k,j}$ and original $s_{k,j} = a'_{k,j} +
ib'_{k,j}$.
\else
\begin{eqnarray}\label{eq:score}
C_\alpha & = & \frac{1}{MN^2} \sum\nolimits_{k=1}^{M'} \left(a_{k,1} \left(a'_{k,1}
\cos(x_1) - b'_{k,1}
\sin(x_1) \right) + \right. \nonumber \\  & & + 2\sum\nolimits_{j=2}^{N'} \left(\cos(x_j)
\left(a_{k,j}a_{k,j}' + b_{k,j}b'_{k,j}\right) + \right. \\ \nonumber & & +
\left. \left. \sin(x_j) \left(a'_{k,j}b_{k,j} - a_{k,j}b'_{k,j}\right)\right)\right)  
\end{eqnarray}
where $t_{k,j} = a_{k,j} + ib_{k,j}$, original $s_{k,j} = a'_{k,j} +
ib'_{k,j}$, $x_j = -(2\pi j \alpha)/N$.
\fi

Returning to Algorithm~\ref{alg3}, we generalize the algorithm to work for
any value of $\lambda$ so that only $O(\log \lambda)$ score computations are
necessary. Let $\lambda = c \cdot 3^k$ for some constants $c \ge 2$ and $k
\ge 1$. Our algorithm proceeds in $k+1$ iterations computing $c$ scores in
the first iteration and 2 scores in each consecutive iteration, which
results in the total of $c+2k$ score computations. In the initial iteration
0, we compute $c$ scores at indices $\lambda = \lceil 3^k/2 \rceil,
{\ldots}, \lceil 3^k(2c-1)/2 \rceil$, then determine their maximum $C_{max}$
and the index of the maximum score $\alpha_{max}$. In iteration $i = 1,
{\ldots}, k$, the algorithm computes two scores at indices $\alpha_{max} \pm
3^{k-i}$, determines the maximum of the computed scores and $C_{max}$, and
updates the value of $\alpha_{max}$ as needed. Compared to
Algorithm~\ref{alg3}, this approach covers 54 shifts using 8 score
computations instead of 9 score computations for 31 shifts. The best
performance is achieved when $\lambda = 2 \cdot 3^k$ for some $k$. If
$\lambda$ is not of that form, we could use a different value of $c$, cover
a wider range of shifts than required, or decrease the step size of the
algorithm by less than a factor of 3 (as was done in
Algorithm~\ref{alg3}), which results in redundant coverage.
\ignore{Let $\lambda = c \cdot 3^k$ for some constants $c \ge 2$ and $k
\ge 1$. Our algorithm proceeds in $k+1$ iterations computing $c$ scores in
the first iteration and 2 scores in each consecutive iteration, which
results in the total of $c+2k$ score computations. In the initial iteration
0, we compute $c$ scores at indices $\lambda = \lceil 3^k/2 \rceil,
{\ldots}, \lceil 3^k(2c-1)/2 \rceil$, then determine their maximum $C_{max}$
and the index of the maximum score $\alpha_{max}$. In iteration $i = 1,
{\ldots}, k$, the algorithm computes two scores at indices $\alpha_{max} \pm
3^{k-i}$, determines the maximum of the computed scores and $C_{max}$, and
updates the value of $\alpha_{max}$ as needed. Compared to
Algorithm~\ref{alg3}, this approach covers 54 shifts using 8 score
computations instead of 9 score computations for 31 shifts. The best
performance is achieved when $\lambda = 2 \cdot 3^k$ for some $k$. If
$\lambda$ is not of that form, we could use a different value of $c$, cover
a wider range of shifts than required, or decrease the step size of the
algorithm by less than a factor of 3 (as was done in
Algorithm~\ref{alg3}), which results in redundant coverage.}

In \cite{haiyan}, parameters $M'$ and $N'$ are chosen to retain most of the
signal's energy after the transformations (e.g., 90\%) and may be dependent
on the data set. The complexity of Algorithm~\ref{alg3} for two fingerprints
after both types of feature reduction is ${O}(M'N'\log\lambda)$.
\ifFull
\else 
\fi
\section{Security Model}
\label{sec:sm}

\ifFull
\subsection{Problem Statement and Security Definitions}
\fi

Because of the variety of settings in which fingerprint recognition may be
used, we distinguish between different computation setups and the
corresponding security settings.

\ifFull
\begin{enumerate}
\else
\begin{compactenum}
\fi
\item There will be circumstances when two entities would like to compare
  fingerprints that they respectively possess without revealing any
  information about their data to each other. This can correspond to the
  cases when both entities own a fingerprint database and would like to find
  entries common to both of them or when one entity would like to search a
  database belonging to a different entity for a specific fingerprint. In
  these settings, the computation takes the form of secure two-party
  computation and the participants can be regarded as semi-honest or
  malicious depending on the application and their trust level.

\item There will also be circumstances when one or more data owners are
  computationally limited and would like to securely offload their work to
  more powerful servers or cloud providers (this applies even if all
  fingerprints used in the computation belong to a single entity). Then
  multi-party settings that allow for input providers to be different from
  the computational parties apply. In such settings, the computational
  parties are typically treated as semi-honest, but stronger security models
  can also be used for added security guarantees.
\ifFull
\end{enumerate}
\else
\end{compactenum}
\fi
{\color{black}Regardless of the setup, the same strong data privacy guarantees
must hold: A data owner possesses one or more fingerprints after feature
extraction (e.g., a fingerprint is represented as a set of minutia
or high curvature points as discussed above). The size of each fingerprint
representation (i.e., the number of elements stored in it) is public, 
but the values themselves are private. For example, a set of minutia points
is used in secure computation as $\langle ([x_1]$, $[y_1]$, $[\theta_1])$, 
{\ldots}, $([x_m]$, $[y_m]$, $[\theta_m])\rangle$, where $[x]$ denotes that
the value of $x$ is private and always protected. Then no information that
depends on private values can be revealed during the computation and the
only information that can be available to the parties running secure
computation is public parameters and data sizes. Once the computation is
complete, the output can be revealed to one or more parties based on their
prior agreement. 

For example, in the algorithm in section~\ref{sec:alg1}, the private inputs
are two sets of minutiae points $T$ and $S$ of public sizes $m$ and $n$,
respectively, fixed thresholds $\lambda$ and $\lambda_\theta$ are public
inputs, and the private output is $[C_{max}]$ and the corresponding alignment.
Similarly, in fingerprint recognition in section~\ref{sec:alg2}, the private
input are two sets of minutiae and high curvature points $T$ and $S$, while
their sizes $m$, $\hat{m}$, $n$, $\hat{n}$ and other fixed parameters (such
as $f$, $\gamma$, etc.) are public; and so forth.
In each protocol that we describe in this work, private inputs are
explicitly marked, and any information derived from them must remain
properly protected (i.e., no information leakage is tolerated), as
formalized in our security definition below.  }

Security of any multi-party protocol (with two or more
participants) can be formally shown according to one of the two
standard security definitions. The first, weaker security model
assumes that the participants are semi-honest, defined as they follow the
computation as prescribed, but might attempt to learn additional
information about the data from the intermediate results. The
second, stronger security model allows dishonest participants to
arbitrarily deviate from the prescribed computation. 
The focus of this work is on the semi-honest adversarial model, although
standard techniques for strengthening security in the presence of fully
malicious participants can be used as well. 
\ifFull
Security in the semi-honest setting is defined as follows.
\begin{definition} \label{def:security}
    Let parties $P_1, {\ldots}, P_n$ engage in a protocol $\Pi$ that
    computes function $f({\sf in}_1, {\ldots}, {\sf in}_n) = ({\sf out}_1,
    {\ldots}, {\sf out}_n)$, where ${\sf in}_i$ and ${\sf out}_i$ denote
    the input and output of party $P_i$, respectively. Let
    $\mathrm{VIEW}_\Pi(P_i)$ denote the view of participant $P_i$ during the
    execution of protocol $\Pi$. More precisely, $P_i$'s view is formed by
    its input and internal random coin tosses $r_i$, as well as messages
    $m_1, {\ldots}, m_k$ passed between the parties during protocol
    execution: $\mathrm{VIEW}_{\Pi}(P_i) = ({\sf in}_i, r_i, m_1, {\ldots},
    m_k).$ Let $I = \{P_{i_1}, P_{i_2}, {\ldots}, P_{i_{\tau}}\}$ denote
    a subset of the participants for $\tau < n$ and $\mathrm{VIEW}_\Pi(I)$
    denote the combined view of participants in $I$ during the execution of
    protocol $\Pi$ (i.e., $\mathrm{VIEW}_\Pi(I) =
    (\mathrm{VIEW}_\Pi(P_{i_1}), {\ldots}, \mathrm{VIEW}_\Pi(P_{i_\tau}))$)
    and $f_I({\sf in}_1, {\ldots}, {\sf in}_n)$ denote the projection of
    $f({\sf in}_1, {\ldots}, {\sf in}_n)$ on the coordinates in $I$ (i.e.,
    $f_I({\sf in}_1, {\ldots}, {\sf in}_n)$ consists of the $i_1$th,
    {\ldots}, $i_{\tau}$th element that $f({\sf in}_1, {\ldots}, {\sf
    in}_n)$ outputs). We say that protocol $\Pi$ is $\tau$-private in the
    presence of semi-honest adversaries if for each coalition of size at
    most $\tau$ there exists a probabilistic polynomial time simulator $S_I$
    such that $\{S_I({\sf in}_I, f_I({\sf in}_1, {\ldots}, {\sf in}_n)),
    f({\sf in}_1, {\ldots}, {\sf in}_n)\} \equiv \{\mathrm{VIEW}_\Pi(I),
    ({\sf out}_1, {\ldots}, {\sf out}_n)\},$ where ${\sf in}_I =
    \bigcup_{P_i \in I} \{{\sf in}_i\}$ and $\equiv$ denotes computational
    or statistical indistinguishability.
\end{definition}
\else
Security in the semi-honest setting is defined according to the simulation
paradigm and the (standard) formal definition is given in Appendix \ref{app:def}. 
\fi

\ignore{
\begin{definition} \label{def2}
Let parties $P_1, {\ldots}, P_n$ engage in a protocol $\Pi$ that
computes function $f({\sf in}_1, {\ldots}, {\sf in}_n) = ({\sf
out}_1, $ ${\ldots}, {\sf out}_n)$. Let $\mathrm{VIEW}_\Pi(P_i)$
denote the view of participant $P_i$
during the execution of protocol $\Pi$ in the real world and
$\mathrm{VIEW}_f(P_i)$ the view of $P_i$ during the execution of
function $f$ in the ideal model with the presence of simulator $S$
setup as described above. Also, let $I = \{P_{i_1}, P_{i_2},
{\ldots}, P_{i_{\tau}}\}$ denote a subset of the participants for
$\tau < n$ and $\mathrm{VIEW}_\Pi(I)$ (resp.,
$\mathrm{VIEW}_f(I)$) denote the combined view of participants in
$I$ in the real (resp., ideal) model. Protocol $\Pi$ is said to
$\tau$-securely compute $f$ in the malicious model if
%
$\mathrm{VIEW}_\Pi(I) \equiv \mathrm{VIEW}_f(I)$.
\end{definition}}

\ifFull
\subsection{Secure Function Evaluation Frameworks}
\fi

To realize our constructions in a variety of settings, two secure
computation frameworks are of interest to us. In the two-party setting, we
build on garbled circuit (GC) evaluation techniques, while in the multi-party
setting, we employ linear secret sharing (SS) techniques. 
These frameworks are briefly described 
\ifFull below.
\else in Appendix~\ref{app:sfe}.
\fi

\ifFull
\medskip \noindent \textbf{Garbled circuit evaluation.}
The use of GCs allows two parties $P_1$ and $P_2$ to
securely evaluate a Boolean circuit of their choice. That is,
given an arbitrary function $f(x_1, x_2)$ that depends on private
inputs $x_1$ and $x_2$ of $P_1$ and $P_2$, respectively, the
parties first represent is as a Boolean circuit. One party, say
$P_1$, acts as a circuit generator and creates a garbled
representation of the circuit by associating both values of each
binary wire with random labels. The other party, say $P_2$, acts
as a circuit evaluator and evaluates the circuit in its garbled
representation without knowing the meaning of the labels that it
handles during the evaluation. The output labels can be mapped to
their meaning and revealed to either or both parties.


An important component of GC evaluation is 1-out-of-2
Oblivious Transfer (OT). It allows the circuit evaluator to obtain
wire labels corresponding to its inputs. In particular, in OT the
sender (i.e., circuit generator in our case) possesses two strings
$s_0$ and $s_1$ and the receiver (circuit evaluator) has a bit
$\sigma$. OT allows the receiver to obtain string $s_\sigma$ and
the sender learns nothing. An OT extension allows
any number of OTs to be realized with small additional overhead
per OT after a constant number of regular more costly OT protocols
(the number of which depends on the security parameter). The
literature contains many realizations of OT and its extensions,
including very recent proposals such as \cite{nao01,ish03,ash13} and
others.

The fastest currently available approach for GC generation
and evaluation we are aware of is by Bellare et al. \cite{bel13}.
It is compatible with earlier optimizations, most notably the
``free XOR'' gate technique \cite{kol08} that allows XOR gates to
be processed without cryptographic operations or communication,
resulting in lower overhead for such gates. A recent half-gates
optimization \cite{zah15} can also be applied to this construction to reduce
communication\ifFull associated with garbled gates.\else.\fi 

\medskip \noindent \textbf{Secret sharing.} SS techniques allow
for private values to be split into random shares, which are distributed
among a number of parties, and perform computation directly on secret shared
values without computationally expensive cryptographic operations. Of a
particular interest to us are linear threshold SS schemes. With
a $(n,\tau)$-secret sharing scheme, any private value is secret-shared among
$n$ parties such that any $\tau + 1$ shares can be used to reconstruct the
secret, while $\tau$ or fewer parties cannot learn any information about the
shared value, i.e., it is perfectly protected in the information-theoretic
sense. In a linear SS scheme, a linear combination of
secret-shared values can be performed by each party locally, without any
interaction, but multiplication of secret-shared values requires
communication between all of them.

In this setting, we can distinguish between the input owner who provide
input data into the computation (by producing secret shares), computational
parties who conduct the computation on secret-shared values, and output
recipients who learn the output upon computation termination (by
reconstructing it from shares). These groups can be arbitrarily overlapping
and be composed of any number of parties as long as there are at least 3
computational parties. 

In the rest of this work, we assume that Shamir SS \cite{shamir} is
used with $\tau < n/2$ in the semi-honest setting for any $n \ge 3$.
\else 
\fi

\section{Secure Building Blocks}
\label{sec:bb}

To be able to build secure protocols for different fingerprint recognition
algorithms, we will need to rely on secure algorithms for performing a
number of arithmetic operations, which we discuss in this section. Most of
the computation 
is performed on real
numbers, while for some of its components integer arithmetic will suffice
(e.g., Algorithm \ref{alg-matching} can be executed on integers when its
inputs are integer values). Complex numbers are represented as a pair (a
real part and an imaginary part) of an appropriate data type.

For the purposes of this work, we choose to use fixed-point representation
for real numbers. Operations on fixed-point numbers are generally faster
than \ifFull operations \else those \fi using floating-point representation (see, e.g.,
\cite{ali13}), and fixed-point representation provides sufficient precision
for this application.
When it is relevant to the discussion, we assume that integers are
represented using $\ell$ bits and fixed-point values are represented using
the total of $\ell$ bits, $k$ of which are stored after the radix point (and
thus $\ell-k$ are used for the integer part).

In the two-party setting based on GCs, complexity of an operation is
measured in the number of all and non-XOR Boolean gates. In the multi-party
setting based on SS, we count the total number of interactive operations as
well as the number of sequential interactive operations (rounds).

\ifFull
In what follows, we start by listing building blocks from prior literature
that we utilize in our solutions (Section~\ref{BB1}) and then proceed with
presenting our design for a number of new building blocks for which we did
not find secure realizations in the literature (Section~\ref{BB2}).

\subsection{Known Building Blocks}
\label{BB1}
\fi

\ifFull
Some of the operations used in the computation are elementary and are
well-studied in the security literature (e.g.,
\cite{catrina-int,catrina-fp,marina4,bla16}), while others are more complex,
but still have presence in prior work (e.g.,
\cite{marina4,catrina-fp,marina6,marina7}).
\else
We provide description of known building blocks (or those that could be
readily assembled from known building blocks) which we use in Appendix
\ref{app:kbb}.
They include arithmetic operations $+$/$-$, $\cdot$, $\sf Div$,
comparisons $\sf LT$, $\sf EQ$, type conversions $\sf Int2FP$, $\sf
FP2Int$, conditional statements, and operations on sets $\sf Min/Max$,
$\sf PreMul$, $\sf Comp$, $\sf Lookup$, and $\sf Sort$. {\color{black}A
construction for square root $\sf Sqrt$ using SS is available
in \cite{liedel1}, but in Appendix~\ref{app:SR} we show how to optimize it
for the GC setting.  {\color{black}In what follows, we describe our design
of new building blocks such as trigonometric and inverse trigonometric
functions and selection. We note that sine and cosine protocols were
recently published in \cite{kerik16} and the structure of our protocols is
similar to that used in \cite{kerik16}, i.e., based on polynomial
approximation. For other protocols (arctangent and selection), we did not
find secure realizations in the literature.} 
\fi 

\ifFull
{
\begin{itemize}
  \item \emph{Addition} $[c] \leftarrow [a] + [b]$ and \emph{subtraction}
    $[c] \leftarrow [a] - [b]$ are considered free (non-interactive) using
    SS using both fixed-point and integer representations
    \cite{catrina-fp}. Their cost is $\ell$ non-free gates for $\ell$-bit
    $a$ and $b$ \cite{kolesnikov} using GCs for both integer
    and fixed-point representations.
    
  \item \emph{Multiplication} $[c] \leftarrow [a] \cdot [b]$ of integers
    involves 1 interactive operation (in 1 round) using SS. For
    fixed-point numbers, truncation of $k$ bits is additionally needed,
    resulting in $2k+2$ interactive operations in 4 rounds (which reduces to
    2 rounds after pre-computation) \cite{catrina-fp}. Using GCs,
    multiplication of $\ell$-bit values (both integer and
    fixed-point) involves $2\ell^2-\ell$ non-free gates using the
    traditional algorithm \cite{kolesnikov}. This can be reduced using the
    Karatsuba's method \cite{kara}, which results in fewer gates when $\ell
    > 19$ \cite{sadeghi1}. Note that truncation has no cost in Boolean
    circuits.
   
  \item \emph{Comparison} $[c] \leftarrow {\sf LT}([a], [b])$ that tests for
    $a < b$ (and other variants) and outputs a bit involves $4\ell-2$
    interactive operations in 4 rounds (which reduces to 3 rounds after
    pre-computation) using SS \cite{catrina-int} (alternatively,
    $3\ell-2$ interactive operations in 6 rounds). This
    operation costs $\ell$ non-free gates using GCs
    \cite{kolesnikov}. Both implementations work with integer and
    fixed-point values of length $\ell$.
  
  \item \emph{Equality testing} $[c] \leftarrow {\sf EQ}([a], [b])$
    similarly produces a bit and costs $\ell+4\log \ell$ interactive 
    operations in 4 rounds using SS \cite{catrina-int}. GC-based
    implementation requires $\ell$ non-free gates
    \cite{kol08}. The implementations work with both integer and fixed-point
    representations.

 \item \emph{Division} $[c] \leftarrow {\sf Div}([a], [b])$ is available in
   the literature based on different underlying algorithms and we are
   interested in the fixed-point version. A fixed-point division based on
   SS is available from \cite{catrina-fp} which uses
   Goldschmidt's method. The algorithm proceeds in $\xi$ iterations, where
   $\xi = \lceil\log _2(\frac{\ell}{3.5})\rceil$. The same underlying
   algorithm could be used to implement division using GCs, but
   we choose to use the readily available solution from \cite{marina4} that
   uses the standard (shift and subtraction) division algorithm. The
   complexities of these implementations are given in \ifFull Table~\ref{tab1} \else Table~\ref{tab2-app}\fi.
 
 \item \emph{Integer to fixed-point conversion} $[b] \leftarrow {\sf
     Int2FP}([a])$ converts an integer to the fixed-point representation by
   appending a number of zeros after the radix point. It involves no
   interactive operations using SS and no gates using GCs.
 
 \item \emph{Fixed-point to integer conversion} $[b] \leftarrow {\sf
     FP2Int}([a])$ truncates all bits after the radix point of its input. It
   involves no gates using GCs and costs $2k+1$ interactive
   operations in 3 rounds to truncate $k$ bits using SS \cite{catrina-fp}.
 
  \item \emph{Conditional statements with private conditions} of the form 
    ``if $[priv]$ then $[a] = [b]$; else $[a] = [c]$;'' are transformed into
    statements $[a] = ([priv] \wedge [b]) \vee (\neg [priv] \wedge [c])$,
    where $b$ or $c$ may also be the original value of $a$ (when only a
    single branch is present). Our optimized implementation of this
    statement using GCs computes $[a] = ([priv] \wedge ([b]
    \oplus [c])) \oplus [c]$ with the number of non-XOR gates equal to the
    bitlength of variables $b$ and $c$. Using SS,
    we implement the statement as $[a] = [priv]\cdot([b]-[c])+[c]$
    using a single interactive operation.
    
  \item \emph{Maximum or minimum} of a set $\langle [a_{max}], [i_{max}]
    \rangle \leftarrow {\sf Max}$ $([a_1], \ldots,$ $ [a_m])$ or $\langle
  [a_{min}], [i_{min}] \rangle \leftarrow {\sf Min}([a_1], \ldots, [a_m])$,
  respectively, is defined to return the maximum/minimum element together
  with its index in the set. The operation costs $2\ell(m-1)+m+1$ non-free
  gates using GCs, where $\ell$ is the bitlength of the
  elements $a_i$ \cite{kolesnikov}. Using SS, the cost is dominated by the
  comparison operations, giving us $4\ell(m-1)$ interactive operations.
  Instead of performing comparisons sequentially, they can be organized into
  a binary tree with $\lceil \log m \rceil$ levels of comparisons. Then in
  the first iteration, $m/2$ comparisons are performed, $m/4$ comparisons in
  the second iteration, etc., with a single comparison in the last
  iteration. This allows the number of rounds to grow \mbox{logarithmically with
  $m$ and give us $4 \lceil \log m \rceil + 1$ rounds.}

  When each record of the set contains multiple fields (i.e., values other
  than those being compared), the cost of the operation increases by $m-1$
  non-free gates for each additional \emph{bit} of the record using GCs
  and by $m-1$ interactive operations for each additional
  \emph{field element} of the record without increasing the number of
  rounds.

\item \emph{Prefix multiplication} $\langle [b_1], {\ldots}, [b_m] \rangle
  \leftarrow {\sf PreMul} $ $([a_1], {\ldots},$ $[a_m])$ simultaneously computes
  $[b_i] = \prod _{j = 1}^i [a_j]$ for $i = 1, \ldots, m$. We also use an
  abbreviated notation $\langle [b_1], {\ldots},$ $[b_m] \rangle \leftarrow
  {\sf PreMul}($ $[a], m)$ when all $a_i$'s are equal. In the SS
  setting, this operation saves the number of rounds (with GCs,
  the number of rounds is not the main concern and multiple multiplications
  can be used instead of this operation). The most efficient constant-round
  implementation of $\sf PreMul$ for integers is available from
  \cite{catrina-int} that takes only 2 rounds and $3m-1$ interactive
  operations. The solution, however, is limited to non-zero integers. For fixed-point values, {\color{black}we use the structure
  of the solution for arbitrary prefix operations from \cite{catrina-int}
  which in this case results in invoking $0.5m\log m$ fixed-point
  multiplications in $\log m$ iterations of $0.5m$ operations each. This
  gives us the total of $0.5m \log m(2k+2)$ interactive operations executed
  in $2\log m +2$ rounds. In the special case when all $a_i$s are equal, we
  can optimize this solution to save many multiplications and execute only
  $m-1$ fixed-point multiplications. This results in $(m -1)(2k+2)$
  interactive operations in $2\log m +2$ rounds.}

  \item \emph{Compaction} $\langle [b_1], {\ldots}, [b_m] \rangle \leftarrow
    {\sf Comp}([a_1], {\ldots}, [a_m])$ pushes all non-zero elements of its
    input to appear before any zero element of the set. We are interested in
    order-preserving compaction that also preserves the order of the
    non-zero elements in the input set. A solution from \cite{bla15esorics}
    (based on data-oblivious order-preserving compaction in \cite{goodrich})
    can work in both GCs and SS settings using any type of input data. In
    this paper we are interested in the variant of compaction that takes a
    set of tuples $\langle a'_i, a''_i \rangle$ as its input, where each
    $a'_i$ is a bit that indicates whether the data item $a''_i$ is zero or
    not (i.e., comparison of each data item to 0 is not needed). The
    complexities of this variant are given in \ifFull Table~\ref{tab1} \else Table~\ref{tab2-app}\fi.

  \item \emph{Array access at a private index} allows to read or write an
    array element at a private location. In this work we utilize only read
    accesses and denote the operation as a lookup $[b] \leftarrow {\sf
    Lookup}(\langle [a_1], {\ldots}, [a_m] \rangle, [ind])$. The array
    elements $a_i$ might be protected or publically known, but the index is
    always private. Typical straightforward implementations of this
    operations include a multiplexer (as in, e.g., \cite{picco}) or
    comparing the index $[ind]$ to all positions of the array and obliviously
    choosing one of them. Both implementations have complexity $O(m \log m)$
    and work with GCs and SS techniques and data of different types. Based
    on our analysis and performance of components of this functionality, a
    multiplexer-based implementation outperforms the comparison-based
    implementation for GCs, while the opposite is true for SS-based
    techniques. We thus report performance of the best option for SS and GC
    settings in \ifFull Table~\ref{tab1} \else Table~\ref{tab2-app}\fi.
    
    Each record $a_i$ can be large, in which case the
    complexity of the operation additionally linearly grows with the size of
    array elements (or the number of field elements that each array stores
    in the SS setting).
    
\item \emph{Oblivious sorting} $\langle [b_1], \ldots, [b_m]\rangle
  \leftarrow {\sf Sort}([a_1],$ $ \ldots, [a_m])$ obliviously sorts an
  $m$-element set. While several algorithms of complexity $O(m \log m)$ are
  known, in practice the most efficient oblivious sorting is often the
  Batcher's merge sort \cite{bat68}.
  According to \cite{bla16}, the algorithm involves $\frac{1}{4}m(\log^2m -
  \log m + 4)$ compare-and-exchange operations that compare two elements and
  conditionally swap
  \ifFull them.
  \else them and the overall complexity is given in Table~\ref{tab2-app}.\fi
\end{itemize}
}
\fi

\ifFull
\noindent The complexities of all building blocks are listed in \ifFull Table~\ref{tab1} \else Table~\ref{tab2-app}\fi, and
notation is explained with each respective protocol. All functions with the
exception of $\sf Int2FP$ and $\sf FP2Int$ the associated integer and 
fixed-point variants, performance of which might differ in the SS sharing.
Because most protocols exhibit the same performance for integer and
fixed-point variants, for the functions with different performance, we list
both variants (integer followed by fixed-point) separated by ``:''.
\fi

\ifFull
\begin{table*}[t] \centering \small \setlength{\tabcolsep}{0.5ex} 
\begin{tabular}{|c|c|c|c|c|} 
\hline
\multirow{2}{*}{Prot.} & \multicolumn{2}{c|}{Secret sharing} & \multicolumn{2}{c|}{Garbled circuits}\\
\cline{2-5}
& Rounds & Interactive operations & XOR gates & Non-XOR gates \\
\hline

${\sf Add}/{\sf Sub}$ & 0 & 0 & $4\ell$  & $\ell$ \\
\hline
${\sf LT}$ & 4 & $4\ell-2$ & $3\ell$ & $\ell$\\
\hline
${\sf EQ}$ & 4 & $\ell+4\log \ell$ & $\ell$ & $\ell$\\
\hline
${\sf Mul}$ & 1 : 4 & 1 : $2k+2$ & $4\ell ^2-4\ell$  & $2\ell ^2-\ell$\\
\hline
\multirow{2}{*}{${\sf Div}$} & \multirow{2}{*}{$-:3\log \ell+2\xi +12$}  & $-:1.5\ell \log \ell+ 2 \ell \xi$ & \multirow{2}{*}{$7\ell ^2+7\ell$}  &  \multirow{2}{*}{$3\ell ^2+3\ell$}\\
 & & $+ 10.5 \ell +4 \xi+6$ & & \\
\hline
${\sf PreMul}$ & 2 : $2 \log m  + 2$ & $3m-1$ : $(m-1)(2k+2)$ & $-$ & $-$\\
\hline
${\sf Max}/{\sf Min}$ & $4
\log m + 1$ & $4\ell (m-1)$ & $5\ell (m-1)$ & $2\ell (m-1)$\\
\hline
${\sf Int2FP}$ & 0 & 0 & 0  & 0\\
\hline
${\sf FP2Int}$ & 3  & $2k+1$ & 0  & 0\\
\hline
\multirow{2}{*}{${\sf Comp}$} & \multirow{2}{*}{$\log m+ \log \log m + 3$} & $m \log m \log \log m+4 m \log m$ & $(\ell +4)m \log m$ & $(2\ell +1)m \log m - 2 \ell m$\\

& & $-m +\log m+2$ & $- m \ell -4 \log m +\ell$ & $+(\ell -1) \log m +2\ell$ \\
\hline
\multirow{2}{*}{${\sf Sort}$} & \multirow{2}{*}{$2 \log m (\log m +1)+1$} & $\ell (m-0.25)(\log
^2m$ & $1.5 m \ell (\log ^2m$ & $0.5 m \ell (\log^2m$\\
& & $+\log m +4)$ & $+\log m +4)$ & $+\log m +4)$\\
\hline 
${\sf Lookup}$ & 5 & $m \log m +4m \log \log m+m$ & $m\ell + \log m -\ell$ & $m \log m + m (\ell -1)$\\
\hline
\end{tabular}
\caption{Performance of known secure building blocks on integer and
  fixed-point values.} \label{tab1}
\end{table*}

\fi

\ifFull
\subsection{New Building Blocks}
\label{BB2}

To build secure fingerprint recognition algorithms, we also need a number of
secure building blocks for rather complex operations that previously have not been
sufficiently treated in the literature. Thus, in this section, we present
secure constructions for a number of operations, including trigonometric
operations, and selection of the $f$th smallest element of a
set.


We explore the available algorithms for computing these functions (e.g.,
Chebyshev polynomials for trigonometric functions) and build protocols that
optimize performance in each of the chosen settings of two-party and
multi-party computation. Our optimizations for designing the building
blocks as well as the overall protocols focus on the specific costs of the
underlying techniques for secure arithmetic and ensure that we achieve
efficiency together with precision and provable security guarantees.
\fi

\ifFull
\subsubsection{Sine, Cosine, and Arctangent}
\else
\subsection{Sine, Cosine, and Arctangent}
\fi
\label{sec:trig}

Here, we give secure building blocks for three trigonometric 
functions, which are sine, cosine, and arctangent. We denote secure
protocols as $[b] \leftarrow {\sf Sin}([a])$, $[b] \leftarrow {\sf
Cos}([a])$, and $[b] \leftarrow {\sf Arctan}([a])$ for each respective
function. The output $b$ is a fixed-point value, while the input $a$ (for
sine and cosine) represented in degrees can be either integer or
fixed-point. For generality, we will assume that $a$ used with trigonometric
functions is also a fixed-point value, while slight optimizations are
possible when $a$ is known to be an integer (as in the fingerprint
algorithms considered in this work). 

There are a variety of different approximation algorithms that can be used
for trigonometric functions, many of which take the form of polynomial
evaluation.
The best option should combine high precision with low
computational complexity. Chebyshev polynomials are often used for simple
operations including trigonometric functions \cite{chev}, but based on the
application and desired precision different approaches can be used. For
instance, \cite{kerik16} proposed a secure sine protocol using Taylor
series for floating-point numbers. 

Upon examining the options, we chose to proceed with the
polynomials described in~\cite[Chapter~6]{computer-app}, as they achieve
good precision using only small degree polynomials. \ignore{Note that Taylor series
of trigonometric functions achieve an approximation when the input is very
close to some fixed point. However, in secure computation, we can only
assume that an input value belongs to a known range (e.g., $[0, 2\pi]$).
Therefore, we choose polynomial approximation over a range for inputs
from \cite{computer-app} instead of Taylor series.} The
polynomials used in \cite{computer-app} for trigonometric functions take the
form $P(x^2)$ or $xP(x^2)$ for some polynomial $P$ over variable $x$ (i.e.,
use only odd or even powers), which requires one to compute only half as
many multiplications as in a polynomial with all powers present.

Another very different approach is to precompute the values of these
functions for the desired precision and the range of input \ifFull values \fi and use
private lookup to select the output based on private input. This
approach was recently pursued in \cite{nav16}, but has scaling issues for
general function evaluation. \ifFull We elaborate on the possibility of
using this approach in Appendix~\ref{sec:trig2}, while this section treats
the more general solution based on polynomial evaluation. \else
Thus we treat the general solution based on polynomial evaluation.
\fi

The approach used in \cite{computer-app} offers two types of polynomial
approximations. The first type uses a regular polynomial $P(x)$ of degree
$N$ to approximate the desired function. The second type uses a rational
function of the form $P(x)/Q(x)$, where $P$ and $Q$ are polynomials of
degree $N$ and $M$, respectively. For the same desired precision, the second
option will yield lower degrees $N$ and $M$ and thus fewer multiplications
to approximate the function. This option, however, is undesirable when the
division operation is much costlier than \ifFull the multiplication
operation. \else multiplication. \fi In
our setting, the rational form is preferred in the case of GC
evaluation (at least for higher precisions) where multiplication and division have similar costs. However,
with SS, the cost of division is much higher than that of
multiplication, and we use the first option with regular polynomial
evaluation. 

It is assumed in \cite{computer-app} that the initial range reduction to
$[0, \pi/2]$ is used for the input to trigonometric functions. This means
that if input $a$ to sine or cosine is given in degrees, it needs to be
reduced to the range $[0, 90]$ and normalized to the algorithm's
expectations. Note that it is straightforward to extend the result of the
computation to cover the entire range $[0, 2\pi]$ given the output of this
function and the original input.

Furthermore, because evaluating trigonometric functions on a smaller range
of inputs offers higher precision, it is possible to apply further range
reduction and evaluate the function on even a smaller range of inputs, after
which the range is expanded using an appropriate formula or segmented
evaluation. We refer the reader to \cite{computer-app} for additional
detail. For that reason, the tables of polynomial coefficients provided in
\cite{computer-app} are for $\sin(\alpha \pi x)$ and $\cos(\beta \pi x)$,
where $0 \le x \le 1$ and $\alpha$ and $\beta$ are fixed constants equal to 
1/2 or less. To see the benefit provided by smaller $\alpha$ or $\beta$, for
example, consider the sine function. Then by using $\alpha = 1/2$ and
approximating the function using a regular polynomial of degree 7, we obtain
precision to 15.85 decimal places, while with $\alpha = 1/6$ and the same
polynomial degree 25.77 decimal places of the output can be computed.
However, for simplicity, in this work we suggest to choose $\alpha$ and
$\beta$ equal to 1/2,
as this option will not require non-trivial post-computation to compensate
for the effect of computing the function on a different angle.

Based on the desired precision, one can retrieve the
minimum necessary polynomial degree used in the approximation to achieve the
desired precision, then look up the polynomial coefficients and evaluate the
polynomial or polynomials on the input. As mentioned before, for
trigonometric functions only polynomials with even or odd powers are used,
and we have that sine is approximated as $xP(x^2)$ or $xP(x^2)/Q(x^2)$ (odd
powers) and cosine as $P(x^2)$ or $P(x^2)/Q(x^2)$ (even powers). \ifFull Thus, in
when a rational function is used to evaluate sine, we obtain the following
secure protocol. It assumes that the input is given in degrees in the range
[0, 360).

{\small \noindent \ifFull \line(1,0){470} \else \line(1,0){240} \fi \\ [-0.02in]
$[b] \leftarrow {\sf Sin}([a])$ \\[-0.09in]
\ifFull \line(1,0){470} \else \line(1,0){240} \fi
\begin{compactenum}
  \item Compute $[s] = {\sf LT}(180, [a])$.
  \item If ($[s]$) then $[a] = [a] - 180$.
    
  \item If (${\sf LT}(90, [a])$) then $[a] = 180 - [a]$.
    
  \item Compute $[x] = \frac{1}{90} [a]$ and then $[w] = [x]^2$.
    
  \item Lookup the minimum polynomial degrees $N$ and $M$ using $\alpha =
    1/2$ for which precision of the approximation is at least $k$ bits.
  \item Lookup polynomial coefficients $p_0, {\ldots}, p_N$ and $q_0,
    {\ldots}, q_M$ for sine approximation.
    
  \item Compute \mbox{$([z_1], \ldots, [z_{\max(N, M)}]) \leftarrow {\sf
      PreMul}([w], \max(N, M))$.}
    
  \item Set $[y_P] = p_0 + \sum_{i=1}^N p_i [z_i]$.
    
  \item Set $[y_Q] = q_0 + \sum_{i=1}^M q_i [z_i]$.
    
  \item Compute $[y] \leftarrow {\sf Div}([y_P], [y_Q])$.
    
  \item If $([s])$ then $[b] = 0 - [x]$ else $[b] = [x]$.
    
  \item Compute and return $[b] = [b] \cdot [y]$.
\end{compactenum}
\vspace{-0.1in}
\ifFull \line(1,0){470} \else \line(1,0){240} \fi}

\noindent Recall that we recommend using a rational function in the form of
$P(x)/Q(x)$ as in the above protocol for GCs-based implementation of high
precision. With SS, we modify the protocol to evaluate only a single
polynomial $P$ of (a different) degree $N$ and skip the division operation
in step 10. Also, based on our calculations, the single polynomial variant
is also slightly faster in the GC setting when precision is not high (e.g.,
with 16 bits of precision).

As far as optimizations go, we, as usual, simultaneously execute independent
operations in the SS setting. Also, note that because coefficients $p_i$'s
and $q_j$'s are not integers, evaluations of polynomials in steps 8 and 9 is
not local in the SS setting and requires truncation. We, however, can reduce
the cost of truncating $N$ (respectively, $M$) values to the cost of one
truncation. This is because we first add the products and truncate the sum
once. This also applies to polynomial evaluation in other protocols such as
cosine and arctangent.

When input $a$ is known to lie in the range $[0, 90]$ (as in the fingerprint
algorithms we consider in this work), steps 1, 2, 3, 11, and 12 are not
executed and instead we return $[y] \cdot [x]$ after evaluating step 10.
Lastly, when input $a$ is known to be an integer, the comparisons in steps 1
and 3 and the multiplication of $[a]$ and a fixed-point constant
$\frac{1}{90}$ in step 4 become more efficient in both SS and
GCs-based implementations (i.e., both settings benefit from the
reduced bitlength of $a$ if only integer part is stored and the cost of the
multiplication in the SS setting becomes 1 interactive
operation).

\ifFull
To show security of this protocol, we need to build simulators 
according to Definition~\ref{def:security}. 
The main argument here is that because we
only use secure building blocks that do not reveal any private information,
we apply Canetti's composition theorem~\cite{can00} to result in security of
the overall construction. More precisely, to simulate the adversarial view,
we invoke simulators corresponding to the building blocks. In more detail,
if we assume an implementation based on a $(n, \tau)$-threshold linear
secret sharing, our protocols inherit the same security guarantees as those
of the building blocks (i.e., perfect or statistical security in the
presence of secure channels between the parties with at most $\tau$ corrupt
computational parties) because no information about private values is
revealed throughout the computation. More formally, to comply with the
security definition, it is rather straightforward to build a simulator for
our protocols by invoking simulators of the corresponding building blocks to
result in the environment that will be indistinguishable from the real
protocol execution by the participants. The same argument applies to other
protocols presented in this work and we do not include explicit analysis
(with the exception of selection). 
\fi
\fi

\ifFull
The cosine function is evaluated similarly to sine. The main difference is
in the way the input is pre- and post-processed for polynomial evaluation
due to the behavior of this function. When cosine is evaluated using a
rational function, we have the following secure protocol:

{\small \noindent \ifFull \line(1,0){470} \else \line(1,0){240} \fi \\ [-0.02in]
$[b] \leftarrow {\sf Cos}([a])$ \\[-0.09in]
\ifFull \line(1,0){470} \else \line(1,0){240} \fi
\begin{compactenum} 
  \item If (${\sf LT}(180, [a])$) then $[a] = 360 - [a]$.
  
  \item $[s] = {\sf LT}(90, [a])$.
  \item If ($[s]$) then $[a] = 180- [a]$. 
    
  \item Compute $[x] = \frac{1}{90} [a]$ and then $[x] = [x]^2$.
    
  \item Lookup the minimum polynomial degrees $N$ and $M$ using $\beta =
    1/2$ for which precision of the approximation is at least $k$ bits.
    
  \item Lookup polynomial coefficients $p_0, {\ldots}, p_N$ and $q_0,
    {\ldots}, q_M$ for cosine approximation.
    
  \item Compute $([z_1], \ldots, [z_{\max(N, M)}]) \leftarrow {\sf
      PreMul}([x], \max(N, M))$. 
    
  \item Set $[y_P] = p_0 + \sum_{i=1}^N p_i [z_i]$.
    
  \item Set $[y_Q] = q_0 + \sum_{i=1}^M q_i [z_i]$.
    
  \item Compute $[y] \leftarrow {\sf Div}([y_P], [y_Q])$.
    
  \item If $([s])$ then $[b] = 0 - [y]$ else $[b] = [y]$.
  
  \item Return $[b]$.
\end{compactenum}
\vspace{-0.1in}
\ifFull \line(1,0){470} \else \line(1,0){240} \fi}

The same optimizations as those described for the sine protocol apply to the
cosine computation as well. Furthermore, to use a single polynomial to
evaluate cosine, we similarly lookup coefficients for a single polynomial of
(a different) degree $N$ and skip steps 9 and 10 in the $\sf Cos$ protocol. 

\else
{\color{black} Because the structure of our sine and cosine protocols is
  similar to the protocols in \cite{kerik16}, we omit further description
  due to space considerations and we refer the reader to the full version
  \cite{FingerprintFull} for additional details.}
\fi
Complexities of ${\sf Sin}$ and ${\sf Cos}$ are provided in
\ifFull Table~\ref{tab2} \else Table~\ref{tab2-short}\fi as function of
parameters $N$ and $M$. To achieve 16, 32, or 64 bits of precision,
$N$ needs to be set to 3, 5, or 9, respectively, for both ${\sf Sin}$ and
${\sf Cos}$ using a single polynomial, and $(N, M)$ need to be set to (2,
1), (3, 2), and (6, 2), respectively, for ${\sf Sin}$ using the rational
function approach.


\ignore{ Their details are:
$${\sf Sin}(\alpha x) = 2 \sum _{k = 0}^{\infty} (-1)^k J_{2k+1}(\alpha) T_{2k+1}(x).$$

$${\sf Cos}(\alpha x) = 2 \sum _{k = 0}^{\infty} J_{2k}(\alpha) T_{2k}(x).$$

$${\sf Arctan}(\rho x) = 2 \sum _{k = 0}^{\infty} \frac{-1^k}{2k+1}\tan^{2k+1}(\alpha) T_{2k+1}(x).$$

Where $T_k(x)$ is the Chebyshev polynomial of the first kind of order $k$, and $J_k(\alpha)$ is the Bessel function of the first kind. Also, the Bessel functions and $\tan$ can be precomputed because they do not depend on the variable $x$. In addition, $T_{2k}(x)=2 (T_{k}(x))^{2}-1$ as well as $T_{2k+1}(x)=2 T_{k+1}(x) T_{k}(x)-x$.
}

\ignore{Furthermore, in Chebyshev's approximation of ${\sf Cos}$, $T_{2k}(x)$ can be written as $2 (T_{k}(x))^{2}-1$ and therefore, ${\sf Cos}$ is approximated as $P(x^2)$ (even function). Similarly, $T_{2k+1}(x)$ in Chebyshev's approximation of ${\sf Sin}$ can be written as $2 T_{k+1}(x) T_{k}(x)-x$ and therefore, ${\sf Sin}$ is approximated as $x P(x^2)$ (odd function).}

\ifFull
\begin{table*}[t] \centering \small \setlength{\tabcolsep}{0.5ex} 
\begin{tabular}{|c|c|c|c|c|} 
\hline
\multirow{2}{*}{Prot.} & \multicolumn{2}{c|}{Secret sharing} & \multicolumn{2}{c|}{Garbled circuits}\\
\cline{2-5}
& Rounds & Interactive operations & XOR gates & Non-XOR gates \\
\hline

\multirow{3}{*}{${\sf Sin}$} & \multirow{3}{*}{$2 \log N +16$} &
\multirow{3}{*}{$2Nk+8 \ell +2N + 6k + 4$} & $(\max(N,M)+N+M+2)$ &
$(\max(N,M)+N+M+2)$\\
 & & & $\times(4\ell^2-4\ell)+7 \ell ^2$ & $\times(2\ell ^2-\ell)+3 \ell ^2$\\
& & & $+4\ell (N+M) + 31\ell$ & $+\ell (N+M) + 11\ell$ \\
\hline
\multirow{3}{*}{${\sf Cos}$} & \multirow{3}{*}{$2 \log N +15$} &
\multirow{3}{*}{$2Nk+8 \ell + 2N + 4k + 2$} & $(\max(N,M)+N+M+1)$ & $(\max(N,M)+N+M+1)$\\
 & & & $\times(4\ell^2-4\ell) + 7\ell^2$ & $\times(2\ell^2-\ell) + 3\ell^2$\\
& & & $+4\ell (N+M) + 31\ell$ & $+ \ell (N+M) + 11\ell$ \\

\ignore{
\multirow{2}{*}{${\sf Sin}$} & \multirow{2}{*}{$15+2 \log N $} & $8 \ell + 2Nk$ & $18 \ell + 12Nk^2- 4Nk$ & $6 \ell + 6Nk^2-Nk^2$\\
 & & $+ 2N + 4k + 4$ & $8k^2 - 8k +2$ & $+4k^2 - k + 1$\\
\hline
\multirow{2}{*}{${\sf Cos}$} & \multirow{2}{*}{$14+2 \log N$} & $8 \ell + 2Nk$ & $18 \ell + 12Nk^2- 4Nk$ & $6 \ell + 6Nk^2-Nk^2$\\
 & & $+ 2N +2k +2 $ &  $8k^2 - 8k +2$ & $+4k^2 + 1$\\
\hline
${\sf Sin}$ & $3\log k$ & $8 \ell + 2N'k+ 2N'$ & $18 \ell + 12N'k^2$ & $6 \ell + 6N'k^2$\\
(rati- & $+2 \log N'$ & $+ 14.5k + 1.5 k \log k$ &  $- 4N'k+15k^2$ & $-N'k^2+7k^2$\\
onal) & $+2\xi+27$ & $+2k\xi+4\xi+10$ & $ - k +2$ & $ +2k + 1$\\
\hline
${\sf Cos}$ & $3\log k$ & $8 \ell + 2N'k+ 2N'$ & $18 \ell + 12N'k^2$ & $6 \ell + 6N'k^2$\\
(rati- & $+2 \lceil \log N'\rceil$ & $+ 12.5k + 1.5 k \log k$ & $- 4N'k+15k^2$ & $-N'k^2+7k^2$\\
onal) & $+2\xi+26$ & $+2k\xi+4\xi+8$ & $ - k +2$ & $ +3k + 1$\\
}
\hline
\multirow{2}{*}{${\sf Arctan}$} & $2\log N + 3\log\ell$ & $1.5\ell\log\ell
+ 2 \ell \log(\frac{\ell}{3.5}) + 2Nk$ & \multirow{2}{*}{$8N\ell^2+3 \ell^2 -
4N\ell + 43\ell$} & \multirow{2}{*}{$4N\ell^2 + \ell^2 - N \ell + 15\ell$}\\
& $+ 2\log(\frac{\ell}{3.5})+22$ & $+18.5 \ell+2 +4 \log(\frac{\ell}{3.5})+6$
& & \\ \hline
\multirow{2}{*}{${\sf Sqrt}$} & $0.5\ell+2\log \ell$ & $2\ell^2+\ell \log \ell
+3k(\xi +1)$ & $12 \xi \ell^ 2+12.5 \ell^2-k^2 + \ell k$ & $6 \xi \ell^2 +
6.5\ell^2-k^2 + \ell k$\\
& $+6\xi+24$ & $+5\ell +6 \xi +12$ & $ -8 \xi \ell-7.5  \ell +k -2$ & $-2 \xi
\ell-0.5 \ell+k -4$\\ \hline
\multirow{4}{*}{${\sf Select}$} & $c(2\log ^2 t_1$ & $c(\ell (t_1-0.25)
(\log^2t_1+\log t_1 +4)$ & $c(1.5 t_1 \ell (\log^2 t_1+\log t_1 +4)$ & $c(0.5 t_1 \ell (\log ^2 t_1+\log t_1 +4)$\\
& $+2\log ^2 t_2$ & $+\ell (t_2-0.25) (\log^2 t_2+\log t_2 +4)$ & $+ 1.5
t_2 \ell (\log ^2 t_2+\log t_2 +4)$ & $+ 0.5 t_2 \ell (\log^2 t_2 +\log t_2 +4)$ \\
& $+6\log m+2$ & $+2m\log m \log \log m+8m \log m+ $ & $+(2\ell + 20.5)m
\log m+m$ & $+(4\ell +5)m \log m+m$\\
& $ \log \log m+17)$ & $18m\ell-5.5 m+2\log m-8\ell+9)$ & $+12.5m\ell-6\log
m-8\ell)$ & $+4m\ell+(2\ell -1)\log m)$\\ 
\hline
\end{tabular}
\caption{Performance of proposed secure building blocks for fixed-point values.} \label{tab2}
\end{table*}
\fi

In the case of inverse trigonometric functions and arctangent in particular,
the function input domain is the entire $(-\infty, \infty)$ and the range of
output is $(-\pi/2, \pi/2)$. We recall that arctangent is an odd function
(i.e., $\arctan(-x)= - \arctan(x)$) and for all positive inputs we have
$\arctan(x)+ \arctan(\frac{1}{x})= \pi/2$. It is, therefore,
sufficient to approximate arctangent over the interval of $[0,1]$. To this end,
we use the technique introduced by Medina in \cite{medinaArctan} and its
formal proof in \cite{medinaProof}. In particular, \cite{medinaArctan}
defines a sequence of polynomials  over the input
domain of $[0,1]$, denoted as $h_N(x)$, with the property that
\ifFull
$$
\else
$
\fi
|h_N(x) - \arctan(x)| \leq \left(\frac{1}{4^{5/8}}\right)^{\mathrm{deg}(h_N)+1}.
\ifFull
$$
\else
$
\fi
We use this formula to determine the degree $N$ for any $k$-bit precision.
This degree $N$ is logarithmic with respect to the desired precision.
Afterwards, the coefficients of $h_N(x)$ are computed from the recursive
definitions in \cite{medinaArctan}. We choose this approach over other
alternatives such as \cite{computer-app} for its efficiency. As an example,
in \cite{computer-app} the authors propose to divide $[0, \infty)$ into
$s+1$ segments for some value of $s$ and perform a search to discover in
which interval the input falls. Afterwards, the input is transformed into
another variable (involving division) whose value is guaranteed to be within
a small fixed interval and arctangent of this new variable is approximated
using standard polynomial approximation techniques. In this approach, the
search can be performed using binary search trees whose secure realization
is non-trivial or alternatively will involve $s$ steps. This extra
computation makes the overall performance of the solution in
\cite{computer-app} worse than that of \cite{medinaArctan}'s approach used
here. Another candidate for arctangent evaluation is the well-known Taylor
series of arctangent. However, Taylor series provides possibly reasonable
precision if the input is bound to be very close to one point from the input
domain, which we cannot assume. In addition, arctangent Taylor series
converges very slowly. For example, for a three decimal precision on inputs
around 0.95 Taylor series requires a polynomial of degree 57, whereas
\cite{medinaArctan}'s approach requires a 7th-degree polynomial
\cite{medinaProof}.
Our secure protocol to approximate arctangent based on the approach from
\cite{medinaArctan} is given next:

{\small \noindent \ifFull \line(1,0){470} \else \line(1,0){240} \fi \\ [-0.02in]
$[b] \leftarrow {\sf Arctan}([a])$ \\[-0.09in]
\ifFull \line(1,0){470} \else \line(1,0){240} \fi
\begin{compactenum} 
  \item Compute $[s] \leftarrow {\sf LT}([a], 0)$.

  \item If $([s])$ then $[x] = 0 - [a]$ else $[x] = [a]$.

  \item Compute $[c] \leftarrow {\sf LT}(1, [x])$.
  \item If $([c])$ then $[d]=\pi/2$, $[y] \leftarrow {\sf Div}(1, [x])$; else
    $[d]=0$, $[y]=[x]$.
    
  \item Lookup the minimum polynomial degree $N$ for which precision of the approximation is at least $k$ bits.
    
  \item Lookup polynomial coefficients $p_0, {\ldots}, p_N$ for arctangent approximation from \cite{medinaArctan}.
    
  \item Compute $([z_1], \ldots, [z_N]) \leftarrow {\sf
      PreMul}([y], N)$. 
    
  \item Set $[z] = p_0 + \sum_{i=1}^N p_i [z_i]$.
    
  \item If $([c])$ then $[z] = [d] - [z]$ else $[z] = [d] + [z]$.

  \item If $([s])$ then $[b] = 0 - [z]$ else $[b] = [z]$.
  
  \item Return $[b]$.
\end{compactenum}
\vspace{-0.1in}
\ifFull \line(1,0){470} \else \line(1,0){240} \fi}

\noindent
The complexity of this protocol can be found in \ifFull Table~\ref{tab2} \else Table~\ref{tab2-short}\fi.

\textbf{Security analysis.} 
To show security of $\sf Sin$, $\sf Cos$, and $\sf Arctan$,
we build simulators according to Definition~\ref{def:security} \ifFull \else in
Appendix \ref{app:def}\fi that simulate the adversarial view without access to
private inputs of honest parties. The main argument is that because
we only use secure building blocks that do not reveal any information
about private inputs and compute the output in a protected form, we
can apply Canetti's composition theorem~\cite{can00} to result in
security of the overall construction. More precisely, to simulate the
adversarial view, we invoke simulators corresponding to the building
blocks. Suppose a setup based on a $(n, \tau)$-threshold linear SS
is used; our protocols inherit the same security guarantees as
those of the building blocks (i.e., perfect or statistical security in
the presence of secure channels between the parties with at most
$\tau$ corrupt computational parties) because no information about
private values is revealed throughout the computation. More formally,
it is rather straightforward
to build a simulator for our protocols by invoking simulators of the
corresponding building blocks to result in the environment that will
be indistinguishable from the real protocol execution by the
participants. {\color{black}A detailed simulator to prove security of
the \ifFull $\sf Sin$ \else $\sf Arctan$ \fi protocol is given in \ifFull the following.\else Appendix~\ref{app:def}.\fi } The same
argument applies to other protocols presented in this work and we do
not include explicit analysis (with the exception of selection, where
some intermediate values are opened during protocol execution).

\ifFull
{\color{black}
\begin{theorem}Protocol $\sf Sin$ is secure according to
Definition~\ref{def:security}.
\end{theorem}
\textbf{Proof.} For concreteness of the proof, let us assume a SS setting
with $n$ computational parties from which at most $\tau < \frac{n}{2}$ are
corrupt. In the ideal world, $S_I$ simulates the view of the corrupted
parties $I$. It is given access to the corrupted parties' inputs (${\sf
in}_I$) and their outputs ($f_I({\sf in}_1, {\ldots}, {\sf in}_n)$). In the
current formulation of $\sf Sin$, the input $a$ is provided in the form
of secret shares and thus is not available to a participant in the clear and
similarly output $b$ is computed in a shared form and is not given to any
given party. We, however, also consider the variant when the output is
opened to a corrupt participant from $I$, in which case the simulator
obtains access to $b$ and must ensure that the adversary is able to
reconstruct the correct output $b$. 

We build $S_I$ as follows: during step 1 of $\sf Sin$, $S_I$ invokes the simulator for $\sf
LT$. In step 2, $S_I$ invokes simulator for (integer) multiplication
used in executing the conditional statements (note that subtraction is
local). Similarly, in step 3, $S_I$ invokes both simulators for $\sf
LT$ and (integer) multiplication to simulate conditional execution.
In steps 4, 8, and 9, $S_I$ invokes a simulator for floating-point
multiplication several times. In steps 7 and 10, $S_I$ invokes the
simulators for (floating-point) prefix multiplication and divisions,
respectively. Step 11 is simulated in the same way as step 2. In step
12, $S_I$ invokes the simulator for floating-point multiplication. If no party in $I$ is entitled to learning output
$b$, $S_I$ simply communicates random shares to the
corrupted parties. Otherwise, $S_I$ has $b$ and uses it to set shares of 
honest parties' output in such a way that the shares reconstruct to the
correct output $b$ (honest parties' shares will be consequently communicated
to the output recipient). This is always possible because $\tau$ or fewer
shares are independent of a secret and the remaining shares can be set to
enable reconstruction of any desired value.}
\else
\fi

\ifFull
\subsubsection{Square Root}
\label{SR}

We now proceed with the square root computation defined by the interface $[b]
\leftarrow {\sf Sqrt}([a])$, where $a$ and $b$ are fixed-point values to
cover the general case.

Secure multi-party computation of the square root based on SS
has been treated by Liedel in \cite{liedel1}. The approach is based on the
Goldschmidt's algorithm, which is faster than the Newton-Raphson's method.
However, to eliminate the accumulated errors, the last iteration of the
algorithm is replaced with the self-correcting iteration of the
Newton-Raphson's method. For an $\ell$-bit input $a$ with $k$ bits after the
radix point, this protocol uses $2\ell^2 + \ell \log \ell + 3k(\xi+1) +
5\ell + 6\xi + 12$ interactive operations in $0.5\ell + 2\log \ell + 6\xi +
24$ rounds, where $\ell$ is assumed to be a power of 2 and $\xi = \lceil
\log_2(\ell/5.4) \rceil$ is the number of algorithm iterations. For the
purposes of this work, we use the protocol from \cite{liedel1} for the
SS setting. We also optimize the computation to be used with
GCs based on the specifics of that technique as described next. 
\ifFull Furthermore,
because polynomial approximation (similar to the way it was done for
trigonometric functions in section~\ref{sec:trig}) is attractive in the SS
setting, we provide an alternative solution in Appendix~\ref{sec:sqrt2}. \fi

Goldschmidt's method starts by computing an initial approximation for
$\frac{1}{\sqrt{a}}$, denoted by $b_0$, that satisfies $\frac{1}{2} \leq
ab_0^2 < \frac{3}{2}$. It then proceeds in iterations increasing the
precision of the approximation with each consecutive iteration. To
approximate $\frac{1}{\sqrt{a}}$, \cite{liedel1} uses an efficient
method (a linear equation) that expects that input $a$ is in the range
$[\frac{1}{2}, 1)$. Thus, there is a need to first normalize $a$ to $a'$
such that $\frac{1}{2} \le a' < 1$ and $a = a' \cdot 2^w$. Note that
$\frac{1}{\sqrt{a}} = \frac{1}{\sqrt{a'}} \sqrt{2^{-w}}$, therefore, once we
approximate $\frac{1}{\sqrt{a'}}$, we can multiply it by $\sqrt{2^{-w}}$ to
determine an approximation of $\frac{1}{\sqrt{a}}$.

In our $(\ell,k)$-bit fixed-point representation, a normalized input $a' \in
[\frac{1}{2}, 1)$ has the most significant non-zero bit exactly at position
$k-1$. We also express $\sqrt{2^{-w}}$ as $\frac{1}{\sqrt{2}} 2^{-\lfloor
\frac{w}{2} \rfloor}$ when $w$ is odd and as $2^{-\lfloor \frac{w}{2}
\rfloor}$ when $w$ is even. Our normalization procedure $\sf Norm$ that we
present next thus takes input $a$ and returns $\frac{1}{2} \le a' < 1$
(normalized input $a$ in $(\ell,k)$-bit fixed-point representation),
$2^{-\lfloor \frac{w}{2} \rfloor}$ and bit $c$ set to 1 when $w$ is odd and
to 0 otherwise. The protocol is optimized for the GC approach
with cheap XOR gates \cite{kol08}. \mbox{It assumes that $\ell$ and $k$ are even.}

{\small \noindent \ifFull \line(1,0){470} \else \line(1,0){240} \fi\\[-0.02in]
$\langle [a'], [2^{-\lfloor \frac{w}{2} \rfloor}], [c]\rangle \leftarrow {\sf
  Norm}([a])$\\[-0.09in] 
\ifFull \line(1,0){470} \else \line(1,0){240} \fi
\begin{compactenum}
\item $([a_{\ell-1}], \ldots, [a_0]) \leftarrow [a]$.

\item Set $[x_{\ell-1}]= [a_{\ell-1}]$.

\item For $i = \ell-2, \ldots, 0$ do $[x_i] =[a_i] \vee [x_{i+1}]$.

\item Set $[y_{\ell-1}]= [x_{\ell-1}]$.

\item For $i = 0, {\ldots}, \ell-2$ do in parallel $[y_i] =[x_i] \oplus [x_{i+1}]$.

\item For $i = 0, \ldots, \ell-1$ do in parallel $[z^{(i)}] \leftarrow
  ([a_{\ell-1}] \wedge [y_i], \ldots, [a_0] \wedge [y_i])$.

\ifFull 
\item Compute $[a'] = \left(\bigoplus_{i = 0}^{k-1} \left([z^{(i)}] \ll
  (k-1-i)\right)\right) \oplus \left(\bigoplus_{i = k}^{\ell-1}
  \left([z^{(i)}]\gg (i-(k-1))\right)\right)$.
\else
\item Compute $[a'] = \left(\bigoplus_{i = 0}^{k-1} \left([z^{(i)}] \ll
  (k-1-i)\right)\right) \oplus \\ \left(\bigoplus_{i = k}^{\ell-1}
  \left([z^{(i)}]\gg (i-(k-1))\right)\right)$.
\fi

\item Let $[u_0] = [y_0]$, $[u_{\frac{\ell}{2}}] = [y_{\ell-1}]$ and for $i =
    1, {\ldots}, \frac{\ell}{2}-1$ do in parallel $[u_i] =[y_{2i-1}] \oplus [y_{2i}]$.


\item Set $[2^{-\lfloor \frac{w}{2} \rfloor}] \leftarrow ([d_{\ell-1}],
  {\ldots}, [d_0]) = (0^{\ell-\frac{3k}{2}-1}, [u_0], [u_1], {\ldots},
  $ $[u_{\frac{\ell}{2}}], 0^{\frac{3k-\ell}{2}})$, where $0^x$ corresponds to
  $x$ zeros. 
  
\item Set $[c] = \bigoplus_{i =0}^{\frac{\ell}{2}-1} [y_{2i}]$.




\item Return $\langle [a'], [2^{-\lfloor \frac{w}{2} \rfloor}], [c] \rangle$.
\end{compactenum}
\ifFull \line(1,0){470} \else \line(1,0){240} \fi}

\noindent
Here, lines 2--3 preserve the most significant zero bits of $a$ and set the
remaining bits to 1 in variable $x$ (i.e., all bits following the most
significant non-zero bit are 1). Lines 4--5 compute $y$ as a vector with the
most significant non-zero bit of $a$ set to 1 and all other bits set to 0.
On line 6, each vector $z^{(i)}$ is either filled with 0s or set to $a$
depending on the $i$th bit of $y$ (thus, all but one $z^{(i)}$ can be
non-zero). Line 7 computes the normalized value of $a$ by aggregating all
vectors $z^{(i)}$ shifted an appropriate number of positions. Here operation $x
\ll y$ shifts $\ell$-bit representation of $x$ $y$ positions to the left by
discarding $y$ most significant bits of $x$ and appending $y$ 0s in place
of least significant bits. Similarly, $x \gg y$ shifts $x$ to the right by
prepending $y$ 0s in place of most significant bits and discarding $y$ least
significant bits of $x$. Note that we can use cheap XOR gates for this
aggregation operation because at most one $y_i$ can take a non-zero value.

Lines 8 and 9 compute $[2^{-\lfloor \frac{w}{2} \rfloor}]$. Because a pair of
consecutive $i$'s results in the same value of $w$, we first combine the
pairs on line 8 and shift them in place on line 9. As before, we can use
cheap XOR and free shift operations to accomplish this task because at most
one $y_i$ is set. Lastly, line 10 computes the bit $c$, which is set by
combining all flags $y_i$'s at odd distances from $k-1$.

Note that for simplicity of exposition, we AND and XOR all bits in steps 6
and 7. There is, however, no need to compute the AND for the bits discarded
in step 7 or compute the XOR with newly appended or prepended 0s. We obtain
that this protocol can be implemented as a circuit using $0.5\ell^2 - k^2 +
\ell k + 1.5\ell - 4$ non-XOR and $0.5\ell^2 - k^2  + \ell k + 0.5\ell-3$
XOR gates.

Once we determine normalized input $a'$, Liedel \cite{liedel1} approximates
$\frac{1}{\sqrt{a'}}$, denoted by $b'_0$, by a linear equation $b'_0 =
\alpha a' + \beta$, where $\alpha$ and $\beta$ are precomputed. The values
of these coefficients are set to $\alpha = -0.8099868542$ and $\beta =
1.787727479$ in \cite{liedel1} to compute an initial approximation $b'_0$
with almost 5.5 bits of precision (when $a'$ is in the range $[\frac{1}{2},
1)$). We then use $b'_0$ to compute $b_0$ that approximates
$\frac{1}{\sqrt{a}}$ as described above.

After the initialization, Goldschmidt's algorithm sets $g_0 = x b_0$ and
$h_0 = 0.5 b_0$ and proceeds in $\xi$ iterations that compute: $g_{i+1} =
g_i(1.5 - g_i h_i)$, $h_{i+1} = h_i(1.5- g_i h_i).$ The last iteration is
replaced with one iteration of Newton-Raphson's method to eliminate
accumulated errors that computes the following: $h_{i+1} =
h_i(1.5-0.5ah_i^2).$ This gives us the following square root protocol, which
we present optimized for the GC approach.

{\small \noindent \ifFull \line(1,0){470} \else \line(1,0){240} \fi\\[-0.02in]
$[b] \leftarrow {\sf Sqrt}([a])$\\[-0.09in]
\ifFull \line(1,0){470} \else \line(1,0){240} \fi
\begin{compactenum}
\item Let $\xi = \lceil\log _2(\frac{\ell}{5.4})\rceil$.

\item Execute $\langle [a'], [2^{-\lfloor \frac{w}{2} \rfloor}], [c] \rangle
  \leftarrow {\sf Norm}([a])$. 

\item Let $\alpha = -0.8099868542$, $\beta = 1.787727479$ and compute
  $[b'_0] = \alpha \cdot [a'] + \beta$. 
  
\item Compute $[b_0] = \left(\left([c] \wedge \frac{1}{\sqrt{2}}\right)
  \oplus \neg[c]\right) \cdot [2^{-\lfloor \frac{w}{2} \rfloor}] \cdot
  [b'_0]$. 

\item Compute $[g_0] = [a] \cdot [b_0]$ and $[h_0] = ([b_0] \gg 1)$.

\item For $i = 0, \ldots, \xi - 2$ do
\begin{compactenum}
\item $[x] = 1.5 - [g_i] \cdot [h_i]$.

\item if $(i < \xi-2)$ then $[g_{i+1}] = [g_i] \cdot [x]$.

\item $[h_{i+1}] = [h_i] \cdot [x]$.
\end{compactenum}

\item Compute $[h_{\xi}] = [h_{\xi-1}](1.5 - ([a] \cdot
  [h_{\xi-1}]^2 \gg 1))$.

\item Return $[b]=[h_{\xi}]$.
\end{compactenum}
\ifFull \line(1,0){470} \else \line(1,0){240} \fi}

\noindent
Here multiplication by 0.5 is replaced by shift to the right by 1 bit that
has no cost. Complexity of this protocol can be found in \ifFull Table~\ref{tab2} \else Table~\ref{tab2-short}\fi.
\fi

\ifFull
\else
\begin{table*}[t] \centering \small \setlength{\tabcolsep}{0.3ex} 
\begin{tabular}{|c|c|c|c|c|} 
\hline
\multirow{2}{*}{Prot.} & \multicolumn{2}{c|}{Secret sharing} & \multicolumn{2}{c|}{Garbled circuits}\\
\cline{2-5}
& Rounds & Interactive operations & XOR gates & Non-XOR gates \\
\hline
\multirow{3}{*}{${\sf Sin}$} & \multirow{3}{*}{$2 \log N +16$} &
\multirow{3}{*}{$2Nk+8 \ell +2N + 6k + 4$} & $(\max(N,M)+N+M+2)$ &
$(\max(N,M)+N+M+2)$\\
 & & & $\times(4\ell^2-4\ell)+7 \ell ^2$ & $\times(2\ell ^2-\ell)+3 \ell ^2$\\
& & & $+4\ell (N+M) + 31\ell$ & $+\ell (N+M) + 11\ell$ \\
\hline
\multirow{3}{*}{${\sf Cos}$} & \multirow{3}{*}{$2 \log N +15$} &
\multirow{3}{*}{$2Nk+8 \ell + 2N + 4k + 2$} & $(\max(N,M)+N+M+1)$ & $(\max(N,M)+N+M+1)$\\
 & & & $\times(4\ell^2-4\ell) + 7\ell^2$ & $\times(2\ell^2-\ell) + 3\ell^2$\\
& & & $+4\ell (N+M) + 31\ell$ & $+ \ell (N+M) + 11\ell$ \\
\hline
\multirow{2}{*}{${\sf Arctan}$} & $2\log N + 3\log\ell$ & $1.5\ell\log\ell
+ 2 \ell \log(\frac{\ell}{3.5}) + 2Nk$ & \multirow{2}{*}{$8N\ell^2+3 \ell^2 -
4N\ell + 43\ell$} & \multirow{2}{*}{$4N\ell^2 + \ell^2 - N \ell + 15\ell$}\\
& $+ 2\log(\frac{\ell}{3.5})+22$ & $+18.5 \ell+2 +4 \log(\frac{\ell}{3.5})+6$
& & \\ \hline

\multirow{4}{*}{${\sf Select}$} & $c(2\log^2 t_1 +2\log ^2 t_2$ &
$c(\ell (t_1-0.25) (\log^2t_1+\log t_1 +4)$ & $c(1.5 t_1 \ell
(\log^2 t_1+\log t_1 +4)$ & $c(0.5 t_1 \ell (\log ^2 t_1+\log t_1 +4)$\\
& $+6\log m + 2\log\log m$ & $+\ell (t_2-0.25) (\log^2 t_2+\log t_2 +4)$ &
$+ 1.5 t_2 \ell (\log ^2 t_2+\log t_2 +4)$ & $+ 0.5 t_2 \ell (\log^2 t_2 +
\log t_2 +4)$ \\
& $+17)$ & $+2m\log m \log \log m+8m \log m+ $ & $+(2\ell + 20.5)m \log m+m$
& $+(4\ell +5)m \log m+m$\\
& & $18m\ell-5.5 m+2\log m-8\ell+9)$ & $+12.5m\ell-6\log m-8\ell)$ &
$+4m\ell+(2\ell -1)\log m)$\\ \hline
\end{tabular}
\caption{Performance of secure building blocks for fixed-point values.} \label{tab2-short}
\end{table*}
\fi

\ifFull
\subsubsection{Selection}
\else
\subsection{Selection}
\fi

There are multiple ways to find the $f$th smallest element of a set. The
most straightforward method is to obliviously sort the input set and return
the $f$th element of the sorted set. This approach is often beneficial when
the input set is small. As a general solution, it is usually faster to use a
selection algorithm, and the goal of this section is to design an oblivious
selection protocol for finding the $f$th smallest element of a set.
{\color{black}One compelling alternative is to perform a random shuffle
followed by a regular selection algorithm \cite{hamada2012practically} such
as quickselect (similar to the way some sorting algorithms work) and we
discuss this option at the end of this section. The only other publication
that studied selection in the context of SMC was \cite{aggarwal2004secure},
but it assumed that set elements are known to the participants, which is not
applicable to this work.} 

The regular non-oblivious selection algorithms run in $O(m)$ time on
$m$-element sets. We considered both deterministic and probabilistic
algorithms and settled on a probabilistic algorithm that performs the best
in the secure setting in practice. Its complexity is $O(m \log m)$ due to
the use of data-oblivious compaction. Goodrich \cite{goodrich} proposed a
probabilistic oblivious selection algorithm that works in linear time {\color{black}(there
are also selection networks such as \cite{leighton1997probabilistic} of
higher complexity)}. Our starting point, however, is a different simpler
algorithm that performs well in practice. Our algorithm proceeds by scanning
through the input set and selects each element with probability
$\frac{c_1}{\sqrt m}$ for some constant $c_1$. The expected number of
selected elements is $O(\sqrt{m})$. The algorithm then sorts the selected
elements and determines two elements $x$ and $y$ between which the $f$th
smallest element is expected to lie based on $f$, $m$, and the number of
selected elements. Afterwards, we scan the input set again retrieving all
elements with values between $x$ and $y$ and simultaneously computing the
ranks of $x$ and $y$ in the input set; let $r_x$ and $r_y$ denote the ranks.
The expected number of retrieved elements is $O(\sqrt{m})$. We sort the
retrieved elements and return the element at position $f-r_x$ in the sorted
set. Let $t_1$ denote the number of elements randomly selected in the
beginning of the algorithm and $t_2$ denote the \mbox{number of elements
with values between $x$ and $y$.}

Because of the randomized nature of our algorithm, it can fail at multiple
points of execution. In particular, if the number of elements selected
in the beginning is too large or too small (and the
performance guarantees cannot be met), we abort. Similarly, if the number of
the retrieved elements that lie between $x$ and $y$ is too small, we abort.
Lastly, if $f$ does not lie between the ranks of $x$ and $y$, the $f$th
smallest element will not be among the retrieved elements and we abort. It
can be shown using the union bound that all of these events happen with
a probability negligible in the input size. By using the appropriate choice
of constants we can also ensure that faults are very rare in practice (see
below). Thus, in the rare event of failure, the algorithm needs to be
restarted using new random choices, and the expected number of times the
algorithm is to be executed is slightly above 1.

Our protocol ${\sf Select}$ is given below. Compared to the algorithm
structure outlined above, we need to ensure high probability of success
through the appropriate selection of $x$ and $y$ as well as some constant
parameters. We also need to ensure that the execution can proceed
obliviously without compromising private data. To achieve this, we use three
constants $c_1$, $c_2$, and $\hat{c}$. The value of $c_1$ influences the
probability with which an element is selected in the beginning of the
algorithm. The value of $\hat{c}$ influences the distance between $x$ and
$y$ in the set of $t_1$ selected elements. In particular, we first determine
the position where $f$ is expected to be positioned among the $t_1$ selected
elements as $(f t_1)/m$. We then choose $x$ to be $\hat{c}$ elements before
that position and choose $y$ to be $3\hat{c}$ elements after that position.
Lastly, $c_2$ is used to control the size of $t_2$. When $t_2$ is too small
due to the random \mbox{choices made during the execution, we abort.}

To turn this logic into a data-oblivious protocol, we resort to secure and
oblivious set operations. In particular, after we mark each element of the
input set as selected or not selected, we use data-oblivious compaction to
place all selected elements in the beginning of the set. Similarly, we use
data-oblivious sorting to sort $t_1$ and $t_2$ elements.

{\small
\noindent \ifFull \line(1,0){470} \else \line(1,0){240} \fi\\
[-0.02in]
$[b] \leftarrow {\sf Select}(\langle [a_1], \ldots, [a_m]\rangle, f)$\\
[-0.09in]
\ifFull \line(1,0){470} \else \line(1,0){240} \fi
\begin{compactenum}
\item Set constants $c_1$, $c_2$, and $\hat{c}$; also set $n = \sqrt m$,
  $[t_1] = 0$, $[t_2] = 0$.

\item For $i = 1, \ldots, m$ set $[b_i] = 0$. 

\item For $i = 1, \ldots, m$ do in parallel
\begin{compactenum}
\item Generate random $[r_i] \in \mathbb{Z}_n$.
\item If ${\sf LT}([r_i], c_1)$ then $[b_i] = 1$.
\end{compactenum}

\item For $i = 1, \ldots, m$ do $[t_1] = [t_1] + [b_i]$.

\item Open the value of $t_1$.

\item If $((t_1 > 2c_1n) \vee (t_1 < \frac{1}{2}c_1n))$ then abort.

\item Execute $\langle[a'_1], \ldots, [a'_m]\rangle \leftarrow {\sf
    Comp}(\langle [b_1], [a_1] \wedge [b_1] \rangle$, \ldots, $\langle [b_m],
  [a_m] \wedge [b_m] \rangle)$. 

\item Execute $\langle [a''_1], \ldots, [a''_{t_1}] \rangle \leftarrow {\sf
    Sort}([a'_1], \ldots, [a'_{t_1}])$.

\item Compute $k$ as the closest integer to $(ft_1)/m$.

\item If $(k-\hat{c} < 1)$ then $[x] \leftarrow {\sf Min}([a_1], \ldots,
  [a_m])$; else $[x] = [a''_{k-\hat{c}}]$.

\item If $(k+3\hat{c} > t_1)$ then $[y] \leftarrow {\sf Max}([a_1], \ldots,
  [a_m])$; else $[y] = [a''_{k+3\hat{c}}]$.

\item Set $[r_x] = 0$ and $[r_y] = 0$.

\item For $i = 1, \ldots, m$ do in parallel $[g_i] = {\sf LT}([a_i], [x])$
  and $[g'_i] = {\sf LT}([a_i], [y])$. 

\item For $i = 1, \ldots, m$ do in parallel $[b'_i] = \neg [g_i] \wedge [g'_i]$.

\item For $i = 1, \ldots, m$ do $[r_x] = [r_x] + [g_i]$ 
  and $[t_2] = [t_2] + [b'_i]$.

\item Open the values of $r_x$ and $t_2$.

\item If $((t_2 < 4\hat{c}c_2n) \vee ((f-r_x) < 0) \vee ((f-r_x) > t_2))$ then abort.
  
\item Execute $\langle[d_1], \ldots, [d_m]\rangle \leftarrow
  {\sf Comp}(\langle [b'_1], [a_1] \wedge [b'_1] \rangle$, \ldots, $\langle [b'_m], [a_m] \wedge [b'_m]
  \rangle)$. 

\item Execute $\langle[d'_1], \ldots, [d'_{t_2}]\rangle \leftarrow {\sf
    Sort}([d_1], \ldots, [d_{t_2}])$. 

\item Return $[b] = [d'_{f-r_x}]$.
%
%
%
%
\end{compactenum}
\vspace{-0.1in}
\ifFull \line(1,0){470} \else \line(1,0){240} \fi}
%
%

\noindent Note that because maximum and minimum functions are evaluated on
the same set, we reduce the overhead of evaluating them simultaneously
compared to evaluating them individually. In particular, when min and max
functions are evaluated in a tree-like fashion on an $m$-element set, we use
the $m/2$ comparison in the first layer to set both minimum and maximum
elements of each pair using the same cost as evaluating only one of them.
The rest is evaluated as before, and we save about $1/2$ overhead of one of
min or max.

Complexity of this protocol is given in \ifFull Table~\ref{tab2} \else Table~\ref{tab2-short} \fi as a function of
$m$, $t_1$, $t_2$, $c$, and bitlength $\ell$. The expected value
of $t_1$ is $c_1 \sqrt{m}$, the expected value of $t_2$ is $(4 \hat{c}
\sqrt{m})/c_1$, and $c$ indicates the average number of times the algorithm
needs to be invoked, which is a constant slightly larger than 1 (see below).
We experimentally determined the best values for $c_1$, $c_2$, and $\hat{c}$
for different values of $n$ to ensure high success of the algorithm. Based
on the experiments, we recommend to set $c_1 = 2$, $c_2 = 2$, and $\hat{c} =
10$. Using these values, the probability of the protocol's failure on a
single run is at most a few percent up to input size 10,000. With larger
input sizes, a larger value of $\hat{c}$ might be preferred.

{\color{black}As far as alternative constructions go, the goal can be
achieved using oblivious sorting. A random shuffle followed by a
quicksort, where the results of private comparisons are opened, was
determined to be among the fastest sorting algorithms
\cite{bog14}. This solution, however, is not competitive with our
algorithm. A better approach is to run quickselect 
after a shuffle. This solution is not competitive with our
algorithm in the GC setting because of the random shuffle
cost. Furthermore, because many values are opened at intermediate
steps of the computation to determine what comparisons are to be
performed next, the circuit can no longer be garbled at once and needs
to be assembled during the computation.

The situation is different with SS, for which a fast random shuffle
algorithm is known \cite{Laur11}. In its general form it requires
$O(2^n n^{3/2} m\log m)$ interactive operations in
$O(\frac{2^n}{\sqrt{n}})$ rounds, but when the number of parties $n$
is a small constant, it is very efficient. For example, with $n = 3$
and computationally-bounded adversaries it uses $3(2m+\alpha)$
interactive operations in $4$ rounds, where $\alpha$ is proportional to
the computational security parameter. Quickselect additionally performs
on average about $4m$ comparisons, which translates into close to
$16\ell m$ interactive operations in $4\log m$ rounds. Our $\sf
Select$ has an additive term with $18\ell m$ interactive operations
and thus is expected to be somewhat slower for this setting.

We can thus conclude that $\sf Select$ is an attractive choice with
GCs or for the SS setting when the number of computational parties is
not very small or when the high worst-case behavior of quickselect is
unacceptable. In general, our construction is a versatile solution
that exhibits good performance in any setting and has low worst-case
complexity.}

\textbf{Security analysis.} The $\sf Select$ protocol is designed
using secure building blocks listed in \ifFull Section \ref{BB1}. \else Appendix \ref{app:kbb}.\fi ~Unlike all
other protocols in this work, three values $t_1$, $t_2$, and $r_x$ privately
computed in $\sf Select$ are opened during its execution. Thus, to guarantee
security based on Canetti's composition theorem~\cite{can00} {\color{black}similar to
other protocols}, we need to show that these values are independent of the
private input set and result in no information leakage, and as a result the
overall $\sf Select$ protocol is secure. First, observe that the value of
$t_1$ is computed purely based on random choices made in step 3(a). Thus,
its value is data-independent. Second, the rank of $x$ $r_x$ is determined
by random choices, but not the data values themselves. Similarly, the value
of $t_2$ depends only on the ranks of the randomly chosen pivots, but not on
the actual data values. Therefore, execution is data-oblivious and security
is maintained.

\ignore{
Note that, we also proposed another secure $\sf Select$ protocol which has higher complexity based on the popular recursive selection algorithm that is elaborated in Appendix~\ref{sec:select2}.
}

\ifFull
\else 
\fi
\section{\mbox{Secure Fingerprint Recognition}}
\label{sec:protocols}

We are now ready to proceed with describing our solutions for secure
fingerprint recognition using the building blocks introduced in
Section~\ref{sec:bb}. We provide three constructions, one for each
fingerprint recognition approach in Section~\ref{sec:fprelim}.

\ifFull
\else
\fi
\subsection{Secure Fingerprint Recognition using Brute Force Geometrical Transformation}

The easiest way to execute all (non-secure) fingerprint recognition
algorithms in Section~\ref{sec:fprelim} is to use floating-point
representation. We, however, choose to utilize integer and fixed-point
arithmetic in their secure counterparts to reduce overhead associated with
secure execution. In particular, typically minutia coordinates $(x_i, y_i)$
as well as their orientation $\theta_i$ are represented as integers. This
means that all inputs in Algorithm~\ref{alg1}, the first fingerprint matching
algorithm based on brute force geometrical transformation, are integers and computing 
the number of matched minutiae using Algorithm~\ref{alg-matching} can also
be performed on integers. The output of sine and cosine functions in step
2(b) of Algorithm~\ref{alg1}, however, are not integers and we utilize
fixed-point representation for their values. Moreover, after the
minutia
points are transformed, we can truncate the transformed coordinates $x''_i$
and $y''_i$ and use their integer representation in
Algorithm~\ref{alg-matching}.

Our secure implementation of Algorithm~\ref{alg1} is given below as protocol
${\sf GeomTransFR}$. It uses the same logic as
Algorithm~\ref{alg1} with the difference that the computation of the maximum
matching is performed outside the main for-loop to reduce the number of
rounds in the SS setting.

{\small
\noindent \ifFull \line(1,0){470} \else \line(1,0){240} \fi\\[-0.02in]
$([C_{max}], \langle [\Delta x_{max}], [\Delta y_{max}], [\Delta
\theta_{max}] \rangle) \leftarrow {\sf GeomTransFR}$ $(T = \{t_i = ([x_i], [y_i],
 [ \theta_i])\}_{i=1}^m, S = \{s_i = ([x'_i], [y'_i],$ $[\theta'_i])\}_{i=1}^n)$\\
[-0.09in]
\ifFull \line(1,0){470} \else \line(1,0){240} \fi
\begin{compactenum}
\item $[C_{max}] = [0]$;

\item For $i = 1, \ldots, m$ and $j = 1, \ldots, n$, compute in parallel:
\begin{compactenum}

\item $[\Delta x_{i,j}] = [x'_j]- [x_i]$, $[\Delta y_{i,j}] = [y'_j]-[y_i]$,
  and $[\Delta \theta_{i,j}] = [\theta'_j]- [\theta_i]$. 

\item $[c_{i,j}] = {\sf Sin}([\Delta \theta_{i,j}])$ and $[c'_{i,j}] = {\sf
    Cos}([\Delta \theta_{i,j}])$. 

\item For $k = 1, \ldots, n$, compute in parallel $[x_{i,j}^{(k)}] =
  [c'_{i,j}] \cdot [x'_k] + [c_{i,j}] \cdot [y'_k] - [\Delta x_{i,j}]$,
  $[y_{i,j}^{(k)}] = [c'_{i,j}] \cdot [y'_k] - [c_{i,j}] \cdot [x'_k] -
  [\Delta y_{i,j}]$, and $[\theta_{i,j}^{(k)}] = [\theta'_k] - [\Delta
  \theta_{i,j}]$ and save the computed points as $S_{i,j} =
  \{([x_{i,j}^{(k)}], [y_{i,j}^{(k)}], [\theta_{i,j}^{(k)}])\}_{k=1}^n$.

\item Compute the number $[C_{i,j}]$ of matched minutiae between $T$ and
  $S_{i,j}$ using protocol ${\sf Match}$.
\end{compactenum}

\ifFull
\item Compute the largest matching $\langle [C_{max}], [\Delta x_{max}],
  [\Delta y_{max}], [\Delta \theta_{max}] \rangle = {\sf Max}(\langle
  [C_{1,1}], [\Delta x_{1,1}], [\Delta y_{1,1}],$ $[\Delta \theta_{1,1}]
  \rangle, \ldots, \langle [C_{m,n}]), [\Delta x_{m,n}]), [\Delta
  y_{m,n}], [\Delta \theta_{m,n}] \rangle)$. 
\else
\item Compute the largest matching $\langle [C_{max}], [\Delta x_{max}],$ $
  [\Delta y_{max}], [\Delta \theta_{max}] \rangle = {\sf Max}(\langle
  [C_{1,1}], [\Delta x_{1,1}], [\Delta y_{1,1}], $ $[\Delta \theta_{1,1}]
  \rangle$, \ldots, $\langle [C_{m,n}]), [\Delta x_{m,n}]), [\Delta
  y_{m,n}], [\Delta \theta_{m,n}] \rangle)$. 
\fi

\item Return $[C_{max}]$ and the corresponding alignment $\langle [\Delta
  x_{max}],$ $ [\Delta y_{max}], [\Delta \theta_{max}] \rangle$. 
\end{compactenum}
\vspace{-0.1in}
\ifFull \line(1,0){470} \else \line(1,0){240} \fi}

We obtain that all steps use integer arithmetic except step 2(b) (where sine
and cosine have integer inputs, but fixed-point outputs) and multiplications
in step 2(c) take one integer and one fixed-point operand (after which
$x_{i,j}^{(k)}$ and $y_{i,j}^{(k)}$ are truncated to integer
representation). Note that the latter corresponds to some optimization of
the computation, where instead of converting integers $x'_k$ and $y'_k$ to
fixed-point values and multiplying two fixed-point numbers, we can reduce the
overhead by multiplying an integer to a fixed-point value. This eliminates
the cost of truncating the product in the SS setting and reduces
the cost of multiplication in the GC setting. Furthermore, we
can also optimize the point at which the fixed-point values are converted
back to integers in the computation of $x_{i,j}^{(k)}$ and $y_{i,j}^{(k)}$
in step 2(c) of the algorithm. In particular, in the SS setting
we add the fixed-point products and $\Delta x_{i,j}^{(k)}$, $\Delta
y_{i,j}^{(k)}$ converted to fixed-point representation (all of which have 0
cost), after which the sum is converted to an integer (paying the price of
truncation). In the GCs setting, on the other hand, we could
convert the products to integers first (which has 0 cost) and then perform
addition/subtraction on shorter integer values.

Our secure implementation of Algorithm~\ref{alg-matching} is given as
protocol ${\sf Match}$. All computation is performed on integer values.
Compared to the structure of Algorithm~\ref{alg-matching}, there are a few
notable differences. First, note that instead of checking the $\sqrt{(x_i-
x'_j)^2 +(y_i- y'_j)^2} < \lambda$ constraint, the protocol checks the
equivalent constraint $(x_i- x'_j)^2 +(y_i- y'_j)^2 < \lambda^2$ to avoid
using a costly square root operation (and it is assumed that $\lambda^2$ is
supplied as part of the input instead of $\lambda$). Second, to evaluate the
constraint $\min(|\theta_i-\theta'_j|, 360 - |\theta_i - \theta'_j|) <
\lambda_{\theta}$, instead of computing $|\theta_i-\theta'_j|$ the protocol
computes $\theta_i-\theta'_j$ or $\theta'_j-\theta_i$ depending on which
value is positive and uses that value in the consecutive computation. In
addition, instead of computing the minimum, the computation proceeds as
$(|\theta_i-\theta'_j| < \lambda_\theta) \vee (360 - |\theta_i-\theta'_j| <
\lambda_\theta)$ in steps 3(c) and 3(d) of the protocol, which allows us to
make the computation slightly faster. Because this computation is performed
a large number of times, even small improvements can have impact on the
overall performance. 

Similar to restructuring protocol $\sf GeomTransFR$ to maximize parallelism
and lower the number of rounds, in $\sf Match$ all distance and orientation
constraints are evaluated in parallel in step 3. Then step 4 iterates
through all minutiae in $T$ and constructs a matching between a
minutia of
$T$ and an available minutia from $S$ within a close proximity to it (if
any). Variable $l_j$ indicates whether the $j$th minutia of fingerprint $S$
is currently available (i.e., has not yet been paired up with another
minutia from $T$). For each minutia, $s_j$, marked as unavailable, its
distance to the $i$th minutia in $T$ is set to $\lambda^2$ to prevent it
from being chosen for pairing the second time.
Because step 4(d) is written to run all loop iterations in parallel, there
is a single variable $C_j$ for each loop iteration, all values of which are
added together at the end of the loop (which is free using SS).
With GCs this parallelism is often not essential, and if the
loop is executed sequentially, $C$ can be incremented on line 4(d) directly
instead of using $C_j$'s.
\ifFull (And if parallel computation is used, the sum on
line 4(e) will need to be implemented as the (free) XOR of all $C_j$.) \fi

{\small
\noindent \ifFull \line(1,0){470} \else \line(1,0){240} \fi\\[-0.02in]
$[C] \leftarrow {\sf Match}(T = \{t_i = ([x_i], [y_i],
 [ \theta_i])\}_{i=1}^m, S = \{s_i = ([x'_i], [y'_i],
 $ $[\theta'_i])\}_{i=1}^n, \lambda^2, \lambda_{\theta})$\\[-0.09in]
\ifFull \line(1,0){470} \else \line(1,0){240} \fi
\begin{compactenum}
\item Set $[C] = [0]$.

\item For $j = 1, \ldots, n$, set in parallel $[l_j] = [0]$.

\item For $i = 1, \ldots, m$ and $j = 1, \ldots, n$, compute in parallel:
\begin{compactenum}
\item  $[d_{i,j}] = ([x_i]- [x'_j])^2 + ([y_i]- [y'_j])^2$.

\item If ${\sf LT}([\theta_i], [\theta'_j])$, then $[a_{i,j}] = [\theta'_j] -
  [\theta_i]$, else $[a_{i,j}] = [\theta_i] - [\theta'_j]$. 

\item $[c_{i,j}] = {\sf LT}([d_{i,j}], [\lambda^2])$, $[c'_{i,j}] = {\sf
    LT}([a_{i,j}], [\lambda_{\theta}])$, and $[c''_{i,j}] = {\sf
    LT}((360-[a_{i,j}]), [\lambda_{\theta}])$. 

\item $[v_{i,j}] = [c_{i,j}] \wedge ([c'_{i,j}] \vee [c''_{i,j}])$.
\end{compactenum}

\item For $i = 1, \ldots, m$, do
\begin{compactenum}
\item For $j = 1, \ldots, n$, do in parallel: if $([l_j] \vee
  \neg[v_{i,j}])$, then $[d_{i,j}] = \lambda^2$.

\item Execute $([d_{min}], [j_{min}]) = {\sf Min}([d_{i,1}], \ldots,
  [d_{i,n}])$. 

\item Set $[u] = {\sf LT}([d_{min}], \lambda^2)$.
\item For $j = 1, \ldots, n$, compute in parallel: if (${\sf EQ}([j_{min}],
  j) \wedge [u]$), then $[C_j] = [1]$ and $[l_j] = [1]$, else $[C_j] = [0]$.

\item $[C] = [C] + \sum_{j=1}^n [C_j]$.
\end{compactenum}
\item Return $[C]$.
\end{compactenum}
\vspace{-0.1in}
\ifFull \line(1,0){470} \else \line(1,0){240} \fi}

The asymptotic complexity of $\sf GeomTransFR$ and $\sf Match$
remains similar to their original non-secure counterparts. In particular, if
we treat the bitlength of integer and fixed-point values as constants, the
complexity of $\sf GeomTransFR$ is $O(n^2m^2)$ and $\sf Match$ is
$O(nm)$. Their round complexity (for the SS setting) is $O(m
\log n)$ and $O(m \log n)$, respectively. If we wish to include
dependency on the bitlengths $\ell$ and $k$, the complexity increases by at
most a factor of $O(\ell)$ in the SS setting and at most a
factor of $O(\ell^2)$ in the GCs setting due to the differences
in the complexity of the underlying operations.

\ifFull
\else 
\fi
\subsection{Secure Fingerprint Recognition using High Curvature Points}

We next treat our secure realization of fingerprint recognition using high
curvature points from Algorithm~\ref{alg2}. The corresponding secure
computation is provided as protocol $\sf HighCurvatureFR$ below. It makes
calls to 
\ifFull
a secure version of Algorithm~\ref{alg2-trans}, which we
consequently call as protocol $\sf OptimalMotion$. 
\else
$\sf OptimalMotion$ protocol.
\fi
Also, because of the
complexity of the computation associated with finding the closest points in
step 3 of Algorithm~\ref{alg2}, we provide the corresponding secure
computation as a separate protocol $\sf ClosestPoints$. All computation is
performed on fixed-point values with the exception of step 7 of $\sf
HighCurvatureFR$, where a call to protocol $\sf
Match$ is made.

{\small
\noindent \ifFull \line(1,0){470} \else \line(1,0){240} \fi\\
[-0.02in]
$([C], (R, v)) \leftarrow {\sf HighCurvatureFR}(T = (\{t_i = ([x_i], [y_i], [\theta_i])\}_{i=1}^m, $ $ \{\hat{t}_i =
  ([\hat{x}_i], [\hat{y}_i], [\hat{w}_i])\}_{i=1}^{\hat{m}}), S = (\{s_i =
  ([x'_i], [y'_i], [\theta'_i])\}_{i=1}^n$, $\{\hat{s}_i = ([\hat{x}'_i], [\hat{y}'_i],
 $ $[\hat{w}'_i])\}_{i=1}^{\hat{n}}), f, \gamma, (\alpha_1, \ldots,
\alpha_\gamma), \beta, \lambda^2, \lambda_{\theta})$\\
[-0.09in]
\ifFull \line(1,0){470} \else \line(1,0){240} \fi
\begin{compactenum}
\item Set $[S_{LTS}] = [0]$.
\item For $i = 1, \ldots, \hat{n}$, set $\bar{s}_i = \hat{s}_i$.

\item For ${\sf ind} = 1, \ldots, \gamma$ do
\begin{compactenum}
\item Execute $\{([d_i], \tilde{t}_i)\}_{i=1}^{\hat{n}} \leftarrow {\sf
    ClosestPoints}(\{\hat{t}_i\}_{i=1}^{\hat{m}},
  \{\hat{s}_i\}_{i=1}^{\hat{n}},$ $\alpha_{\sf ind})$.
  
\item Execute $[y] \leftarrow {\sf Select}(([d_1],\ldots, [d_{\hat{n}}]), f)$.

\item For $i = 1, \ldots, \hat{n}$ do in parallel $[l_i] = {\sf LT}([d'_i], [y]+1)$.

\item Execute $(([a_1], b_1, c_1), \ldots, ([a_{\hat{n}}], b_{\hat{n}},
  c_{\hat{n}})) \leftarrow {\sf Comp}(([l_1],$ $ \hat{s}_1 \wedge [l_1], \tilde{t}_1 \wedge [l_1]),
  {\ldots}, ([l_{\hat{n}}], \hat{s}_{\hat{n}} \wedge [l_{\hat{n}}], \tilde{t}_{\hat{n}} \wedge [l_{\hat{n}}]))$ using
  $[l_i]$'s as the keys.

\item Compute the optimal motion $(R, v) \leftarrow {\sf
    OptimalMotion}$ $(\{c_i, b_i\}_{i=1}^f)$.

\item For $i = 1, \ldots, \hat{n}$ transform the points in parallel as
  $[x''_i] = [v_1] + [r_{11}] \cdot [\hat{x}'_i]+ [r_{12}] \cdot
  [\hat{y}'_i]+ [r_{13}] \cdot [\hat{w}'_i]$, $[y''_i] = [v_2] + [r_{21}]
  \cdot [\hat{x}'_i]+ [r_{22}] \cdot [\hat{y}'_i]+ [r_{23}] \cdot
  [\hat{w}'_i]$, and $[w''_i] = [v_3] + [r_{31}] \cdot [\hat{x}'_i]+
  [r_{32}] \cdot [\hat{y}'_i]+ [r_{33}] \cdot [\hat{w}'_i]$, then set
  $\hat{s}_i = ([x''_i], [y''_i], [w''_i])$.     
\end{compactenum}

\item Execute $(R, v) \leftarrow {\sf
    OptimalMotion}(\{\bar{s}_i, \hat{s}_i\}_{i=1}^{\hat{n}})$.

\item Compute $[c_1] = {\sf Div}([\bar{y}_2]-[\bar{y}_1], [\bar{x}_2]-[\bar{x}_1])$, $[c_2] = {\sf Div}([\hat{y}'_2]-[\hat{y}'_1], [\hat{x}'_2]-[\hat{x}'_1])$, $[c_3] = {\sf Div}([c_1]-[c_2], 1+[c_1]\cdot [c_2])$, and $[\Delta \theta] = {\sf Arctan}([c_3])$.

\item For $i = 1, \ldots, n$ do in parallel $[x''_i] = [v_1] + [r_{11}]
  \cdot [x'_i]+ [r_{12}] \cdot [y'_i]$ and $[y''_i] = [v_2] + [r_{21}] \cdot
  [x'_i]+ [r_{22}] \cdot [y'_i]$, then set $s_i = ([x''_i], [y''_i], [\theta
  '_i] - [\Delta \theta])$.  
  
 \item Compute the number $[C]$ of matched minutiae by executing ${\sf
     Match}$ $(\{t_i\}_{i=1}^m, \{s_i\}_{i=1}^n, \lambda^2, \lambda_{\theta})$.
 \item Return $[C], (R, v)$. 
\end{compactenum}
\vspace{-0.1in}
\ifFull \line(1,0){470} \else \line(1,0){240} \fi}

Different from Algorithm~\ref{alg2}, our ${\sf
HighCurvatureFR}$ proceeds using the maximum number of iterations $\gamma$,
as not to reveal the actual number of iterations needed (which depends on
private inputs). The remaining algorithm's structure is maintained, where
the computation is replaced with secure and data-oblivious operations. In
particular, in each iteration of step 3, we first determine the closest
high-curvature point from $T$ to each (possibly transformed) high-curvature
point from $S$ and compute $f$ pairs with the smallest distances. Secure
computation of the closest points using protocol $\sf ClosestPoints$ is
described below, while determining the closest $f$ pairs is performed on
steps 3(b)--(d) as follows. After selecting the $f$th smallest element $y$
among the computed distances (step 3(b)), each distance is compared to that
element (step 3(c)). We then proceed with pushing all elements which are
less than $y$ to the beginning of the set using compaction (step 3(d)) and
consequently use the $f$ smallest distances (located in the beginning of the
set) for optimal motion computation.

Once the closest $f$ pairs and the corresponding optimal motion have been
determined, the protocol proceeds with applying the optimal motion to the
high-curvature points in $S$. After executing this computation a necessary
number of times, the protocol computes the overall motion in step 4 and
applies it to the minutia points in the same way as in Algorithm~\ref{alg2}.
One thing to notice is that $\sf HighCurvatureFR$ has additional (public)
inputs compared to Algorithm~\ref{alg2}. The parameters $\alpha_1, {\ldots},
\alpha_\gamma$ specify bounding box sizes for the purposes of closest points
computation in step 3(a) (see below). 

The protocol $\sf ClosestPoints$ below takes two sets of points $t_i$'s and
$s_i$'s (represented using three coordinates each) and for each point $s_i$
returns the closest element among the $t_i$'s and the corresponding
distance. As mentioned in section~\ref{sec:alg2}, the TICP algorithm on
which our construction builds uses the bounding box approach to eliminate
errors during this computation. In particular, only points within a certain
distance from a point are considered, and the maximum allowable distance
(i.e., the bounding box size) is denoted by $\alpha_i$ in the $i$th
iteration of $\sf HighCurvatureFR$ or Algorithm~\ref{alg2}. The sizes of the
bounding boxes become smaller with each iteration as the two sets of points
become closer to each other. When we make a call to $\sf ClosestPoints$, we
pass a single bounding box size for the current iteration and that parameter
is denoted as $\alpha$ in the interface of $\sf ClosestPoints$.

{\small
\noindent \ifFull \line(1,0){470} \else \line(1,0){240} \fi \\[-0.02in] 
$\{([d_i], \tilde{t}_i)\}_{i=1}^{\hat{n}} \leftarrow {\sf ClosestPoints}(\{t_i
  = (x_i, y_i, w_i)\}_{i=1}^{\hat{m}}, \{s_i = (x'_i, y'_i, w'_i)\}_{i=1}^{\hat{n}}, \alpha, \beta)$\\[-0.09in] \ifFull \line(1,0){470} \else \line(1,0){240} \fi 
\begin{compactenum}
    
\item For $i = 1, \ldots, \hat{n}$ and for $j = 1, \ldots, \hat{m}$ do in
  parallel: 
\begin{compactenum}
  \item Set $[d_{(i, j)}] = {\sf Sqrt}([\hat{x}_j]- [\hat{x}'_i])^2 +
    ([\hat{y}_j] - [\hat{y}'_i])^2$. 

\item If $({\sf LT}([\hat{w}_j], [\hat{w}'_i]))$ then $[a_{(i, j)}] =
  [\hat{w}'_i] - [\hat{w}_j]$; else $[a_{(i, j)}] = [\hat{w}_j] -
  [\hat{w}'_i]$. 

\item Set $[d_{(i, j)}] = [d_{(i, j)}] + \beta \cdot [a_{(i, j)}]$.
\end{compactenum}

\item For $i = 1, \ldots, \hat{m}$ do in parallel $[l_i] = 1$.
\item For $i = 1, \ldots, \hat{n}$ do in parallel $[l'_i] = 1$.

\item For $i = 1, \ldots, \hat{n}$ do
\begin{compactenum}
\item For $j = 1, \ldots, \hat{m}$ do in parallel if $(\neg[{l_j}])$ then
  $[d_{(i, j)}] = \infty$. 

\item Execute $([d_i], [j_{min}])  \leftarrow {\sf Min}([d_{(i, 1)}],
  \ldots, [d_{(i, \hat{m})}])$. 

\item $[b] = {\sf LT}([d_i], \alpha)$.

\item For $j = 1, \ldots, \hat{m}$ do in parallel if $({\sf EQ}([j_{min}],
  j) \wedge [b])$ then $[l_j] = 0$, $[l'_i] = 0$, $\tilde{t}_i = t_j$. 
\end{compactenum}

\item For $i = 1, \ldots, \hat{n}$ do
\begin{compactenum}
\item For $j = 1, \ldots, \hat{m}$ do in parallel if $(\neg[{l_j}])$ then
  $[d_{(i, j)}] = \infty$. 

\item Execute $([d_i], [j_{min}])  \leftarrow {\sf Min}([d_{(i, 1)}],
  \ldots, [d_{(i, \hat{m})}])$.


\item For $j = 1, \ldots, \hat{m}$ do in parallel if $({\sf EQ}([j_{min}], j)
  \wedge [l'_i])$ then $[l_j] = 0$, $[l'_i] = 0$, $\tilde{t}_i = t_j$. 
\end{compactenum}
\item Return $\{([d_i], \tilde{t}_i)\}_{i=1}^{\hat{n}}$.
\end{compactenum}
\vspace{-0.1in}
\ifFull \line(1,0){470} \else \line(1,0){240} \fi}

\noindent
Given two sets of points and public parameter $\alpha$, the $\sf
ClosestPoints$ protocol first computes the distances between each pair of
points in parallel in step 1 (according to the formula in step 3 of
Algorithm~\ref{alg2}). Next, we mark each $t_i$ and $s_i$ as available
(steps 2 and 3, respectively). Step 4 iterates through all $s_i$'s and
determines the closest available point $t_j$ to $s_i$ (the distance to the
unavailable points is set to infinity to ensure that they are not chosen).
If the closest point is within the bounding box, $s_i$ is paired up with $t_j$
and both are marked as unavailable. 

At the end of step 4, some points $s_i$'s will be paired up with one of the
$t_j$'s, while others will not be. To ensure that the algorithm produces
enough pairs for their consecutive use in $\sf HighCurvatureFR$, we repeat
the pairing process with the $s_i$'s that remain available at this point and
without enforcing the constraint that the points of the pair must lie in
close proximity of each other. This computation corresponds to step 5. That
is, this step pairs each available $s_i$ with the closest available point
among the $t_j$'s even if it is far away from $s_i$ (and is likely to be an
unrelated point). This is to ensure that enough distances are returned for
their use in the parent protocol. In this step, distances of all unavailable
points are set to infinity and each $s_i$ which is still marked as available
is updated with the closest distance and the corresponding $t_j$.

\ifFull
What remains is to discuss protocol ${\sf OptimalMotion}$ that corresponds
to secure evaluation of the computation in Algorithm \ref{alg2-trans} and is
given next. 
The computation in $\sf OptimalMotion$ follows the steps of
Algorithm~\ref{alg2-trans} and omit its detailed description here. We
re-arrange some operations in this protocol to reduce the number of
expensive operations.

{\small
\noindent \ifFull \line(1,0){470} \else \line(1,0){240} \fi\\
[-0.02in]
$(R, v) \leftarrow {\sf OptimalMotion}(\{(t_i = ([x_i], [y_i], [z_i]), s_i = ([x'_i], [y'_i],
 [z'_i]))\}_{i=1}^n)$\\
[-0.09in]
\ifFull \line(1,0){470} \else \line(1,0){240} \fi
\begin{compactenum}
\item For $i = 1, \ldots, n$ do in parallel
\begin{compactenum}
\item Compute $[k'_i] = [x_i] \cdot [x'_i]+[y_i] \cdot [y'_i]+[z_i] \cdot
  [z'_i]$, $[k''_i] = {\sf Sqrt}(([x_i]^2 + [y_i]^2 + [z_i]^2) ([x'_i]^2 +
  [y'_i]^2 + [z'_i]^2))$, and $[k_i] = {\sf Div}([k'_i], [k''_i])$. 

\item Compute $[p_{(i,1)}] = {\sf Sqrt}(\frac{1}{2} + \frac{1}{2} \cdot
  [k_i])$, $[p_{(i,2)}] = {\sf Sqrt}(\frac{1}{2}-\frac{1}{2}\cdot [k_i])$. 

\item Compute $[b_i] = [y_i] \cdot [z'_i]-[z_i] \cdot [y'_i]$, $[b'_i] =
  [z_i] \cdot [x'_i]-[x_i] \cdot [z'_i]$, and $[b''_i] = [x_i]\cdot
  [y'_i]-[y_i] \cdot [x'_i]$.

\item Compute $[u'_i] = {\sf Div}(1, {\sf Sqrt}([b_i]^2 + [b'_i]^2 +
  [b''_i]^2))$ and $u_i = ([u_{(i, 1)}], [u_{(i, 2)}], [u_{(i, 3)}]) = ([b_i] \cdot
  [u'_i]), ([b'_i] \cdot [u'_i]), ([b''_i] \cdot [u'_i])$.

\item Compute and set $q'_i = ([q'_{(i, 1)}], [q'_{(i, 2)}], [q'_{(i, 3)}],
  [q'_{(i, 4)}]) = ([p_{(i,1)}], [p_{(i,2)}]\cdot [u_{(i, 1)}],
  [p_{(i,2)}]\cdot [u_{(i, 2)}], [p_{(i,2)}]\cdot [u_{(i, 3)}])$.
\end{compactenum}

\item Set $q = ([q_1], [q_2, [q_3], [q_4]) = q'_1$.

\item For $i = 2, \ldots, n$ compute $[q''_1] = [q_1]\cdot [q'_{(i, 1)}]-
  [q_2]\cdot [q'_{(i, 2)}]-[q_3] \cdot [q'_{(i, 3)}]- [q_4] \cdot [q'_{(i,
    4)}]$, $[q''_2] = [q_1]\cdot [q'_{(i, 2)}]+[q'_{(i, 1)}] \cdot [q_2]+
  [q_3]\cdot [q'_{(i, 4)}] - [q_4]\cdot [q'_{(i, 3)}]$, $[q''_3] =
  [q_1]\cdot [q'_{(i, 3)}]+[q'_{(i, 1)}] \cdot [q_3]+ [q_2]\cdot [q'_{(i,
    4)}] - [q_4]\cdot [q'_{(i, 2)}]$, and $[q''_4] = [q_1]\cdot [q'_{(i,
    4)}]+[q'_{(i, 1)}] \cdot [q_4]+ [q_2]\cdot [q'_{(i, 3)}] - [q_3]\cdot
  [q'_{(i, 2)}]$, then set $[q_1] = [q''_1]$, $[q_2] = [q''_2]$, $[q_3] =
  [q''_3]$, and $[q_4] = [q''_4]$.
 
\item Compute $[\hat{q}_{(1,1)}] = [q_1]^2$, $[\hat{q}_{(2,2)}] = [q_2]^2$, $[\hat{q}_{(3,3)}] = [q_3]^2$, $[\hat{q}_{(4,4)}] = [q_4]^2$, $[\hat{q}_{(2,3)}] = [q_2] \cdot [q_3]$, $[\hat{q}_{(1,4)}] = [q_1] \cdot [q_4]$, $[\hat{q}_{(2,4)}] = [q_2] \cdot [q_4]$, $[\hat{q}_{(1,3)}] = [q_1] \cdot [q_3]$, $[\hat{q}_{(3,4)}] = [q_3] \cdot [q_4]$, and $[\hat{q}_{(1,2)}] = [q_1] \cdot [q_2]$.

\item Compute matrix $R = \{[r_{ij}]\}_{i, j = 1}^3$, where $[r_{11}] = [\hat{q}_{(1,1)}] +[\hat{q}_{(2,2)}] - [\hat{q}_{(3,3)}] - [\hat{q}_{(4,4)}]$, $[r_{12}] = 2\cdot([\hat{q}_{(2,3)}] - [\hat{q}_{(1,4)}])$, $[r_{13}] = 2\cdot([\hat{q}_{(2,4)}] + [\hat{q}_{(1,3)}])$, $[r_{21}] = 2\cdot([\hat{q}_{(2,3)}] + [\hat{q}_{(1,4)}])$, $[r_{22}] = [\hat{q}_{(1,1)}] - [\hat{q}_{(2,2)}] + [\hat{q}_{(3,3)}] - [\hat{q}_{(4,4)}]$, $[r_{23}] = 2\cdot([\hat{q}_{(3,4)}] - [\hat{q}_{(1,2)}])$, $[r_{31}] = 2\cdot([\hat{q}_{(2,4)}] - [\hat{q}_{(1,3)}])$, $[r_{32}] = 2\cdot([\hat{q}_{(3,4)}] - [\hat{q}_{(1,2)}])$, and $[r_{33}] = [\hat{q}_{(1,1)}] - [\hat{q}_{(2,2)}] - [\hat{q}_{(3,3)}] + [\hat{q}_{(4,4)}]$;
 
 \item Compute $[t''_1] = \frac{1}{n}\cdot \Sigma _{i= 1}^n [x_i]$, $[t''_2] = \frac{1}{n}\cdot \Sigma _{i= 1}^n [y_i]$, $[t''_3] = \frac{1}{n}\cdot \Sigma _{i= 1}^n [z_i]$;

\item Compute $[s''_1] = \frac{1}{n}\cdot \Sigma _{i= 1}^n [x'_i]$, $[s''_2] = \frac{1}{n}\cdot \Sigma _{i= 1}^n [y'_i]$, $[s''_3] = \frac{1}{n}\cdot \Sigma _{i= 1}^n [z'_i]$;
 
 \item Compute vector $v = \{[v_i]\}_{i = 1}^3$, where $[v_1] = [t''_1] + [r_{11}] \cdot [s''_1]+ [r_{12}] \cdot [s''_2]+ [r_{13}] \cdot [s''_3]$, $[v_2] = [t''_2] + [r_{21}] \cdot [s''_1]+ [r_{22}] \cdot [s''_2]+ [r_{23}] \cdot [s''_3]$, $[v_3] = [t''_3] + [r_{31}] \cdot [s''_1]+ [r_{32}] \cdot [s''_2]+ [r_{33}] \cdot [s''_3]$.
 
 \item Return $(R, v)$.
\end{compactenum}
\vspace{-0.1in}
\ifFull \line(1,0){470} \else \line(1,0){240} \fi} 

Note that here many steps are independent of each other and can be carried
out in parallel. For examples steps 1(a)--(b) and 1(c)--(d) correspond to
independent branches of computation. Furthermore, some (rather cheap)
redundant operations are retained in the protocol for readability, while an
implementation would execute them only once. Additional small optimizations
are also possible. For example, a number of multiplications in step 1
correspond to multiplication of integer and fixed-point operands, which can
be implemented faster than regular fixed-point multiplication.

\else 
What remains is to discuss protocol ${\sf OptimalMotion}$.
Because its computation is straightforward and because of page limitations,
we defer it to the full version.
\fi

{\color{black} \mbox{The complexity of $\sf HighCurvatureFR$ (using} $\sf
ClosestPoints$ and $\sf OptimalMotion$ as sub-protocols) is
$O(\ell \log \ell+mn\ell+\gamma(\hat{n} \hat{m} \ell^2+c(\hat{n} \log \hat{n} \log \log \hat{n} +\ell \sqrt{\hat{n}} \log^2 \hat{n})))$ interactive operations in
$O(\min(m, n)$ $\log \max(m, n) +\gamma (\ell+\hat{n} \log \hat{m} +c\log \hat{n}))$ rounds using SS and it is
$O(mn\ell^2+N\ell^2+\gamma \ell \hat{n} (\xi \hat{m} \ell +c \log \hat{n}))$ gates using GC.}

\ifFull
\else
\fi

\subsection{Secure Fingerprint Recognition based on Spectral Representation}

\ifFull
In this section, we present our third secure fingerprint recognition
protocol based on spectral minutia representation called ${\sf SpectralFR}$.
The construction builds on Algorithm~\ref{alg3} and incorporates both types
of feature reduction (CPCA and LDFT) not included in Algorithm~\ref{alg3}.
\fi
Recall that in the second type of feature reduction, LDFT, (or when both
types of feature reduction are applied) rotation of fingerprint/matrix $S$
is performed by multiplying each cell of $S$ by value
$e^{-i\frac{2\pi}{N}j\alpha}$, where $j$ is the cell's row and $\alpha$ is
the amount of rotation. While it is possible to implement rotations by
providing $2\lambda+1$ different copies of $S$ rotated by different amounts
as input into the protocol where $\lambda$ is the maximum amount of
rotation, we perform any necessary rotation inside the protocol to avoid the
price of significantly increasing the input size. We were also able to
maintain performing only $O(\log \lambda)$ score computations instead of all
$2\lambda+1$ scores in our secure and oblivious protocol, which is an
important contribution of this work. In addition, because we can avoid
increasing the input size to have a linear dependency on $\lambda$, this
allows us to 
achieve low asymptotic complexity and high efficiency. Low asymptotic
complexity is important because the size of fingerprint representation is
already large in this approach.

In what follows, we treat the case when $\lambda = 15$ (with the total of 31
rotations to be considered) as in the original algorithm \cite{xu09}, but it
is not difficult to generalize the computation to any $\lambda$. We apply
our generalization and optimization of Algorithm~\ref{alg3} described in
section~\ref{sec:alg3} with $4 \cdot 3^2 = 36$ different rotations to cover
at least 31 necessary alignments. Note that this approach computes 8
similarity scores instead of 9 in the original algorithm. We number all 36
rotations as $\alpha = -17, \ldots, 18$. The rotation constants 
$e^{-i\frac{2\pi}{N}j\alpha} = \cos(-\frac{2\pi}{N}j\alpha) +
i\sin(-\frac{2\pi}{N}j\alpha)$ are fixed and can be precomputed for each $j
= 1, \ldots, N'$ and $\alpha = -17, {\ldots}, 18$ by each party. We denote
these \mbox{values by public matrix $Z = \{z_{\alpha,j}\}_{\alpha = -17, j =
1}^{18, N'}$ which each} party stores locally. Each $z_{\alpha,j}$ is a tuple
$(z_{\alpha,j}^{(1)} = \cos(-\frac{2\pi}{N}j\alpha), z_{\alpha,j}^{(2)} =
\sin(-\frac{2\pi}{N}j\alpha))$, and $Z$ is specified as part of the input in $\sf SpectralFR$.

To be able to compute only $O(\log \lambda)$ scores in the protocol, we need
to obliviously determine the correct amount of rotation in steps 3 and 4 of
Algorithm~\ref{alg3} without revealing any information about $k$ or $k'$. 
We securely realize this functionality by placing the values
associated with each possible $k$ or $k'$ in an array and retrieving the
right elements at a private index. In more detail, the protocol first
computes four scores that correspond to rotations by $-13$, $-4$, 3, and 12
positions and computes the best among them (steps 3 and 4 below). Because
the location of the best score cannot be revealed, in the next step of the
algorithm we put together an array consisting of four vectors and one of
them is privately selected using $\sf Lookup$ (steps 4--5 of the protocol).
The selected vector consists of $2N'$ values that allow us to compute two
new scores and the maximum score for the next iteration of algorithm. We
repeat the process of private retrieval of the necessary rotation
coefficients, this time using an array consisting of twelve vectors. After
computing two more scores and determining the best score, the algorithm
outputs the best score together with the amount of rotation that resulted in
that score.

The protocol $\sf SpectralFR$ is given next. Because it 
corresponds to the computation with both types of feature reduction, input
matrices $T$ and $S$ are composed of complex values. Thus, we represent each
cell of $T$ as a pair $(a_{i,j}, b_{i,j})$ with the real and imaginary parts,
respectively, and each cell of $S$ as a pair $(a'_{i,j}, b'_{i,j})$. All
computations proceed over fixed-point values.

{\small
\noindent \ifFull \line(1,0){470} \else \line(1,0){240} \fi\\ [-0.02in] $([C_{max}], [\alpha
_{max}]) \leftarrow {\sf SpectralFR}(T = \{[t_{i,j}] = ([a_{i, j}],$ $[b_{i,
j}])\}_{i=1,j=1}^{M',N'},$ $S = \{[s_{i,j}] = ([a'_{i, j}], [b'_{i, j}])\}_{i
= 1, j = 1}^{M',N'},$ $Z = \{z_{i, j}\}_{i = -17, j= 1}^{18, N'}, \lambda
=15)$\\ [-0.09in] \ifFull \line(1,0){470} \else \line(1,0){240} \fi
\begin{compactenum}
 \item Set $[C_{max}] = [0]$.
 
 \item For $i = 1, \ldots, M'$ and $j = 1, \ldots, N'$ do in parallel if $(j
   \neq 1)$ then $[x_{i,j}] = 2([a_{i,j}] \cdot [a'_{i,j}] + [b_{i,j}] \cdot
   [b'_{i,j}])$, $[y_{i,j}] = 2 ([a'_{i,j}] \cdot [b_{i,j}] - [a_{i,j}]
   \cdot [b'_{i,j}])$, else $[x_{i, j}] = [a_{i,j}] \cdot [a'_{i,j}]$,
   $[y_{i,j}] = - [a_{i,j}] \cdot [b'_{i,j}]$. 
 
 \item For $k = 0, \ldots, 3$, do in parallel
 \begin{compactenum}
   \item $[C_{-13+9k}] = 0$.
   \item For $i = 1, \ldots, M'$ and $j = 1, \ldots, N'$ do $[C_{-13+9k}] =
     [C_{-13+9k}] + z_{-13+9k,j}^{(1)} \cdot [x_{i, j}] + z_{-13+9k,j}^{(2)}
     \cdot [y_{i,j}]$.
\end{compactenum}
 
\item Compute the maximum as $([C_{max}], [\alpha_{max}]) \leftarrow {\sf
    Max}([C_{-13}],[C_{-4}],$ $[C_{5}], [C_{14}])$. 

\item For $i = 0, \ldots, 3$ and $j = 1, \ldots, N'$ do in parallel
  $z'_{i,j} = z_{-16+9i, j}$ and $z'_{i, N'+j} = z_{-10+9i, j}$. 
 
\item Let $Z'_i = (z'_{i,1}, {\ldots}, z'_{i,2N'})$ for $i = 0, {\ldots}, 3$
  \mbox{and execute $[Z'_{max}]$} \ifFull \else\mbox{\fi $\leftarrow {\sf Lookup}((Z'_0, {\ldots}, Z'_3),
  [\alpha_{max}])$; let $[Z'_{max}] = ([\hat{z}_1],$ $ {\ldots}, [\hat{z}_{2N'}])$.\ifFull \else }\fi

 \item For $i = 1, \ldots, M'$ and $j = 1, \ldots, N'$ do
   $[C_{\alpha_{max}-3}] = [C_{\alpha_{max}-3}] + [\hat{z}^{(1)}_{j}] \cdot
   [x_{i,j}] + [\hat{z}^{(2)}_j] \cdot [y_{i,j}]$, $[C_{\alpha_{max}+3}] =
   [C_{\alpha_{max}+3}] + [\hat{z}^{(1)}_{N'+j}] \cdot [x_{i,j}] +
   [\hat{z}^{(2)}_{N'+j}] \cdot [y_{i,j}]$. 

\item Compute $([C_{max}], [\alpha_{max}]) \leftarrow {\sf
    Max}([C_{\alpha_{max}-3}], [C_{max}],$ $ [C_{\alpha_{max}+3}])$. 

\item For $i = 0, \ldots, 11$ and $j = 1, \ldots, N'$ do in parallel
  $z'_{i, j} = z_{-17+3i, j}$ and $z'_{i, N'+j} = z_{-15+3i, j}$. 
 
\item Let $Z'_i = (z'_{i,1}, {\ldots}, z'_{i,2N'})$ for $i = 0, {\ldots}, 11$
  and execute $[Z'_{max}] \leftarrow {\sf Lookup}((Z'_0, {\ldots}, Z'_{11}),
  [\alpha_{max}])$; let $[Z'_{max}] = ([\hat{z}_1], {\ldots}, [\hat{z}_{2N'}])$.

 \item For $i = 1, \ldots, M'$ and $j = 1, \ldots, N'$ do
   $[C_{\alpha_{max}-1}] = [C_{\alpha_{max}-1}] + [\hat{z}^{(1)}_{j}] \cdot
   [x_{i,j}] + [\hat{z}^{(2)}_{j}] \cdot [y_{i,j}]$, $[C_{\alpha_{max}+1}] =
   [C_{\alpha_{max}+1}] + [\hat{z}^{(1)}_{N'+j}] \cdot [x_{i,j}] +
   [\hat{z}^{(2)}_{N'+j}] \cdot [y_{i,j}]$.

\item Compute $([C_{max}], [\alpha_{max}]) \leftarrow {\sf
    Max}([C_{\alpha_{max}-1}], [C_{max}],$ $[C_{\alpha_{max}+1}])$.

\item Return $[C_{max}]$ and $[\alpha_{max}]$.
\end{compactenum}
\vspace{-0.1in}
\ifFull \line(1,0){470} \else \line(1,0){240} \fi}

\noindent
For efficiency reasons, we perform all multiplications associated with the
score computation without any rotation in the beginning of the protocol
(step 1). This computation (i.e., multiplications of the real and imaginary
components of each $t_{i,j}$ and $s_{i,j}$) is common to all score
computations (see equation~\ref{eq:score}) and is reused later in the
protocol. Then to compute a score between $T$ and $S$ rotated by $\alpha$
positions, the coefficients from $Z$ are multiplied to the computed
products. In particular, the computation takes form of $\{[z^{(1)}_{\alpha,
j}]\cdot [x_{i, j}]\}_{i=1,j=1}^{M',N'}$ and $\{[z^{(2)}_{\alpha, j}]\cdot
[y_{i, j}]\}_{i=1,j=1}^{M',N'}$ according to equation~\ref{eq:score}, which
are added together to get the score. The rest of the protocol proceeds as
described above by using private lookups twice. Note that the result
of each lookup is an array as opposed to a single element. Also note
that the score computation uses public $z_{i,j}$'s in step 1(b), but the
coefficients become private in steps 7 and 11 because they depend of private
data. Finally, the factor $\frac{1}{MN^2}$ present in
equation~\ref{eq:score} is not included in the computation because it is
public. The computed score can be scaled down by this factor by an output
recipient if necessary.

\ifFull
Recall that ${\sf SpectralFR}$ uses fixed-point arithmetic, but performance
of some operations can be optimized. For example, with SS
we can skip truncation after each multiplication in step 7 or 11
and instead truncate the sum after adding all products.  In addition, we can
restrict variables to shorter bitlength representations when the range of
values they store is known to be small. This is relevant to multiplications
in step 2 for GCs, where $t_{i,j}$ and $s_{i,j}$
are known not to exceed $N+1$ and can be represented using a short bitlength for
the integer part. 

\else
Some optimizations are possible as described in the full version of this work.
\fi
{\color{black}The complexity of the $\sf SpectralFR$ protocol is 
$O(N'M'k+\ell)$ interactive operations in $O(1)$ rounds using SS, and it is
$O(N'M' \ell^2)$ gates using GC. Therefore, the computation is expected to
be faster using SS.}

\ifFull
\else 
\fi
\section{Performance Evaluation}
\label{sec:perf}

\ifFull
\begin{figure}[t]
\centering \includegraphics[scale=0.375]{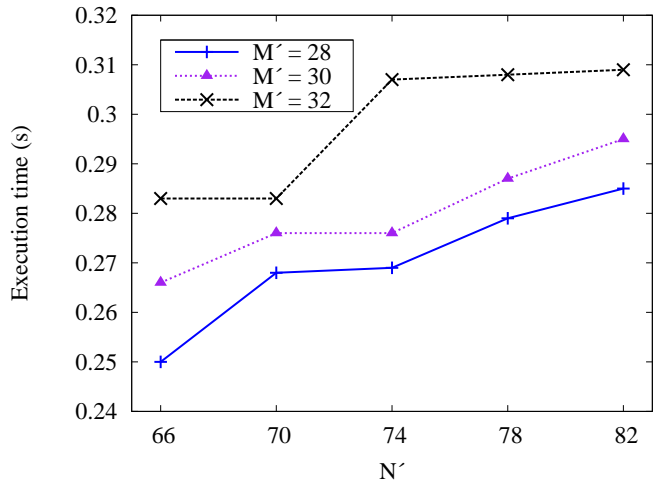}
\caption{Execution time of $\sf SpectralFR$ in seconds in the secret sharing multi-party setting.}
\label{SS-fig}
\end{figure}

\begin{figure}[t]
\centering \includegraphics[scale=0.375]{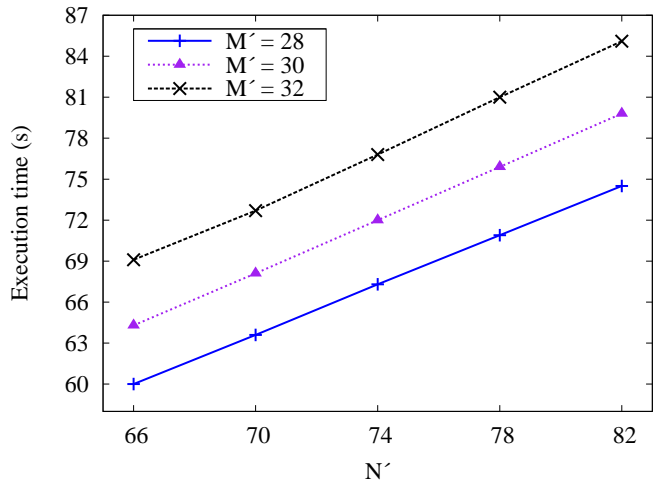}
\caption{Execution time of $\sf SpectralFR$ in seconds in the garbled circuits two-party setting.}
\label{GC-fig}
\end{figure}
\fi

In this section we evaluate performance of the spectral minutia
representation fingerprint recognition protocol $\sf SpectralFR$ to
demonstrate that the proposed protocols are practical. We provide
experimental results for both two-party and multi-party settings. The
implementations were written in C/C++. In the two-party setting, our
implementation uses the JustGarble library \cite{bel13,JG} for circuit
garbling and GC evaluation which we modified to support
the half-gates optimization \cite{zah15}. In the multi-party settings, we
use PICCO \cite{picco} with three computational parties, which utilized
linear threshold SS. {\color{black}The statistical security parameter $\kappa$
is set in PICCO to 48 and communication channels are protected using 128-bit
AES.} 

In our implementation, we assume that the inputs $T$ and $S$ are represented
with 32-bit precision after the radix point. Our GC
implementation maintains 56-bit fixed-point representation (24 and 32 bits
before and after the radix point), while in the SS setting we
let the numbers grow beyond the 56 bits to avoid the cost of truncation, but
perform truncation prior to returning the result. All machines used in the
experiments had identical hardware with four-core 3.2GHz Intel i5-3470
processors running Red Hat Linux 2.6.32 and were connected via a 1Gb LAN.
Only one core was used during the experiments. We ran each experiment 10
times and report the mean value. 

\ifFull
\begin{table}
\center \small
\begin{tabular}{ |c|c|c|c|c|c| } 
\hline
\multirow{2}{*}{$M'$} & \multicolumn{5}{| c |}{$N'$} \\
\cline{2-6}
 & $66$ & $70$ & $74$ & $78$ & $82$ \\
 \hline
$28$ & $0.250$ & $0.268$ & $0.269$ & $0.279$ & $0.285$\\
\hline
$30$ & $0.266$ & $0.276$ & $0.276$ & $0.287$ & $0.295$\\
\hline
$32$ & $0.283$ & $0.283$ & $0.307$ & $0.308$ & $0.309$\\
\hline
\end{tabular}
\caption{Execution time of protocol $\sf SpectralFR$ in seconds using secret
  sharing.} 
  \label{SS}
\end{table}

\begin{table*}[t] \centering \small
\center
\begin{tabular}{ |c|c|c|c|c|c|c|c|c|c|c| }
\hline
\multirow{3}{*}{$M'$} & \multicolumn{10}{| c |}{$N'$} \\
\cline{2-11}
& \multicolumn{2}{| c |}{$66$} & \multicolumn{2}{| c |}{$70$} & \multicolumn{2}{| c |}{$74$} & \multicolumn{2}{| c |}{$78$} & \multicolumn{2}{| c |}{$82$} \\
\cline{2-11}
& Time & Gates & Time & Gates & Time & Gates & Time & Gates & Time & Gates \\
\hline
28 & 60.0 & 454.2M & 63.6 & 481.8M & 67.3 & 509.4M & 70.9 & 537.0M & 74.5 & 564.5M \\
\hline
30 & 64.3 & 486.7M & 68.1 & 516.2M & 72.0 & 545.8M & 75.9 & 575.3 & 79.8 & 604.8M \\
\hline
32 & 69.1 & 519.1M & 72.7 & 550.6M & 76.8 & 582.1M & 81.0 & 613.6 & 85.1 & 645.1M \\
\hline
\end{tabular}
\caption{Execution time of protocol $\sf SpectralFR$ in seconds and the
  total number of gates in millions in its implementation using GCs.}
\label{GC}
\end{table*}

\else

\begin{table*}[t] \centering \small
\center\hspace*{-0.2in}
\begin{tabular}{ |c|c|c|c|c|c|c|c|c|c|c|c|c|c|c|c| }
\hline
\multirow{4}{*}{$M'$} & \multicolumn{10}{| c |}{Garbled Circuit} & \multicolumn{5}{| c |}{Secret Sharing} \\
\cline{2-16}
& \multicolumn{2}{| c |}{66} & \multicolumn{2}{| c |}{70} & \multicolumn{2}{| c |}{74} & \multicolumn{2}{| c |}{78} & \multicolumn{2}{| c |}{82} & 66 & 70 & 74 & 78 & 82\\
\cline{2-16}
& Time & Gates & Time & Gates & Time & Gates & Time & Gates & Time & Gates & \multicolumn{5}{| c |}{Time} \\
\hline
28 & 60.0 & 454.2M & 63.6 & 481.8M & 67.3 & 509.4M & 70.9 & 537.0M & 74.5 & 564.5M & 0.250 & 0.268 & 0.269 & 0.279 & 0.285\\
\hline
30 & 64.3 & 486.7M & 68.1 & 516.2M & 72.0 & 545.8M & 75.9 & 575.3 & 79.8 & 604.8M  & 0.266 & 0.276 & 0.276 & 0.287 & 0.295\\
\hline
32 & 69.1 & 519.1M & 72.7 & 550.6M & 76.8 & 582.1M & 81.0 & 613.6 & 85.1 & 645.1M & 0.283 & 0.283 & 0.307 & 0.308 & 0.309\\
\hline
\end{tabular}
\caption{Execution time of protocol $\sf SpectralFR$ in seconds and the
  total number of gates in millions.}
\label{GC}
\end{table*}
\fi

\ifFull
\else

\fi 

The results of $\sf SpectralFR$ performance evaluation can be found in 
Table \ifFull \ref{SS} \else \ref{GC} for the two-party and the three-party cases. \fi \ifFull and Figure \ref{SS-fig} for the three-party case
and in Table \ref{GC} and Figure \ref{GC-fig} for the two-party
case. \fi ~We report performance for a range of parameters $N'$ and $M'$ that
might be used in practice. All numbers correspond to the overall runtime
including communication. In the SS setting, computation uses 73--79\% of the total time. 
\ifFull
Because in the GC setting the overall runtime is
composed of multiple components, we provide a breakdown of the total time
according to its constituents. We have that garbling uses 26--27\% of the
overall time, GC evaluation takes 15--16\%, oblivious transfer
takes about 1\% of the time, and communication time is about 57\%.
\else 
In the GC setting, garbling uses 26--27\%, GC evaluation 15--16\%, OT 
about 1\%, and communication about 57\% of the total time. 
\fi
{\color{black}Table~\ref{GC} also provides the total number of gates, and}
the portion of non-XOR gates was about 27.6\% in all circuits. 
The above numbers tell
us that communication takes most of the time (even with the half-gates
optimization that reduces communication). Because circuit garbling can be
performed offline, the work associated with circuit garbling (a quarter of
the overall time) and GC transmission can be done in advance
saving most of the total time.


{\color{black} Figure~\ref{SS-f} explores scalability of the computation
and reports performance of $\sf SpectralFR$ when multiple executions
of it are carried out in parallel. It is clear that substantial
performance improvement per comparison is achieved in this case.}

As one can see, secure fingerprint recognition takes a fraction of a second
in the multi-party setting and tens of seconds in the two-party setting.
This shows that the protocol is efficient for practical use. The computation
used in $\sf SpectralFR$ is a rare example of a functionality that can be
implemented significantly more efficiently using SS techniques
than GC evaluation due to its heavy use of multiplications.

\ifFull
\begin{figure}[t]
\centering \includegraphics[scale=0.375]{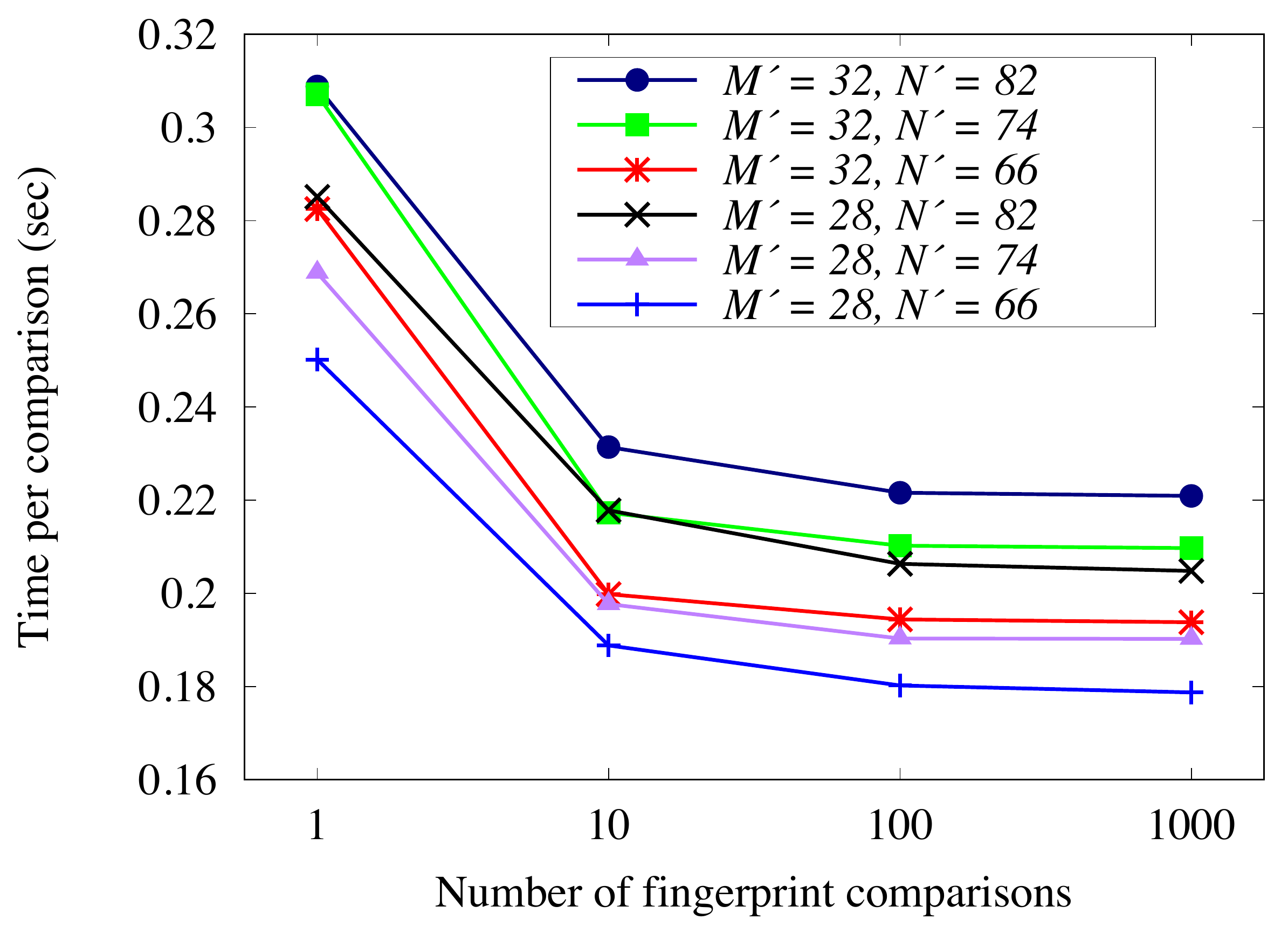}
\caption{Execution time of $\sf SpectralFR$ in seconds using SS.}
\label{SS-f}
\end{figure}
\else
\begin{figure}[t]
\centering \includegraphics[scale=0.265]{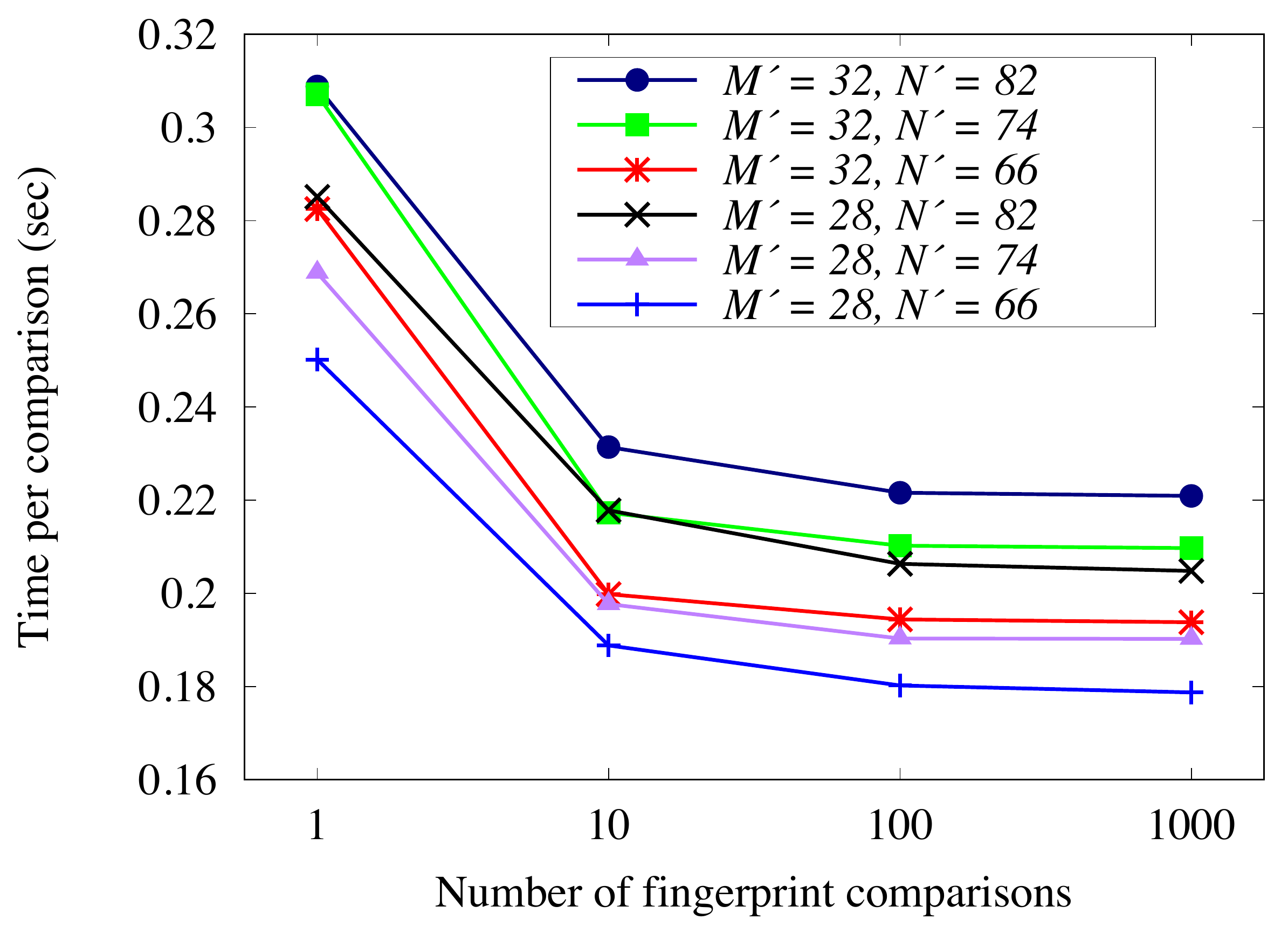}
\caption{Execution time of $\sf SpectralFR$ in seconds using SS.}
\label{SS-f}
\end{figure}
\fi

{\color{black} Because this is the first work that treats secure
fingerprint alignment, its performance is not directly comparable to
other constructions from the literature that consider only the
matching step. The closest to this work are the fingerprint matching
techniques from \cite{bla15esorics,bla15chapter} which have
corresponding implementations. \cite{bla15chapter} combines GCs with
homomorphic encryption, while \cite{bla15esorics} provides a
data-oblivious algorithm that can be instantiated with SS or GC, but
the implementation is based on GCs. Performance reported in these
publications is on the order of 1--3 seconds per comparison and
in \cite{bla15chapter} there are also one time costs that can be
amortized across multiple fingerprint comparisons.

Our $\sf SpectralFR$ is interesting because it uses only a small
number of matching computations (8), but the computation associated
with each matching step is relatively high due to the value to which
parameters $N'$ and $M'$ need to be set. Its performance is well below
1 second in the SS setting, but we cannot necessarily say that it
outperforms other approaches (without alignment). Based on the results
of this work, we anticipate that the runtime of the algorithm
in \cite{bla15esorics} can be substantially improved if multi-party SS
is employed instead of two-party GCs. This is
because the algorithm in \cite{bla15esorics} is multiplication-heavy,
similar to our $\sf SpectralFR$.}

\ifFull
\else 
\fi
\section{Conclusions}
\label{sec:conclusions}

\ifFull
In this work, we design three secure and efficient protocols for fingerprint
alignment and matching for two-party and multi-party settings. They are
based on popular algorithms in the computer vision literature and employ new
non-trivial techniques. The constructions are presented in the semi-honest
setting and known results can be applied to strengthen the security to
sustain malicious actors. We believe that this is the first work that treats
the problem of secure fingerprint alignment. 

Another main contribution of this work is designing secure constructions for
fundamental numeric and set operations. We present novel secure protocols
for sine, cosine, arctangent, and square root for fixed-point numbers, as
well as a novel secure and data-oblivious protocol for selection of the
$f$th smallest element of a set (for any type of data). The techniques are
applicable to both two-party and multi-party settings. Our hope is that
these techniques find their applicability well beyond fingerprint
recognition and that this work will facilitate secure processing of
biometric data in practice.
\else
In this work, we design three secure and efficient protocols for fingerprint
alignment and matching for two-party and multi-party settings using new
non-trivial techniques and based on popular algorithms in the computer
vision literature. We believe that this is the first work that treats the
problem of secure fingerprint alignment. 

Another main contribution of this work is designing secure constructions for
fundamental numeric and set operations. We present novel secure protocols
for sine, cosine, and arctangent for fixed-point numbers, as
well as a data-oblivious protocol for selection of the
$f$th smallest element of a set (for any type of data). We hope that
these techniques find their applicability well beyond fingerprint
recognition and that this work will facilitate secure processing of
biometric data in practice.
\fi
\ifFull
\else
\fi
\ignore{
\ifFull
\else
\begin{table}
\center \small
\begin{tabular}{ |c|c|c|c|c|c| } 
\hline
\multirow{2}{*}{$M'$} & \multicolumn{5}{| c |}{$N'$} \\
\cline{2-6}
 & $66$ & $70$ & $74$ & $78$ & $82$ \\
 \hline
$28$ & $0.250$ & $0.268$ & $0.269$ & $0.279$ & $0.285$\\
\hline
$30$ & $0.266$ & $0.276$ & $0.276$ & $0.287$ & $0.295$\\
\hline
$32$ & $0.283$ & $0.283$ & $0.307$ & $0.308$ & $0.309$\\
\hline
\end{tabular}
\caption{Execution time of $\sf SpectralFR$ in seconds using SS.} 
  \label{SS}
\end{table}
\fi}

\ifFull
\section*{Acknowledgments}

The authors would like to thank Karthik Nandakumar for his help with making
the source code of the fingerprint recognition algorithm from \cite{gain}
available. This work was supported in part by grants 1223699, 1228639,
1319090, and 1526631 from the National Science Foundation and
FA9550-13-1-0066 from the Air Force Office of Scientific Research, as well
as DARPA agreement no.~AFRL FA8750-15-2-0092.  Any opinions, findings, and
conclusions or recommendations expressed in this publication are those of
the authors and do not necessarily reflect the views of the funding agencies
or the U.S. Department of Defense.
\fi

\bibliographystyle{abbrv}
\ifFull
\bibliography{Fingerprint-Full}
\else
\bibliography{Fingerprint-ACM}
\fi

\ifFull
\appendix

\section{Alternative Design of Building Blocks}

\subsection{Trigonometric Functions}
\label{sec:trig2}

Recall that in section~\ref{sec:trig} we provide a method for securely
evaluating trigonometric and inverse trigonometric functions using
polynomial evaluation. In this section we present an alternative approach
based on private lookup. When the input range is constrained (as in
the fingerprint recognition algorithms we are considering), this mechanism
could result in improved performance, but in general applications with
arbitrary precision it is likely to run into scalability issues.

Consider a trigonometric function (such as sine or cosine used in this work)
that needs to be evaluated on private input $a$. Let the value of
$a$ be in the range $[a_{min}, a_{max}]$ with $N$ denoting the number of the
elements in the range. Then the approach consists of precomputing the
function on all possible inputs and storing the result in an array $Z = (z_0,
{\ldots}, z_{N-1})$. Consequently, evaluation of the function on private $a$
corresponds to privately retrieving the needed element of the array $Z$
using $a$ to determine the index. This procedure is formalized in the
protocol $\sf Trig$ below.

{\small \noindent \ifFull \line(1,0){470} \else \line(1,0){240} \fi \\ [-0.02in]
$[b] \leftarrow {\sf Trig}([a], Z = \{z_i\}_{i=0}^{N-1}, a_{min})$ \\[-0.09in]
\ifFull \line(1,0){470} \else \line(1,0){240} \fi
\begin{compactenum}
\item Compute $[b] \leftarrow {\sf Lookup}(\langle z_0, {\ldots}, z_{N-1}
  \rangle, [a]-a_{min})$.

\item Return $[b]$.
\end{compactenum}
\ifFull \line(1,0){470} \else \line(1,0){240} \fi}

\noindent Complexity of this protocol is heavily dominated by that of $\sf
Lookup$ on an array of size $N$, and the overall complexity is thus $O(N
\log N)$. In our application, when we compute image rotation on a small
angle expressed as an integer in degrees, the value of $N$ is small. For
example, with rotations by at most 10 degrees in each direction, we have $N
= 21$ and thus this approach becomes very efficient.

The situation, however, is different with inverse trigonometric functions.
The input to such functions is now a real number, which changes the way a
lookup is to be implemented; but more importantly, it is desirable to
have the function support finer granularity than outputting integer angles
expressed in degrees. This will require the size of the array used for
private lookup increase substantially, mitigating the benefits of this
approach. Thus, for inverse trigonometric functions such as arctangent we
recommend our main solution described in section~\ref{sec:trig}.

\subsection{Square Root}
\label{sec:sqrt2}

As an alternative approach to securely computing the square root of $[a]$,
we suggest to use approximations from~\cite[Chapter~6]{computer-app}.
\cite{computer-app} uses polynomial approximations for square root based on
Heron's and Newton's iterations and has polynomial coefficients precomputed
for different input ranges and precision values. 

To work correctly, this method requires that the range of input $[a]$ is
reduced to a specific range. Range reduction of $a$ follows the formula:
\begin{equation} \label{eq:sqrt}
\sqrt{x} = \beta^v \sqrt{a\beta^{-2v}} = \beta^v \sqrt{\beta} \sqrt{a
  \beta^{-(2v+1)}}
\end{equation}
Based on \cite[Chapter~6]{computer-app}, the constant $\beta$ can be set to
the base of the computer number system (such as 2) and the range of the
input will be reduced to $\frac{1}{\beta} \leq a' \leq 1$, where $a' =
a\beta^{-2v}$ or $a' = a \beta^{-(2v+1)}$ is the normalized input.
Alternatively, if a longer range can give acceptable accuracy, the constant
$\beta$ can be set to a power of the computer number system base (e.g.,
$2^2$, etc.) and the range will still be reduced to $\frac{1}{\beta} \leq a'
\leq 1$. The trade-off here is that the larger the value of $\beta$ is, the
smaller precision can be obtained using polynomials of the same degree.
Thus, to achieve equivalent precision, polynomials of larger degrees will
need to be used with larger $\beta$ compared to lower values of $\beta$. We
suggest to set $\beta = 2$ and thus reduce the range of the input to
$\frac{1}{2} \leq a' \leq 1$, which provides high precision for given
polynomial degrees.

Next, note that if we let $\frac{1}{2} \le a' < 1$, the normalized value
$a'$ lies in the same range as the normalized value in Section~\ref{SR} and
we can apply the normalization procedure $\sf Norm$ described in that
section. Furthermore, if we let $\beta = 2$ and adopt notation $a' = a \cdot
2^{-w}$ used in Section~\ref{SR}, then based on equation~\ref{eq:sqrt},
$2^{-w}$ is either $2^{-2v}$ or $2^{-(2v+1)}$ depending on whether $w$ is
even or odd. Recall that the $\sf Norm$ protocol outputs $[a']$ together
with $[2^{-\lfloor \frac{w}{2} \rfloor}]$ and bit $[c]$ which is set when
$w$ is odd. This is very similar to what want here. That is, $2^v =
2^{\lfloor \frac{w}{2} \rfloor}$ and $c$ is set when $w$ is odd to indicate
that multiplication by $\sqrt{2}$ of the result is needed (as in
equation~\ref{eq:sqrt}). This means that we can utilize a variant of the
$\sf Norm$ protocol from Section~\ref{SR} that outputs $[2^{\lfloor
\frac{w}{2} \rfloor}]$ instead of $[2^{-\lfloor \frac{w}{2} \rfloor}]$ but
otherwise operates in the same way. To make this change to the $\sf Norm$
protocol of Section~\ref{SR}, we only need to replace step 9 with
$[2^{\lfloor \frac{w}{2} \rfloor}] \leftarrow (d_{\ell-1}, {\ldots}, [d_0])
= (0^{\frac{\ell-k}{2}-1}, [u_{\frac{\ell}{2}}], [u_{\frac{\ell}{2}-1}],
{\ldots}, [u_{0}], 0^{\frac{k}{2}})$ and output $[2^{\lfloor \frac{w}{2}
\rfloor}]$ in step 11. It is also not difficult to modify the normalization
protocol in~\cite{liedel1} for the SS setting. In particular, we need
to replace lines 8--10 of $\sf NormSQ$ in \cite{liedel1} by the
following computations:

\medskip 
\begin{compactenum}
\item[8.] Let $[u_{0}] = [z_{0}]$, $[u_{\frac{k}{2}}] = [z_{k-1}]$ and for $i =
    1, {\ldots}, \frac{k}{2}-1$ do in parallel $[u_i] =[z_{2i-1}] + [z_{2i}]$.

\item[9.] Compute $[2^{\lfloor \frac{w}{2} \rfloor}] = \sum_{i =
0}^{k/2} 2^{i+f/2} [u_i]$.

\item[10.] Set $[d] = \sum_{i = 0}^{\frac{k}{2}-1} [z_{2i}]$.
\end{compactenum}

\medskip \noindent 
The protocol then returns $\langle [c], [2^{\lfloor \frac{w}{2}
\rfloor}], [d]\rangle$, which corresponds to $\langle [a'],
[2^{\lfloor \frac{w}{2} \rfloor}], [c] \rangle$ using our notation
from $\sf Norm$. Note that here a fixed-point value is represented
using the total of $k$ bits, $f$ of which are stored after the radix point.

Once the input is normalized to the desired range, evaluation of the square
root function amounts to evaluating polynomials on the normalized input.
Given the desired precision of the computation in bits, we determine the
minimum degrees of the polynomials and the coefficients corresponding to the
polynomials can be looked up from~\cite{computer-app}. Similar to the
polynomial evaluation approach described for trigonometric functions in
Section~\ref{sec:trig}, there are two types of polynomial approximations:
One type uses a regular polynomial $P(x)$ of degree $N$ and the other type
uses a rational function $P(x)/Q(x)$, where $P$ and $Q$ are polynomials of
degrees $N$ and $M$, respectively. Based on the cost of multiplication and
division in the selected setting, either a regular polynomial or rational
function might be preferred. We recommend to use a rational function for
GCs-based implementations and regular polynomial for SS-based
implementations. The overall square root protocol $\sf Sqrt2$ is given next.

{\small \noindent \ifFull \line(1,0){470} \else \line(1,0){240} \fi \\ [-0.02in]
$[b] \leftarrow {\sf Sqrt2}([a])$ \\[-0.09in]
\ifFull \line(1,0){470} \else \line(1,0){240} \fi
\begin{compactenum} 
  \item Execute $\langle [a'], [2^{\lfloor \frac{w}{2}\rfloor}], [c]\rangle
    \leftarrow {\sf Norm}([a])$.
  
  \item Lookup the minimum polynomial degrees $N$ and $M$ using range
    $[\frac{1}{2}, 1)$ for which precision of the approximation is at least
    $k$ bits.
    
  \item Lookup polynomial coefficients $p_0, {\ldots}, p_N$ and $q_0, {\ldots},
    q_M$ for square root approximation.
    
  \item Compute $([z_1], \ldots, [z_{\max(N, M)}]) \leftarrow {\sf
      PreMul}([a'], \max(N, M))$. 
    
  \item Compute $[y_P] = p_0 + \sum_{i=1}^N p_i [z_i]$.
    
  \item Compute $[y_Q] = q_0 + \sum_{i=1}^M q_i [z_i]$.
    
  \item Compute $[y] \leftarrow {\sf Div}([y_P], [y_Q])$.
  
    
  \item Compute and return $[b] = [y] \cdot [2^{\lfloor
      \frac{w}{2}\rfloor}] \cdot (([c] \wedge \sqrt{2}) \oplus \neg[c])$.
\end{compactenum}
\ifFull \line(1,0){470} \else \line(1,0){240} \fi} 

\noindent
In step 8 of this protocol we use (cheap) XOR in the GC setting, while in
the SS setting, that operation is replaced with addition. As before, this
protocol is written using a rational function for approximation. When a
single polynomial is used instead, the protocol evaluates only a single
polynomial $P$ of (a different) degree $N$ and the division operation is
skipped in step 7.

\else
\newpage
\appendix

\subsection{Secure Function Evaluation Frameworks}
\label{app:sfe}

\noindent \textbf{Garbled circuit evaluation.}
The use of GCs allows two parties $P_1$ and $P_2$ to
securely evaluate a Boolean circuit of their choice. That is,
given an arbitrary function $f(x_1, x_2)$ that depends on private
inputs $x_1$ and $x_2$ of $P_1$ and $P_2$, respectively, the
parties first represent is as a Boolean circuit. One party, say
$P_1$, acts as a circuit generator and creates a garbled
representation of the circuit by associating both values of each
binary wire with random labels. The other party, say $P_2$, acts
as a circuit evaluator and evaluates the circuit in its garbled
representation without knowing the meaning of the labels that it
handles during the evaluation. The output labels can be mapped to
their meaning and revealed to either or both parties.


An important component of GC evaluation is 1-out-of-2
Oblivious Transfer (OT). It allows the circuit evaluator to obtain
wire labels corresponding to its inputs. In particular, in OT the
sender (i.e., circuit generator in our case) possesses two strings
$s_0$ and $s_1$ and the receiver (circuit evaluator) has a bit
$\sigma$. OT allows the receiver to obtain string $s_\sigma$ and
the sender learns nothing. An OT extension allows
any number of OTs to be realized with small additional overhead
per OT after a constant number of regular more costly OT protocols
(the number of which depends on the security parameter). The
literature contains many realizations of OT and its extensions,
including very recent proposals such as \cite{nao01,ish03,ash13} and
others.

The fastest currently available approach for GC generation
and evaluation we are aware of is by Bellare et al. \cite{bel13}.
It is compatible with earlier optimizations, most notably the
``free XOR'' gate technique \cite{kol08} that allows XOR gates to
be processed without cryptographic operations or communication,
resulting in lower overhead for such gates. A recent half-gates
optimization \cite{zah15} can also be applied to this construction to reduce
communication\ifFull associated with garbled gates.\else.\fi 

\smallskip \noindent \textbf{Secret sharing.} SS techniques allow
for private values to be split into random shares, which are distributed
among a number of parties, and perform computation directly on secret shared
values without computationally expensive cryptographic operations. Of a
particular interest to us are linear threshold SS schemes. With
a $(n,\tau)$-secret sharing scheme, any private value is secret-shared among
$n$ parties such that any $\tau + 1$ shares can be used to reconstruct the
secret, while $\tau$ or fewer parties cannot learn any information about the
shared value, i.e., it is perfectly protected in the information-theoretic
sense. In a linear SS scheme, a linear combination of
secret-shared values can be performed by each party locally, without any
interaction, but multiplication of secret-shared values requires
communication between all of them.

In this setting, we can distinguish between the input owner who provide
input data into the computation (by producing secret shares), computational
parties who conduct the computation on secret-shared values, and output
recipients who learn the output upon computation termination (by
reconstructing it from shares). These groups can be arbitrarily overlapping
and be composed of any number of parties as long as there are at least 3
computational parties. 

In the rest of this work, we assume that Shamir SS \cite{shamir} is
used with $\tau < n/2$ in the semi-honest setting for any $n \ge 3$.

\subsection{Known Building Blocks}
\label{app:kbb}

Some operations used in the computation are elementary and
well-studied in the security literature (e.g.,
\cite{catrina-int,catrina-fp,marina4,bla16}), while others are more complex,
but still have presence in prior work (e.g.,
\cite{marina4,catrina-fp,marina6,marina7}). Below we define building blocks
that we use in this work; their overhead is listed in Table~\ref{tab2-app}.

\begin{itemize}
  \item \emph{Addition} $[c] \leftarrow [a] + [b]$ and \emph{subtraction}
    $[c] \leftarrow [a] - [b]$ are considered free (non-interactive) using
    SS in both fixed-point and integer representations
    \cite{catrina-fp}, while their cost for $\ell$-bit
    $a$ and $b$ using GCs is from \cite{kolesnikov}.
    
  \item \emph{Multiplication} $[c] \leftarrow [a] \cdot [b]$ of
    integers is a single elementary operation using SS, while its
    complexity for fixed-point numbers is
    from \cite{catrina-fp}. Using GCs, the cost (for integer and
    fixed-point values) using the traditional algorithm is given
    in \cite{kolesnikov}, which can be reduced using the Karatsuba's
    method \cite{kara} for relatively large
    bitlengths \cite{sadeghi1}.
   
  \item \emph{Comparison} $[c] \leftarrow {\sf LT}([a], [b])$ tests for
    $a < b$ (and other variants) and outputs a bit. Its cost for SS is from 
    \cite{catrina-int} and for GCs from \cite{kolesnikov}. 
  
  \item \emph{Equality testing} $[c] \leftarrow {\sf EQ}([a], [b])$
    similarly produces a bit. Its cost can be found in 
    \cite{catrina-int} (SS) and \cite{kol08} (GCs).

 \item \emph{Division} $[c] \leftarrow {\sf Div}([a], [b])$ of SS
   fixed-point numbers is available from \cite{catrina-fp} using
   Goldschmidt's method that proceeds in $\xi =
   \lceil\log _2(\frac{\ell}{3.5})\rceil$ iterations. For GCs,
   we use the standard (shift and subtraction solution from
   \cite{marina4}.
 
 \item \emph{Integer to fixed-point conversion} $[b] \leftarrow {\sf
   Int2FP}([a])$ appends a number of zeros after the radix point and is
   very cheap in both SS and GCs.
 
 \item \emph{Fixed-point to integer conversion} $[b] \leftarrow {\sf
     FP2Int}([a])$ truncates all bits of $a$ after the radix point. It
   involves no gates using GCs and a solution for SS can be found in 
   \cite{catrina-fp}.
 
  \item \emph{Conditional statements with private conditions} of the form 
    ``if $[priv]$ then $[a] = [b]$; else $[a] = [c]$;'' are transformed into
    $[a] = ([priv] \wedge ([b] \oplus [c])) \oplus [c]$ using GCs and
    $[a] = [priv]\cdot([b]-[c])+[c]$ using SS.
    For a single branch, replace one of the assignments with $[a] = [a]$.
    
  \item \emph{Maximum or minimum} of a set $\langle [a_{max}], [i_{max}]
    \rangle \leftarrow {\sf Max}([a_1], \ldots,$ $ [a_m])$ or $\langle
  [a_{min}], [i_{min}] \rangle \leftarrow {\sf Min}([a_1]$, $\ldots, [a_m])$
  returns the maximum/minimum element together with its index in the set.
  Its solution for GCs is from \cite{kolesnikov}. For SS, the comparisons
  are organized into a binary tree with $m/2$ comparisons performed in the
  first iteration, $m/4$ comparisons in the second iteration, etc. 

\item \emph{Prefix multiplication} $\langle [b_1], {\ldots}, [b_m] \rangle
  \leftarrow {\sf PreMul}([a_1], {\ldots}$, $[a_m])$ (or ${\sf PreMul}([a],
  m)$ when all $a_i$s are equal) simultaneously computes $[b_i] = \prod_{j =
  1}^i [a_j]$ for $i = 1, \ldots, m$. It saves the number of rounds with SS
  (and the operation is not used with GCs). An implementation of $\sf
  PreMul$ for integers is available from \cite{catrina-int} (non-zero
  integers only). For fixed-point values, {\color{black}we use the structure
  of the solution for arbitrary prefix operations from \cite{catrina-int}
  which in this case results in invoking $0.5m\log m$ fixed-point
  multiplications in $\log m$ iterations of $0.5m$ operations each. This
  gives us the total of $0.5m \log m(2k+2)$ interactive operations executed
  in $2\log m +2$ rounds. In the special case when all $a_i$s are equal, we
  can optimize this solution to save many multiplications and execute only
  $m-1$ fixed-point multiplications. This results in $(m -1)(2k+2)$
  interactive operations in $2\log m +2$ rounds.}

  \item \emph{Compaction} $\langle [b_1], {\ldots}, [b_m] \rangle \leftarrow
    {\sf Comp}([a_1], {\ldots}, [a_m])$ pushes all non-zero elements of its
    input set to appear before any zero element. We use an order-preserving
    compaction \cite{bla15esorics} (based on \cite{goodrich}) that works with
    both GCs and SS and any type of data, and the variant that takes a
    set of tuples $\langle a'_i, a''_i \rangle$ as its input, where each
    $a'_i$ is a bit that indicates whether the data item $a''_i$ is zero or
    not (i.e., comparison of each data item to 0 is not needed).

  \item \emph{Array access at a private index} allows one to read or write an
    array element at a private location. We use only read accesses and
    denote the operation as $[b] \leftarrow {\sf Lookup}(\langle [a_1]$,
    ${\ldots}, [a_m] \rangle, [ind])$, where $a_i$s might be protected or
    public. Typical simple implementations of this operations include a
    multiplexer (as in, e.g., \cite{picco}) and comparing the index $[ind]$
    to all positions of the array and obliviously choosing one of them, both
    requiring $O(m \log m)$ work. Based on our analysis, a multiplexer-based
    solution is faster for GCs, while it is the comparison-based solution for
    SS, and thus we use these options.
    
\item \emph{Oblivious sorting} $\langle [b_1], \ldots, [b_m]\rangle
  \leftarrow {\sf Sort}([a_1], \ldots, [a_m])$ obliviously sorts an
  $m$-element set. Despite suboptimal complexity, Batcher's merge sort
  \cite{bat68} is often faster than other algorithms and this is what we
  employ.

\item {\color{black}\emph{Square Root} $[b] \leftarrow {\sf Sqrt}([a])$ computes
the square root of a fixed-point value $a$. We are aware only of a
construction for SS \cite{liedel1} and optimize it for GCs based on the
specifics of that technique in Appendix \ref{app:SR} below. The solution in
\cite{liedel1} assumes that bitlength $\ell$ is a power of 2 and the number
of algorithm iterations is $\xi = \lceil \log_2(\ell/5.4) \rceil$.}
\end{itemize}

\ifFull
\else
\begin{table*}[t] \centering \small \setlength{\tabcolsep}{0.3ex} 
\begin{tabular}{|c|c|c|c|c|} 
\hline
\multirow{2}{*}{Prot.} & \multicolumn{2}{c|}{Secret sharing} & \multicolumn{2}{c|}{Garbled circuits}\\
\cline{2-5}
& Rounds & Interactive operations & XOR gates & Non-XOR gates \\
\hline
${\sf Add}/{\sf Sub}$ & $0$ & $0$ & $4\ell$  & $\ell$ \\
\hline
${\sf LT}$ & $4$ & $4\ell-2$ & $3\ell$ & $\ell$\\
\hline
${\sf EQ}$ & $4$ & $\ell+4\log \ell$ & $\ell$ & $\ell$\\
\hline
${\sf Mul}$ & $4$ & $2k+2$ & $4\ell ^2-4\ell$  & $2\ell ^2-\ell$\\
\hline
${\sf Div}$ & $3\log\ell + 2\xi +12$ & $1.5\ell\log\ell+ 2 \ell\xi + 10.5\ell
+ 4\xi + 6$ & $7\ell^2 + 7\ell$ & $3\ell^2 + 3\ell$ \\
\hline
${\sf PreMul}$ & $2 \log m  + 2$ & $(m-1)(2k+2)$ & $-$ & $-$\\
\hline
${\sf Max}/{\sf Min}$ & $4
\log m + 1$ & $4\ell (m-1)$ & $5\ell (m-1)$ & $2\ell (m-1)$\\
\hline
${\sf Int2FP}$ & $0$ & $0$ & $0$  & $0$\\
\hline
${\sf FP2Int}$ & $3$  & $2k+1$ & $0$  & $0$\\
\hline
\multirow{2}{*}{${\sf Comp}$} & \multirow{2}{*}{$\log m+ \log \log m + 3$} & $m \log m \log \log m+4 m \log m$ & $(\ell +4)m \log m$ & $(2\ell +1)m \log m - 2 \ell m$\\

& & $-m +\log m+2$ & $- m \ell -4 \log m +\ell$ & $+(\ell -1) \log m +2\ell$ \\
\hline
${\sf Sort}$ & $2 \log m (\log m +1)+1$ & $\ell (m-0.25)(\log^2 m +\log m +4)$
& $1.5 m \ell (\log^2 m +\log m + 4)$ & $0.5 m \ell (\log^2 m +\log m + 4)$\\
\hline 
${\sf Lookup}$ & $5$ & $m \log m +4m \log \log m+m$ & $m\ell + \log m -\ell$ & $m \log m + m (\ell -1)$\\
\hline
\multirow{2}{*}{${\sf Sqrt}$} & $0.5\ell+2\log \ell$ & $2\ell^2+\ell \log \ell
+3k(\xi +1)$ & $12 \xi \ell^ 2+12.5 \ell^2-k^2 + \ell k$ & $6 \xi \ell^2 +
6.5\ell^2-k^2 + \ell k$\\
& $+6\xi+24$ & $+5\ell +6 \xi +12$ & $ -8 \xi \ell-7.5  \ell +k -2$ & $-2 \xi
\ell-0.5 \ell+k -4$\\ \hline
\end{tabular}
\caption{Performance of secure building blocks for fixed-point values.} \label{tab2-app}
\end{table*}

\fi

\subsection{Square Root}
\label{app:SR}

{\color{black}
We describe the square root computation defined by the interface $[b]
\leftarrow {\sf Sqrt}([a])$, where $a$ and $b$ are fixed-point values to
cover the general case. Secure computation of this function using SS
appeared in \cite{liedel1} and here we optimize it for GCs. The approach is
based on the Goldschmidt's algorithm, which is faster than the
Newton-Raphson's method. However, to eliminate the accumulated errors, the
last iteration of the algorithm is replaced with the self-correcting
iteration of the Newton-Raphson's method. } 

Goldschmidt's method starts by computing an initial approximation for
$\frac{1}{\sqrt{a}}$, denoted by $b_0$, that satisfies $\frac{1}{2} \leq
ab_0^2 < \frac{3}{2}$. It then proceeds in iterations increasing the
precision of the approximation with each consecutive iteration. To
approximate $\frac{1}{\sqrt{a}}$, \cite{liedel1} uses an efficient method (a
linear equation) that expects that input $a$ is in the range $[\frac{1}{2},
1)$. Thus, there is a need to first normalize $a$ to $a'$ such that
$\frac{1}{2} \le a' < 1$ and $a = a' \cdot 2^w$. Note that
$\frac{1}{\sqrt{a}} = \frac{1}{\sqrt{a'}} \sqrt{2^{-w}}$, therefore, once we
approximate $\frac{1}{\sqrt{a'}}$, we can multiply it by $\sqrt{2^{-w}}$ to
determine an approximation of $\frac{1}{\sqrt{a}}$.

In our $(\ell,k)$-bit fixed-point representation, a normalized input $a' \in
[\frac{1}{2}, 1)$ has the most significant non-zero bit exactly at position
$k-1$. We also express $\sqrt{2^{-w}}$ as $\frac{1}{\sqrt{2}} 2^{-\lfloor
\frac{w}{2} \rfloor}$ when $w$ is odd and as $2^{-\lfloor \frac{w}{2}
\rfloor}$ when $w$ is even. Our normalization procedure $\sf Norm$ that we
present next thus takes input $a$ and returns $\frac{1}{2} \le a' < 1$
(normalized input $a$ in $(\ell,k)$-bit fixed-point representation),
$2^{-\lfloor \frac{w}{2} \rfloor}$ and bit $c$ set to 1 when $w$ is odd and
to 0 otherwise. The protocol is optimized for the GC approach
with cheap XOR gates \cite{kol08}. \mbox{It assumes that $\ell$ and $k$ are even.}

{\small \noindent \ifFull \line(1,0){470} \else \line(1,0){240} \fi\\[-0.02in]
$\langle [a'], [2^{-\lfloor \frac{w}{2} \rfloor}], [c]\rangle \leftarrow {\sf
  Norm}([a])$\\[-0.09in] 
\ifFull \line(1,0){470} \else \line(1,0){240} \fi
\begin{compactenum}
\item $([a_{\ell-1}], \ldots, [a_0]) \leftarrow [a]$.

\item Set $[x_{\ell-1}]= [a_{\ell-1}]$.

\item For $i = \ell-2, \ldots, 0$ do $[x_i] =[a_i] \vee [x_{i+1}]$.

\item Set $[y_{\ell-1}]= [x_{\ell-1}]$.

\item For $i = 0, {\ldots}, \ell-2$ do in parallel $[y_i] =[x_i] \oplus [x_{i+1}]$.

\item For $i = 0, \ldots, \ell-1$ do in parallel $[z^{(i)}] \leftarrow
  ([a_{\ell-1}] \wedge [y_i], \ldots, [a_0] \wedge [y_i])$.

\ifFull 
\item Compute $[a'] = \left(\bigoplus_{i = 0}^{k-1} \left([z^{(i)}] \ll
  (k-1-i)\right)\right) \oplus \left(\bigoplus_{i = k}^{\ell-1}
  \left([z^{(i)}]\gg (i-(k-1))\right)\right)$.
\else
\item Compute $[a'] = \left(\bigoplus_{i = 0}^{k-1} \left([z^{(i)}] \ll
  (k-1-i)\right)\right) \oplus \\ \left(\bigoplus_{i = k}^{\ell-1}
  \left([z^{(i)}]\gg (i-(k-1))\right)\right)$.
\fi

\item Let $[u_0] = [y_0]$, $[u_{\frac{\ell}{2}}] = [y_{\ell-1}]$ and for $i =
    1, {\ldots}, \frac{\ell}{2}-1$ do in parallel $[u_i] =[y_{2i-1}] \oplus [y_{2i}]$.


\item Set $[2^{-\lfloor \frac{w}{2} \rfloor}] \leftarrow ([d_{\ell-1}],
  {\ldots}, [d_0]) = (0^{\ell-\frac{3k}{2}-1}, [u_0], [u_1], {\ldots},
  $ $[u_{\frac{\ell}{2}}], 0^{\frac{3k-\ell}{2}})$, where $0^x$ corresponds to
  $x$ zeros. 
  
\item Set $[c] = \bigoplus_{i =0}^{\frac{\ell}{2}-1} [y_{2i}]$ and return $\langle [a'], [2^{-\lfloor \frac{w}{2} \rfloor}], [c] \rangle$.




\end{compactenum}
\ifFull \line(1,0){470} \else \line(1,0){240} \fi}

\noindent
Here, lines 2--3 preserve the most significant zero bits of $a$ and set the
remaining bits to 1 in variable $x$ (i.e., all bits following the most
significant non-zero bit are 1). Lines 4--5 compute $y$ as a vector with the
most significant non-zero bit of $a$ set to 1 and all other bits set to 0.
On line 6, each vector $z^{(i)}$ is either filled with 0s or set to $a$
depending on the $i$th bit of $y$ (thus, all but one $z^{(i)}$ can be
non-zero). Line 7 computes the normalized value of $a$ by aggregating all
vectors $z^{(i)}$ shifted an appropriate number of positions. Here operation $x
\ll y$ shifts $\ell$-bit representation of $x$ $y$ positions to the left by
discarding $y$ most significant bits of $x$ and appending $y$ 0s in place
of least significant bits. Similarly, $x \gg y$ shifts $x$ to the right by
prepending $y$ 0s in place of most significant bits and discarding $y$ least
significant bits of $x$. Note that we can use cheap XOR gates for this
aggregation operation because at most one $y_i$ can take a non-zero value.

Lines 8 and 9 compute $[2^{-\lfloor \frac{w}{2} \rfloor}]$. Because a pair of
consecutive $i$'s results in the same value of $w$, we first combine the
pairs on line 8 and shift them in place on line 9. As before, we can use
cheap XOR and free shift operations to accomplish this task because at most
one $y_i$ is set. Lastly, line 10 computes the bit $c$, which is set by
combining all flags $y_i$'s at odd distances from $k-1$.

Note that for simplicity of exposition, we AND and XOR all bits in steps 6
and 7. There is, however, no need to compute the AND for the bits discarded
in step 7 or compute the XOR with newly appended or prepended 0s. We obtain
that this protocol can be implemented as a circuit using $0.5\ell^2 - k^2 +
\ell k + 1.5\ell - 4$ non-XOR and $0.5\ell^2 - k^2  + \ell k + 0.5\ell-3$
XOR gates.

Once we determine normalized input $a'$, Liedel \cite{liedel1} approximates
$\frac{1}{\sqrt{a'}}$, denoted by $b'_0$, by a linear equation $b'_0 =
\alpha a' + \beta$, where $\alpha$ and $\beta$ are precomputed. The values
of these coefficients are set to $\alpha = -0.8099868542$ and $\beta =
1.787727479$ in \cite{liedel1} to compute an initial approximation $b'_0$
with almost 5.5 bits of precision (when $a'$ is in the range $[\frac{1}{2},
1)$). We then use $b'_0$ to compute $b_0$ that approximates
$\frac{1}{\sqrt{a}}$ as described above.

After the initialization, Goldschmidt's algorithm sets $g_0 = x b_0$ and
$h_0 = 0.5 b_0$ and proceeds in $\xi$ iterations that compute: $g_{i+1} =
g_i(1.5 - g_i h_i)$, $h_{i+1} = h_i(1.5- g_i h_i).$ The last iteration is
replaced with one iteration of Newton-Raphson's method to eliminate
accumulated errors that computes the following: $h_{i+1} =
h_i(1.5-0.5ah_i^2).$ This gives us the following square root protocol, which
we present optimized for the GC approach.

{\small \noindent \ifFull \line(1,0){470} \else \line(1,0){240} \fi\\[-0.02in]
$[b] \leftarrow {\sf Sqrt}([a])$\\[-0.09in]
\ifFull \line(1,0){470} \else \line(1,0){240} \fi
\begin{compactenum}
\item Let $\xi = \lceil\log _2(\frac{\ell}{5.4})\rceil$.

\item Execute $\langle [a'], [2^{-\lfloor \frac{w}{2} \rfloor}], [c] \rangle
  \leftarrow {\sf Norm}([a])$. 

\item Let $\alpha = -0.8099868542$, $\beta = 1.787727479$ and compute
  $[b'_0] = \alpha \cdot [a'] + \beta$. 
  
\item Compute $[b_0] = \left(\left([c] \wedge \frac{1}{\sqrt{2}}\right)
  \oplus \neg[c]\right) \cdot [2^{-\lfloor \frac{w}{2} \rfloor}] \cdot
  [b'_0]$. 

\item Compute $[g_0] = [a] \cdot [b_0]$ and $[h_0] = ([b_0] \gg 1)$.

\item For $i = 0, \ldots, \xi - 2$ do
\begin{compactenum}
\item $[x] = 1.5 - [g_i] \cdot [h_i]$.

\item if $(i < \xi-2)$ then $[g_{i+1}] = [g_i] \cdot [x]$.

\item $[h_{i+1}] = [h_i] \cdot [x]$.
\end{compactenum}

\item Compute $[h_{\xi}] = [h_{\xi-1}](1.5 - ([a] \cdot
  [h_{\xi-1}]^2 \gg 1))$.

\item Return $[b]=[h_{\xi}]$.
\end{compactenum}
\ifFull \line(1,0){470} \else \line(1,0){240} \fi
}

\noindent
Here multiplication by 0.5 is replaced by shift to the right by 1 bit that
has no cost. Complexity of this protocol is in Table~\ref{tab2-app}.

\subsection{Security Model and Security Proofs}
\label{app:def}

Security of a multi-party protocol is defined in the semi-honest setting as follows.

\begin{definition} \label{def:security}
    Let parties $P_1, {\ldots}, P_n$ engage in a protocol $\Pi$ that
    computes function $f({\sf in}_1, {\ldots}, {\sf in}_n) = ({\sf out}_1,
    {\ldots}, {\sf out}_n)$, where ${\sf in}_i$ and ${\sf out}_i$ denote
    the input and output of party $P_i$, respectively. Let
    $\mathrm{VIEW}_\Pi(P_i)$ denote the view of participant $P_i$ during the
    execution of protocol $\Pi$. More precisely, $P_i$'s view is formed by
    its input and internal random coin tosses $r_i$, as well as messages
    $m_1, {\ldots}, m_k$ passed between the parties during protocol
    execution: $\mathrm{VIEW}_{\Pi}(P_i) = ({\sf in}_i, r_i, m_1, {\ldots},
    m_k).$ Let $I = \{P_{i_1}, P_{i_2}, {\ldots}, P_{i_{\tau}}\}$ denote
    a subset of the participants for $\tau < n$ and $\mathrm{VIEW}_\Pi(I)$
    denote the combined view of participants in $I$ during the execution of
    protocol $\Pi$ (i.e., $\mathrm{VIEW}_\Pi(I) =
    (\mathrm{VIEW}_\Pi(P_{i_1}), {\ldots}, \mathrm{VIEW}_\Pi(P_{i_\tau}))$)
    and $f_I({\sf in}_1, {\ldots}, {\sf in}_n)$ denote the projection of
    $f({\sf in}_1, {\ldots}, {\sf in}_n)$ on the coordinates in $I$ (i.e.,
    $f_I({\sf in}_1, {\ldots}, {\sf in}_n)$ consists of the $i_1$th,
    {\ldots}, $i_{\tau}$th element that $f({\sf in}_1, {\ldots}, {\sf
    in}_n)$ outputs). We say that protocol $\Pi$ is $\tau$-private in the
    presence of semi-honest adversaries if for each coalition of size at
    most $\tau$ there exists a probabilistic polynomial time simulator $S_I$
    such that $\{S_I({\sf in}_I, f_I({\sf in}_1, {\ldots}, {\sf in}_n)),
    f({\sf in}_1, {\ldots}, {\sf in}_n)\} \equiv \{\mathrm{VIEW}_\Pi(I),
    ({\sf out}_1, {\ldots}, {\sf out}_n)\},$ where ${\sf in}_I =
    \bigcup_{P_i \in I} \{{\sf in}_i\}$ and $\equiv$ denotes computational
    or statistical indistinguishability.
\end{definition}

\ifFull
\appendix
\section{Security Proofs}
\label{sec:def}
\else 
\fi
{\color{black}
\begin{theorem}Protocol $\sf Arctan$ is secure according to
Definition~\ref{def:security}.
\end{theorem}
\textbf{Proof.} For concreteness of the proof, let us assume a SS setting
with $n$ computational parties from which at most $\tau < \frac{n}{2}$ are
corrupt. In the ideal world, $S_I$ simulates the view of the corrupted
parties $I$. It is given access to the corrupted parties' inputs (${\sf
in}_I$) and their outputs ($f_I({\sf in}_1, {\ldots}, {\sf in}_n)$). In the
current formulation of $\sf Arctan$, the input $a$ is provided in the form
of secret shares and thus is not available to a participant in the clear and
similarly output $b$ is computed in a shared form and is not given to any
given party. We, however, also consider the variant when the output is
opened to a corrupt participant from $I$, in which case the simulator
obtains access to $b$ and must ensure that the adversary is able to
reconstruct the correct output $b$. 

We build $S_I$ as follows: during step 1 of $\sf Arctan$, it invokes the
simulator for $\sf LT$. In step 2, $S_I$ invokes the simulator for (integer)
multiplication used in executing the conditional statements (note that
subtraction is local). Similarly, in step 3, $S_I$ invokes the simulator for
$\sf LT$. In step 4, $S_I$ invokes the simulator for $\sf Div$ and the
simulator for (integer) multiplication several times to simulate conditional
execution. In step 7, $S_I$ invokes the simulators for (floating-point)
prefix multiplication. In step 8, $S_I$ invokes the simulator for floating-point
multiplication $N$ times. Step 9 is simulated in the
same way as step 2. Similarly, in step 10, simulation of integer
multiplications is used. If no party in $I$ is entitled to learning output
$b$, $S_I$ simply communicates random shares to the
corrupted parties. Otherwise, $S_I$ has $b$ and uses it to set shares of 
honest parties' output in such a way that the shares reconstruct to the
correct output $b$ (honest parties' shares will be consequently communicated
to the output recipient). This is always possible because $\tau$ or fewer
shares are independent of a secret and the remaining shares can be set to
enable reconstruction of any desired value.}\fi

\end{document}